\documentclass{jfm}
\usepackage{graphicx}
\usepackage{caption, subcaption}
\usepackage{dcolumn}
\usepackage{epstopdf, epsfig}
\usepackage{bm}
\usepackage{multirow}
\usepackage{scalerel,stackengine}
\usepackage{booktabs}
\usepackage{natbib}
\usepackage[normalem]{ulem}
\usepackage{array} 
\usepackage[dvipsnames]{xcolor}
\usepackage{setspace}

\newcommand{\mybar}[2][3]{{}\mkern#1mu\overline{\mkern-#1mu#2}}
\newcommand{\myhat}[2][3]{{}\mkern#1mu\widehat{\mkern-#1mu#2}}

\newcommand{\mybarb}[2][3]{{}\mkern4mu\overline{\mkern-#1mu#2}}
\newcommand{\myhatb}[2][3]{{}\mkern4mu\widehat{\mkern-#1mu#2}}

\newcommand\mybarhat[1]{\ThisStyle{%
		\setbox0=\hbox{$\SavedStyle\mybarb{#1}$}%
		\ht0=\dimexpr\ht0-.15ex\relax
		\myhatb{\copy0}%
}}

\DeclareRobustCommand\mybarhato[1]{\ThisStyle{%
		\setbox0=\hbox{$\SavedStyle\overline{#1}$}%
		\ht0=\dimexpr\ht0-.15ex\relax
		\widehat{\copy0}%
}}

\newcommand{\codename}[1]{\texttt{#1}}
\newcommand*{\xhat}[1]{#1\kern-0.65em\hat{\phantom{#1}}}
\graphicspath{ {./figs/} }
\usepackage{tikz}
\DeclareRobustCommand\sampleline[1]{%
	\tikz\draw[#1][line width=0.5mm] (0,0) (0,\the\dimexpr\fontdimen22\textfont2\relax)
	-- (2em,\the\dimexpr\fontdimen22\textfont2\relax);%
}

\stackMath
\newcommand\reallywidehat[1]{%
	\savestack{\tmpbox}{\stretchto{%
			\scaleto{%
				\scalerel*[\widthof{\ensuremath{#1}}]{\kern-.6pt\bigwedge\kern-.6pt}%
				{\rule[-\textheight/2]{1ex}{\textheight}}
			}{\textheight}%
		}{0.5ex}}%
	\stackon[1pt]{#1}{\tmpbox}%
}
\title{Neural-network-based mixed subgrid-scale model for turbulent flow}

\author{Myeongseok Kang, Youngmin Jeon$\ddagger$
 \and Donghyun You\corresp{\email{dhyou@postech.ac.kr}}}

\affiliation{Department of Mechanical Engineering, Pohang University of Science and Technology, \\77 Cheongam-ro, Nam-gu, Pohang, Gyeongbuk 37673, Republic of Korea}

\begin{document}

\maketitle
\footnotetext{$\ddagger$ Co-first author.}

\begin{abstract}
An artificial neural-network-based subgrid-scale model, which is capable of predicting turbulent flows at untrained Reynolds numbers and on untrained grid resolution is developed.
Providing the grid-scale strain-rate tensor alone as an input leads the model to predict a subgrid-scale stress tensor aligns with the strain-rate tensor, and the model performs similarly to the dynamic Smagorinsky model. On the other hand, providing the resolved stress tensor as an input in addition to the strain-rate tensor is found to significantly improve the prediction of the subgrid-scale stress and dissipation, thereby the accuracy and stability of the solution. In an attempt to apply the neural-network-based model trained for turbulent flows with a limited range of the Reynolds number and grid resolution to turbulent flows at untrained conditions on untrained grid resolution, special attention is given to the normalisation of the input and output tensors. It is found that successful generalisation of the model to turbulence for various untrained conditions and resolution is possible if distributions of the normalised inputs and outputs of the neural-network remain unchanged as the Reynolds number and grid resolution vary.
In \emph{a posteriori} tests of the forced and the decaying homogeneous isotropic turbulence and turbulent channel flows, the developed neural-network model is found to predict turbulence statistics more accurately, maintain the numerical stability without \emph{ad-hoc} stabilisation such as clipping of the excessive backscatter, and to be computationally more efficient than the algebraic dynamic SGS models. 
\end{abstract}

\begin{keywords}
Authors should not enter keywords on the manuscript, as these must be chosen by the author during the online submission process and will then be added during the typesetting process (see http://journals.cambridge.org/data/\linebreak[3]relatedlink/jfm-\linebreak[3]keywords.pdf for the full list)
\end{keywords}

\section{Introduction}
\label{sec:one}
Large-eddy simulation (LES) is a technique in which large-scale turbulent motions are directly resolved on the computational grid while the effect of filtered subgrid-scale (SGS) turbulent motions are modelled using an SGS model. In order to model the dissipative nature of the SGS stress tensor, which is known to be the essential component of the SGS modelling~\citep{liu1994}, eddy-viscosity SGS models are often utilised~\citep{smagorinsky1963,lilly67,germano91,vreman2004sgs,silvis2017}. In eddy-viscosity SGS models, the deviatoric part of the SGS stress tensor $\tau_{ij}$ is modelled as an algebraic function of the resolved strain-rate tensor $\mybar{S}_{ij}$ such that $\tau_{ij} - \frac{1}{3}\tau_{kk}\delta_{ij} = - 2\nu_t \mybar{S}_{ij}$, where $\overline{\left(\,\cdot\,\right)}$ denotes the spatial filter operation, $S_{ij}$ is the strain-rate tensor, $\delta_{ij}$ is the Kronecker delta, and $\nu_t$ denotes the eddy viscosity~\citep{silvis2017}. While $\nu_t$ is modelled as $C\Pi^g$, where $C$ and $\Pi^g$ are the model coefficient and the SGS kernel at the grid-filter level, respectively, the dynamic procedure developed by~\citet{germano91} enabled dynamic determination of $C$.

However, eddy-viscosity SGS models have the shortcoming that they correlate poorly with the true SGS stress tensor since eddy-viscosity models are aligned with $\mybar{S}_{ij}$~\citep{bardinaphd,anderson1999,dasilva2002} and are not able to predict backscatter to the resolved scales~\citep{zang1993}.
To alleviate the problems, various mixed SGS models that combine eddy-viscosity models with the modified Leonard term~\citep{zang1993,bardinaphd,germano86}, the resolved stress $L_{ij} (=\myhat{\mybar{u}_i\mybar{u}_j}-\myhat{\mybar{u}}_i\myhat{\mybar{u}}_j)$~\citep{anderson1999, liu1994}, and the Clark model~\citep{anderson1999} were proposed. The dynamic procedure~\citep{germano91} was also applied to one-parameter~\citep{zang1993} and two-parameter~\citep{liu1994,salvetti1995,anderson1999} mixed models, where the \emph{parameter} refers to the model coefficients.
Although dynamic mixed models are reported to exhibit smaller fluctuations in model coefficients, dynamic eddy-viscosity models and dynamic mixed models require \emph{ad-hoc} procedures such as averaging of model coefficients in statistically homogeneous directions and clipping of negative model coefficients~\citep{zang1993,salvetti1995} to avoid numerical instability. Despite this issue, dynamic eddy-viscosity and dynamic mixed models have been successfully applied to various turbulent flows~\citep{germano91,vreman2004sgs,silvis2017,zang1993,salvetti1995,anderson1999}.

Recently, there have been studies to develop SGS models using artificial-neural-network (ANN)~\citep{xie2020nonlinear, xie2020spatial, yuan2020, xie2019comp, xie2019sgsforce, wang2018}. 
An ANN constructs a nonlinear mapping between a set of resolved flow variables and unresolved SGS stress using a series of matrix multiplications and nonlinear activation functions, while the conventional models represents the SGS stress in an algebraic function of resolved flow variables.
As a result, ANN-based models are often expected to provide more accurate flow description than algebraic dynamic SGS models~\citep{xie2020spatial, xie2019comp, xie2019sgsforce}. For instance, \citet{xie2020spatial} showed from an \emph{a posteriori} test of forced homogeneous isotropic turbulence that their ANN-based model predicted the energy spectrum and probability density functions (PDFs) of the vorticity and velocity increment more accurately than the dynamic Smagorinsky model (DSM) and the dynamic mixed model. 
ANN-based SGS models are also often known to be free from \emph{ad-hoc} procedures such as averaging and clipping of model coefficients for a certain set of input variables, which will be discussed later~\citep{xie2020nonlinear, xie2020spatial, yuan2020, xie2019comp, xie2019sgsforce, wang2018}. 
Therefore, ANN-based SGS models have the potential to be not only free from the requirement of a statistically homogeneous direction but also provide improved prediction of the SGS stress.

Despite such potentially favourable features, ANN-based SGS models raise three issues. The first issue is that although the performance of ANN-based models is known to be sensitive to the types of variables and the number of data points used for inputs, there is still no general consensus on which input variables and how many data points for each input variable are appropriate~\citep{wang2018,xie2020spatial,park2021,xie2019comp,xie2019sgsforce}. \citet{wang2018} performed an \emph{a priori} study of decaying homogeneous isotropic turbulence to test the utility of single-point input variables such as the filtered velocity vector $\mybar{u}_i$, the velocity gradient tensor $\partial\mybar{u}_i/\partial x_j$, and the second-order velocity derivative $\partial^2\mybar{u}_i/(\partial x_j \partial x_k)$. They found that $\partial\mybar{u}_i/\partial x_j$ and $\mybar{u}_i$ were the most and least important, respectively, for predictive accuracy. \citet{xie2020spatial} found that the multi-point input $\partial\mybar{u}_i/\partial x_j$ led to higher correlation coefficients between the true and the predicted SGS stresses than the single-point input $\partial\mybar{u}_i/\partial x_j$ from an \emph{a priori} study of forced homogeneous isotropic turbulence. 

\citet{park2021}, on the other hand, tested both single-point and multi-point resolved strain-rate tensors $\mybar{S}_{ij}$ and $\partial\mybar{u}_i/\partial x_j$ as inputs for ANN-based LES of turbulent channel flow and found from an \emph{a posteriori} test that the use of multi-point inputs required backscatter-clipping due to numerical instabilities while the use of single-point $\mybar{S}_{ij}$ resulted in the best performance. These are interesting results because, firstly, the use of a single-point input was essential for numerical stability, at least in case of near-wall turbulence; secondly, the elaborate selection of input variables such as $\mybar{S}_{ij}$ led to better performance of ANN-based SGS models than the use of general input variables such as $\mybar{u}_i$ or $\partial\mybar{u}_i/\partial x_j$. 

Based on the knowledge from the algebraic SGS modelling in the framework of mixed modelling and from the recent reports on the selection of input variables and the number of input points, it is expected that an improved ANN-based SGS model can be designed. For example, the drawback associated with misalignment of the SGS stress tensor can be overcome in a numerically stable manner by forming an ANN-based mixed model which utilises both the resolved strain-rate tensor and the resolved stress $L_{ij} (=\myhat{\mybar{u}_i\mybar{u}_j}-\myhat{\mybar{u}}_i\myhat{\mybar{u}}_j)$~\citep{liu1994, anderson1999} as inputs at a single point. 

The second issue for ANN-based models is the generalisation to untrained flow conditions, untrained grid resolution, and untrained types of flow. A few research has reported application of an ANN-based model which was trained for a certain flow at low Reynolds numbers to the same flow at  untrained higher Reynolds numbers~\citep{maulik2018,maulik2019,park2021}. However, generalisation to flow on untrained grid resolution has not been successful. \citet{park2021} reported that an ANN-based model trained at a certain grid resolution did not accurately predict the turbulent statistics on untrained coarser or finer resolution, while it was possible to predict flow on untrained grid resolution when the network was trained on coarser and finer resolution than the target resolution.  Similarly, \citet{zhou2019} reported that predicting turbulent flow on different grid resolution was difficult using an ANN-based model trained on other than the target grid resolution.

Most studies on ANN-based models addressed application of an ANN trained for a particular type of flow to the same type of flow. Although \citet{xie2020spatial,xie2020nonlinear} briefly discussed application of ANN-based models trained with forced homogeneous isotropic turbulence to weakly compressible homogeneous shear flow, both flows are statistically stationary at the same Reynolds number and on the same grid resolution. Therefore, it is important to identify conditions of inputs and outputs of an ANN to achieve better generalisation to flow on different grid resolution and eventually to a different type of flow from flows trained.

The last issue is that the computational cost of ANN-based models has been reported to be higher than that of the algebraic dynamic eddy-viscosity and the dynamic mixed models. 
\citet{park2021}, \citet{wang2018}, \citet{yuan2020}, \citet{xie2019comp}, and \citet{xie2020spatial} reported $1.3$, $1.8$, $2.4$, $15$, and $256$ times higher computational cost than that of the dynamic Smagorinsky model for simulations of turbulent channel flow~\citep{park2021} and homogeneous isotropic turbulence~\citep{wang2018,yuan2020,xie2019comp,xie2020spatial}, respectively.  \citet{yuan2020}, \citet{xie2019comp}, and \citet{xie2020spatial} also reported that the computational cost of their ANN-based models is higher than that of the dynamic mixed model~\citep{anderson1999}. 
To make ANN-based SGS models be practical, both accuracy and computational efficiency which are superior to those of the conventional algebraic SGS models should be achieved.

The primary objective of the present study is to develop a new ANN-based mixed SGS model which predicts the SGS turbulence more accurately and stably. Analogously to the philosophy of the algebraic mixed SGS models, the present model is designed to consider both the resolved stress and strain-rate tensors as inputs and thereby produce the SGS stress associated with the both inputs as an output. The new model is also designed to be applicable to turbulent flow at untrained Reynolds numbers and on untrained grid resolution, and, at the same time, to be computationally more efficient than the conventional algebraic SGS models. 
As noted by~\citet{park2021}, the numerical stability is sought by conducting the input-output data sampling on a single grid point. The single-point data sampling is also beneficial in minimizing the computational cost. 
To achieve the goal, extensive analyses for finding optimal input-output scalings are conducted through \emph{a-priori} and \emph {a-posteriori} tests of homogeneous isotropic turbulence. 
The predictive capability, accuracy, and stability of the ANN-based mixed SGS model for LES of homogeneous isotropic turbulence at untrained Reynolds numbers as well as on untrained grid resolution are evaluated in detail. 
The applicability to other type of turbulent flow, especially the wall-bounded turbulent flow is investigated by performing LES of turbulent channel flow. 

This paper is organised as follows: numerical methods for direct numerical simulation (DNS), LES, and ANN are presented in Section~\ref{sec:Method}. In Section~\ref{sec:Results}, the performance and the characteristics of ANN-based SGS models with different input sets are discussed based on results of \emph{a-priori} and \emph{a-posteriori} tests of forced isotropic turbulence. Application of the present ANN-based mixed SGS model to decaying isotropic turbulence at untrained Reynolds numbers and on untrained grid resolution and untrained turbulent channel flow are presented in Section~\ref{sec:Results}. Comparison of the predictive capability as well as the computational cost of the developed ANN-based mixed SGS model with those of algebraic dynamic mixed models are also discussed in Section~\ref{sec:Results}, followed by concluding remarks in Section~\ref{sec:Conclusion}.

\section{Numerical methods}\label{sec:Method}

\subsection{DNS of forced homogeneous isotropic turbulence}\label{sec:DNS}
The incompressible Navier--Stokes equations for DNS are as follows:
\begin{equation}
	\frac{\partial{u_{i}}}{\partial{x_{i}}} = 0,
	\label{eq:INS_cont}
\end{equation}
\begin{equation}
	\frac{\partial{u_{i}}}{\partial{t}} + \frac{\partial{u_{i}u_{j}}}{\partial{x_{j}}} = -\frac{1}{\rho}\frac{\partial{p}}{\partial{x_{i}}} + \nu\frac{\partial^2{u_{i}}}{\partial{x_{j}}\partial{x_{j}}},
	\label{eq:INS_mom}
\end{equation}
where $x_{i}$ $(=x,y,z)$ are the Cartesian coordinates, $u_{i}$ $(=u,v,w)$ are the corresponding velocity components, $t$ is time, $p$ is pressure, $\rho$ is density, and $\nu$ is kinematic viscosity.

Forced homogeneous isotropic turbulence at $\Rey_{\lambda}=106$, $164$, and $286$, which were simulated by~\citet{mohan2017}, \citet{langford1999}, and \citet{chumakov2008}, respectively, are simulated in this study with the computational parameters shown in table~\ref{tab:table_dns}. A pseudospectral code \codename{HIT3D}~\citep{chumakov2008,chumakov2007} is used to solve the incompressible Navier--Stokes equations in a periodic cubic box with the length of $2\pi$ with $N$ grid points in each direction. A combination of spherical truncation and phase shifting~\citep{canuto88} is used for dealiasing. The second-order Adams--Bashforth scheme is used for time-integration~\citep{kangphd}.

The initial flow field of the Gaussian distribution and random phases is fully developed, and then instantaneous fields are sampled at every eddy turnover time. The number of samples $N_S$ for each case is summarised in table~\ref{tab:table_dns}. Validation of the present DNS is discussed in the following section.

\begin{table}
	\begin{center}
		\def~{\hphantom{0}}
		\begin{tabular}{cp{0.3cm}cp{0.3cm}cp{0.3cm}cp{0.3cm}cp{0.3cm}cp{0.3cm}cp{0.3cm}c}
			$\Rey_{\lambda}$&&
			$N$&&
			$\nu$&&
			$\varepsilon_{p}$&&
			$k_f$&&
			$\eta$&&
			$\eta k_{max}$&&
			$N_S$\\[3pt]
			106 && 192 && 1/263 && 1.0 && 2.0 && $1.5\times10^{-2}$ && 1.4 && 37\\[2pt]
			164 && 256 && 1/150 && 62.9 && 3.0 && $8.4\times10^{-3}$ && 1.0 && 50\\[2pt]
			286 && 512 && 1/2225 && 0.12 && 1.5 && $5.3\times10^{-3}$ && 1.3 && 50\\
		\end{tabular}
		\caption{Parameters for DNS of forced homogeneous isotropic turbulence. $\Rey_{\lambda}$ is the Taylor-scale Reynolds number, $N$ is the number of grid points in each direction, $\nu$ is the viscosity, $\varepsilon_{p}$ is the prescribed dissipation rate, $k_f$ is the upper bound of the forcing wavenumber, $\eta$ is the Kolmogorov length scale, and $k_{max}$ is the maximum resolved wavenumber.}
		\label{tab:table_dns}
	\end{center}
\end{table}

\subsection{Validation of DNS results}\label{sec:validation}

\citet{mohan2017} and \citet{langford1999} used a negative viscosity forcing~\citep{jimenez93,mohan2017} for forced homogeneous isotropic turbulence at $\Rey_{\lambda}=106$ and $164$, respectively, and \citet{chumakov2008} used a deterministic forcing~\citep{machiels1997} for forced homogeneous isotropic turbulence at $\Rey_{\lambda}=286$. The negative viscosity forcing is given as follows:
\begin{equation}
	\myhat{f}^V_{i}(\textbf{k},t) = \left\{
	\begin{array}{ll}
		\varepsilon_{p} \myhat{u}_{i}|\textbf{k}|^2(\textbf{k},t)/\left[2E^V_f(t)\right], & \quad 0 < k \le k_f ,\\[2pt]
		0,         & \quad k_f < k,
	\end{array} \right.
	\label{eq:forcing_vis}
\end{equation}
where $\varepsilon_{p}$ denotes the prescribed mean dissipation rate, $k$ is the spherical wavenumber defined as $k=|\textbf{k}|$, $k_f$ is the upper bound of the forcing wavenumber, and $E^V_f(t)=\int_{0}^{k_f} |\textbf{k}|^2 E(k,t)dk$, where $E(k,t)$ is the energy spectrum at time $t$. The combined forcing and viscous term in the Fourier space becomes $-(\nu-\alpha)|\textbf{k}|^2\myhat{u}_{i}(\textbf{k})$, where $\alpha = \varepsilon_{p}/\left[2E^V_f(t)\right]$. The deterministic forcing of \citet{machiels1997} is given as follows:
\begin{equation}
	\myhat{f}^M_{i}(\textbf{k},t) = \left\{
	\begin{array}{ll}
		\varepsilon_{p} \myhat{u}_{i}(\textbf{k},t)/\left[2E^M_f(t)\right],      & \quad 0 < k \le k_f ,\\[2pt]
		0,  & \quad k_f < k,
	\end{array} \right.
	\label{eq:forcing}
\end{equation}
where $E^M_f(t)=\int_{0}^{k_f} E(k,t)dk$. 

\begin{figure}
	\centering
	\begin{subfigure}[b]{0.328\textwidth}
		\includegraphics[width=\textwidth]{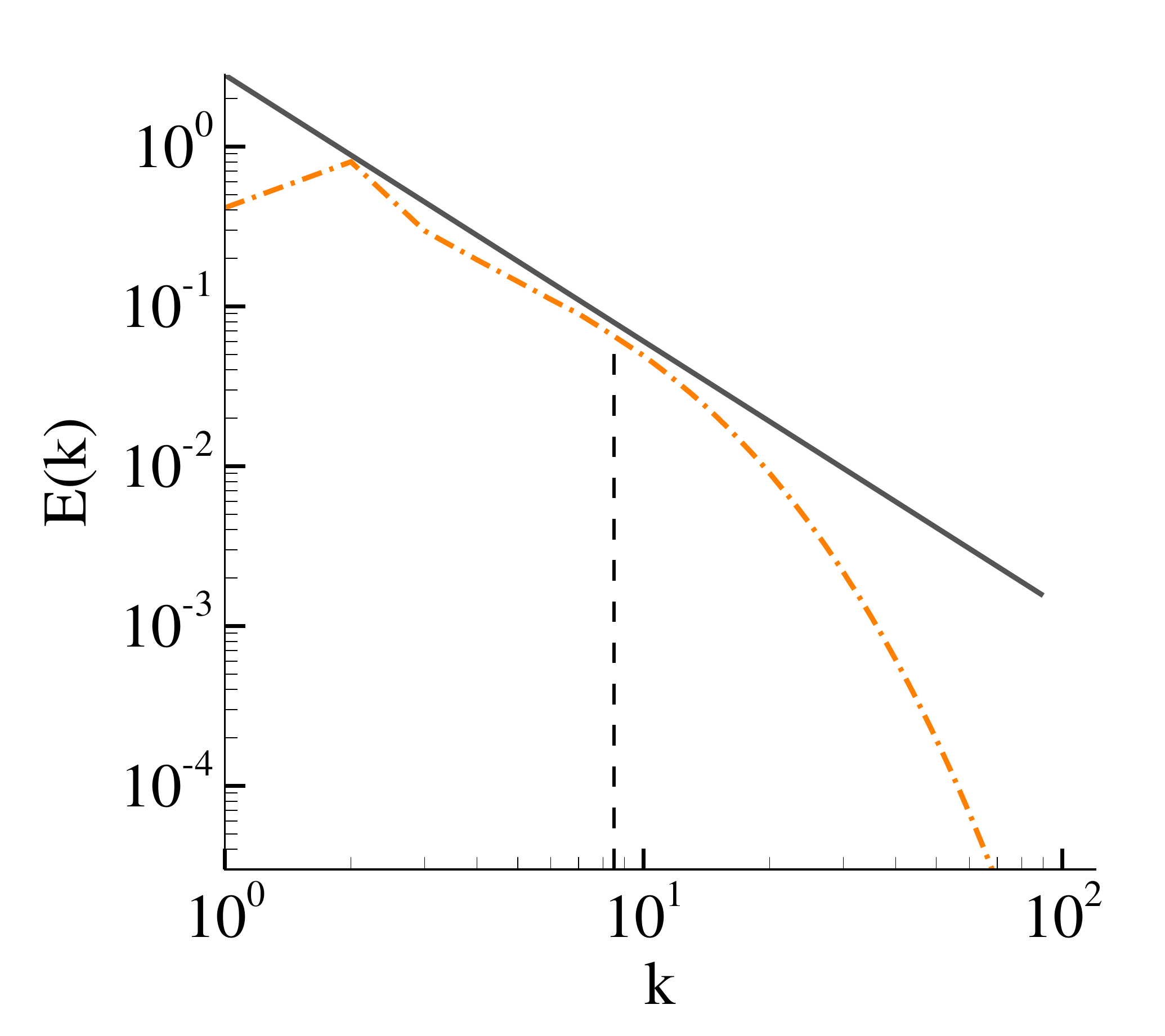}
		\caption{}
		\label{fig:FHIT_ek_a}
	\end{subfigure}
	\begin{subfigure}[b]{0.328\textwidth}
		\includegraphics[width=\textwidth]{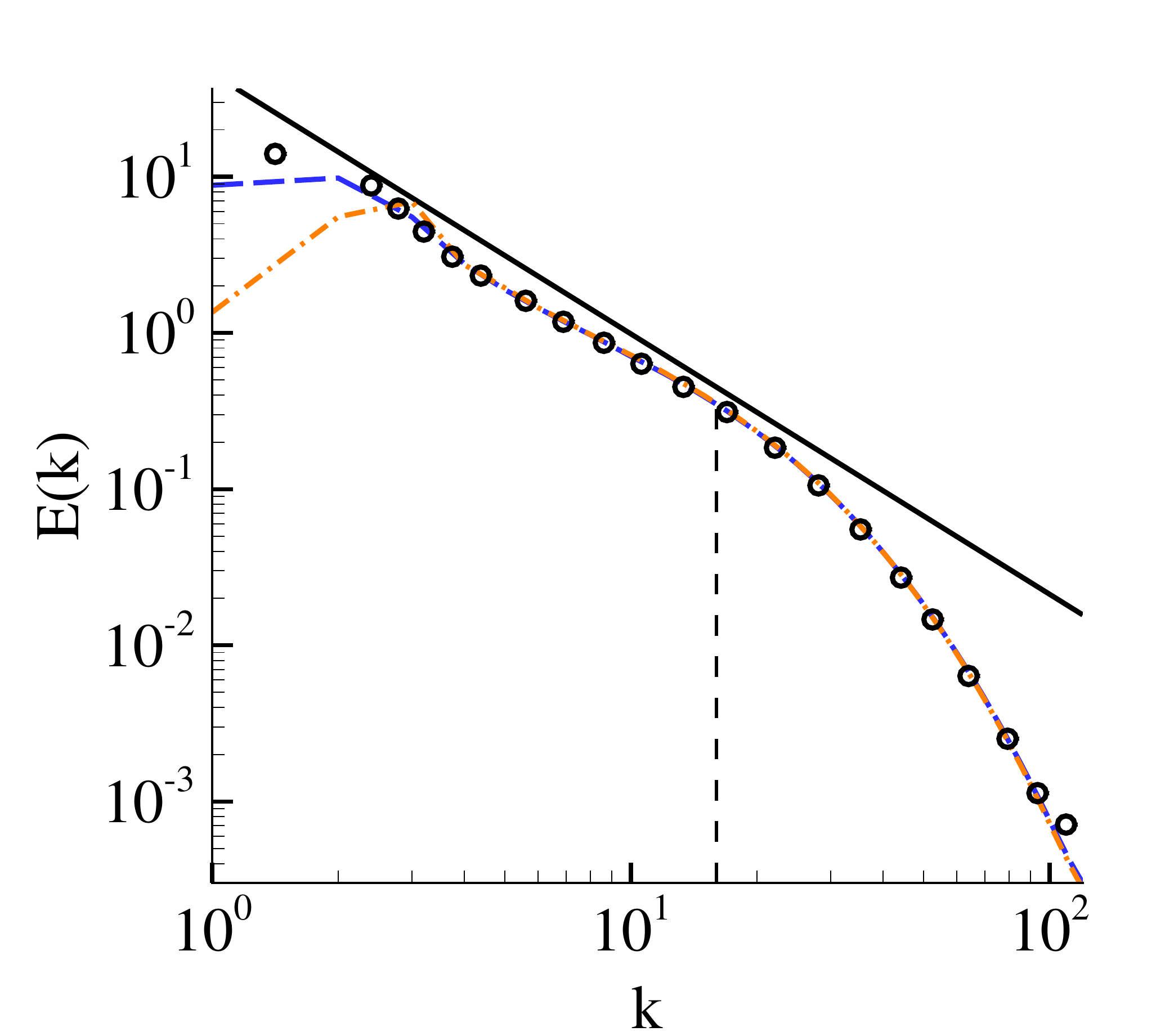}
		\caption{}
		\label{fig:FHIT_ek_b}
	\end{subfigure}
	\begin{subfigure}[b]{0.328\textwidth}
		\includegraphics[width=\textwidth]{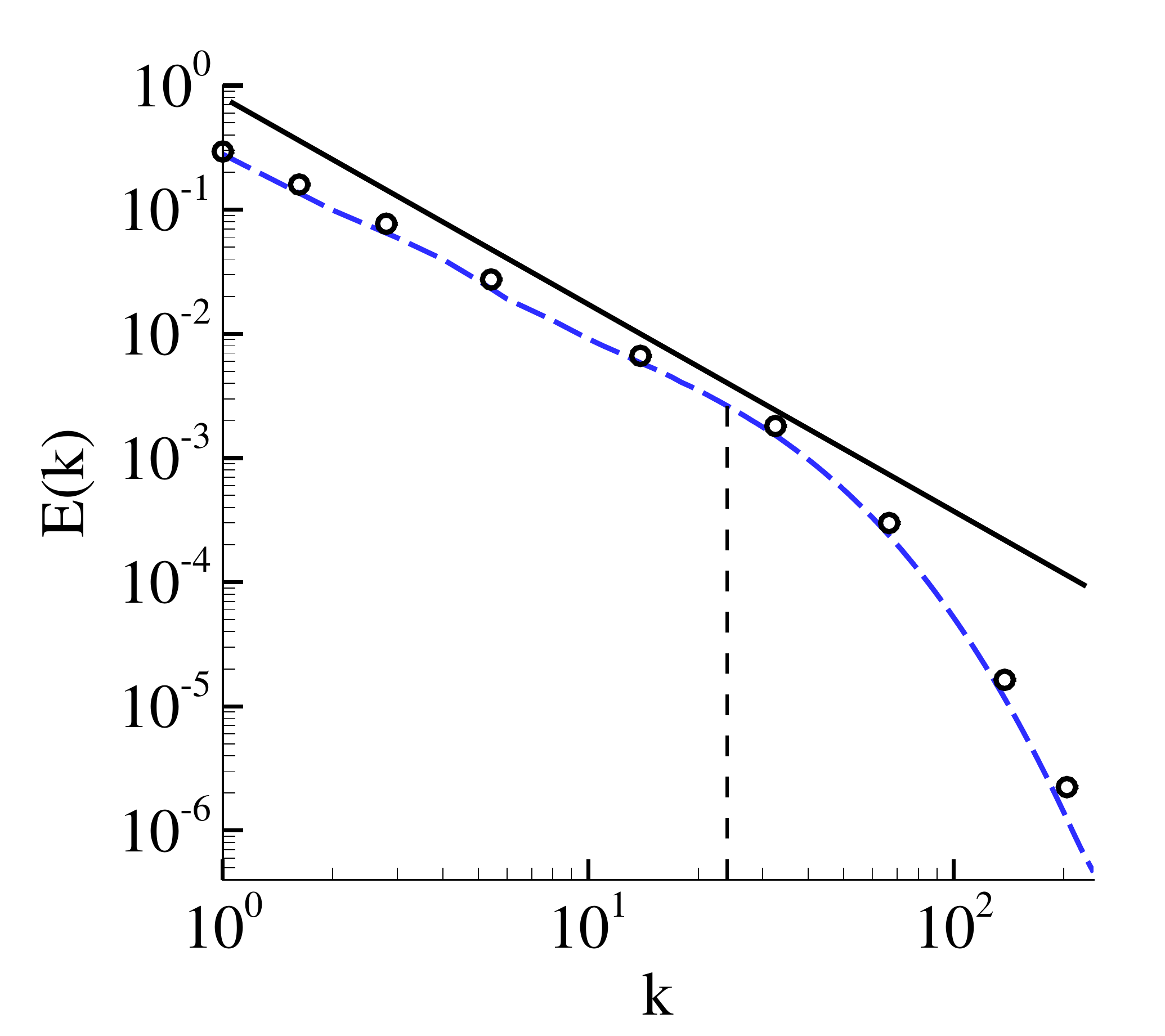}
		\caption{}
		\label{fig:FHIT_ek_c}
	\end{subfigure}
	\caption{Energy spectra from DNS of forced homogeneous isotropic turbulence at (a) $\Rey_{\lambda}=106$, (b) $\Rey_{\lambda}=164$, and (c) $\Rey_{\lambda}=286$. $\bm{\circ}$, DNS by (b) Langford and Moser (1999) and (c) Chumakov (2008); {\color{YellowOrange}\sampleline{dash pattern=on .7em off .2em on .2em off .2em}}, the present DNS with the negative viscosity forcing; {\color{blue}\sampleline{dash pattern=on 6pt off 2pt}}, the present DNS with the Machiels forcing; \sampleline{}, $k^{-5/3}$ line; Vertical dashed lines indicate the grid-filter cutoff wavenumbers in the inertial subrange in which \emph{a\ priori} tests are performed.}
	\label{fig:FHIT_ek}
\end{figure}

Three-dimensional energy spectra of forced homogeneous isotropic turbulence are compared with the reference DNS results in figure~\ref{fig:FHIT_ek}, except for the case at $\Rey_{\lambda}=106$ for which a reference energy spectrum is not available. The case at $\Rey_{\lambda}=164$ is firstly simulated using the negative viscosity forcing following \citet{langford1999}. The resulting energy spectrum shown in figure~\ref{fig:FHIT_ek}(b) exhibits an undershoot in the forcing wavenumber range. Another simulation at $\Rey_{\lambda}=164$ with the Machiels forcing, on the other hand, shows improved agreement with the reference DNS data in the forcing range. While the differences caused by the use of different forcing schemes are further discussed in Appendix~\ref{app:app1-1}, DNS data obtained using the Machiels forcing is used for analyses in the present study, as it better reproduces DNS results of \citet{langford1999}. For the case of $\Rey_{\lambda}=286$, slight underprediction of the energy spectrum is observed and it is discussed in Appendix~\ref{app:app1-2} to conclude that the present DNS is statistically well converged. 

\subsection{Large-eddy simulation and filter operations}\label{sec:filter}

By applying the spatial filter operation $\overline{\left(\,\cdot\,\right)}$ to (\ref{eq:INS_cont}) and (\ref{eq:INS_mom}), the filtered Navier--Stokes equations for LES are obtained as follows:
\begin{equation}
	\frac{\partial{\mybar{u}_{i}}}{\partial{x_{i}}} = 0,
	\label{eq:fNS_cont}
\end{equation}
\begin{equation}
	\frac{\partial{\mybar{u}_{i}}}{\partial{t}} + \frac{\partial{\mybar{u}_{i}\mybar{u}_{j}}}{\partial{x_{j}}} = -\frac{1}{\rho}\frac{\partial{\mybar{p}}}{\partial{x_{i}}} + \nu\frac{\partial^2{\mybar{u}_{i}}}{\partial{x_{j}}\partial{x_{j}}}-\frac{\partial{\tau_{ij}}}{\partial{x_{j}}},
	\label{eq:fNS_mom}
\end{equation}
where $x_{i}$ $(=x,y,z)$ are the Cartesian coordinates, $\mybar{u}_{i}$ $(=\mybar{u},\mybar{v},\mybar{w})$ are the corresponding filtered velocity components, $t$ is time, $\mybar{p}$ is filtered pressure, $\rho$ is density, $\nu$ is kinematic viscosity, and $\tau_{ij} = \mybar{u_{i}u_{j}}-\mybar{u}_{i}\mybar{u}_{j}$ is the SGS stress.

Low-pass spatial filter operations are defined at the grid- and test-filter levels. A grid-filter operation on a scalar variable $\mybar{\phi}$ is defined as follows:
\begin{equation}
	\mybar{\phi}\left(\textbf{x}\right) = \int_{\Omega} G\left(\textbf{x},\bm{\xi}\right) \phi\left(\bm{\xi}\right)d\bm{\xi},
	\label{eq:def_filter}
\end{equation}
where $\bm{x}$ and $\bm{\xi}$ are the spatial coordinate vectors in the flow domain of $\Omega$. The filter kernel $G$ satisfies the normalisation condition and depends on the filter width defined as $\mybar{\Delta}$ for the grid-filter. The filter width in three-dimensional space is calculated using the expression given by \citet{deardorff70}:
\begin{equation}
	\Delta=\sqrt[3]{\Delta_1 \Delta_2 \Delta_3},
	\label{eq:def_filter_width}
\end{equation} 
where $\Delta_i$ denotes the filter width in the $i$ direction.
In line with the definition of the grid-filter in (\ref{eq:def_filter}), a test-filter operator $\widehat{(\,\cdot\,)}$ is similarly defined but with the test-filter width $\widehat{\Delta}$ instead of $\mybar{\Delta}$.

The filter-to-grid ratio is defined as $\mybarhat{\Delta}/\mybar{\Delta}$, where $\mybarhat{\Delta}$ is the filter width associated with the filtering operator $\mybarhato{\left(\,\cdot\,\right)}$. Since the Gaussian filter is adopted for both grid- and test-filters, the filter-to-grid ratio is set to $\sqrt{5}$~\citep{pope2001}.

\subsection{Artificial neural-network for subgrid-scale modelling}\label{sec:Artificial neural network for subgrid-scale modelling}

\begin{figure}
\centering \
\includegraphics[width=0.85\textwidth]{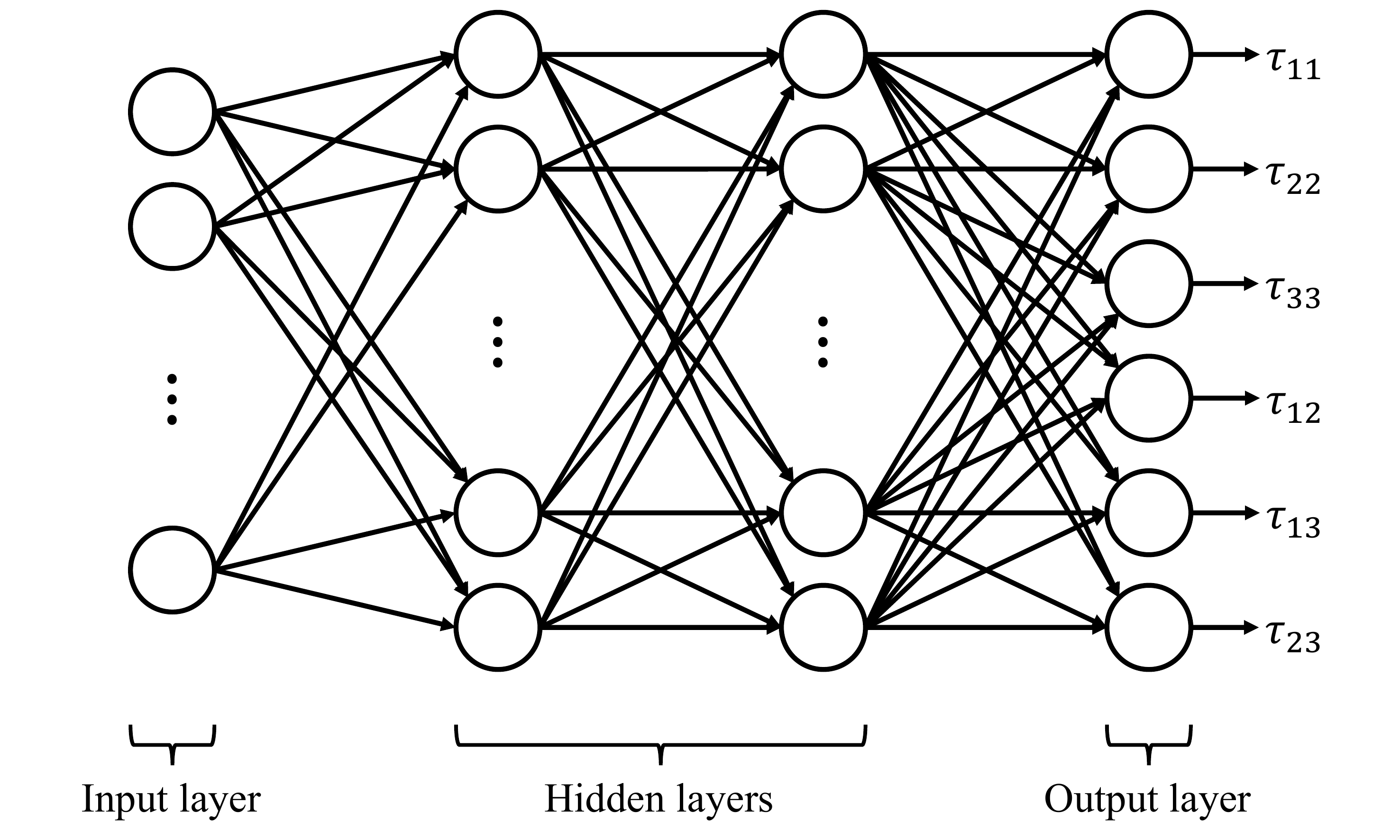}
\caption{Schematic diagram of the ANN to predict the SGS stress (2 hidden layers and 12 neurons per hidden layer).}
\label{fig:ANN}
\end{figure}

A fully connected neural-network, also known as a multi-layer perceptron (MLP), is adopted in the present study. Six components of the SGS stress tensor $\tau_{ij}$ are predicted by an ANN from a set of filtered input variables. The ANN is trained using input and output data from the filtered-DNS (fDNS) fields, which obtained by applying a spatial filter to the instantaneous fields from DNS of forced homogeneous isotropic turbulence in Section~\ref{sec:DNS}. The filter width of the fDNS dataset is the same as the grid size of LES with $N^3=48^3$ cells at $Re_{\lambda} = 106$, and $N^3=128^3$ cells at $Re_{\lambda} = 286$. Through sufficient training, the ANN constructs a nonlinear mapping between a set of inputs and a target SGS stress tensor using a series of linear matrix multiplications and nonlinear activation functions. The mathematical operation from the $(n-1)$-th layer to $n$-th layer takes the form of
\begin{equation}
	X_{i}^{n} = \sigma \left ( \sum_{j}^{} W_{ij}^{n} X_{j}^{n-1} + b_{i}^{n} \right),
	\label{eq:layer}
\end{equation}
where $\sigma\left (\cdot \right )$ is the activation function, and $W_{ij}^{n}$ and $b_{i}^{n}$ are weights and biases, respectively. The output layer is linearly activated as $X_{i}^{N} = \sum_{j}^{} W_{ij}^{N} X_{j}^{N-1} + b_{i}^{N}$, where $X_{i}^{N}$ is the output of the ANN. 

The present ANN, which is shown in figure~\ref{fig:ANN}, consists of an input layer, two hidden layers with 12 neurons per hidden layer, and an output layer. Since the number of trainable parameters of an ANN directly affects its computational cost, a parameter study on the number of neurons per hidden layer is conducted in Section~\ref{sec:DHIT_cost}. Based on the results in Section~\ref{sec:DHIT_cost}, it is found that an ANN with more than 12 neurons per hidden layer does not improve the performance of the ANN. A Leaky-ReLu activation function ($\sigma (x) = max[-0.02x, x] $) is applied at each hidden layer. Leaky-ReLu is known to perform better than ReLu~\citep{xu2015empirical}, as it is capable of resolving the gradient vanishing problem of the ReLu. During the training process, weights $W_{ij}^{n}$ and biases $b_{i}^{n}$ are optimised to minimise the mean-squared error loss function defined as
\begin{equation}
    \textit{L} = \frac{1}{N_{batch}} \frac{1}{6} \sum_{n=1}^{N_{batch}} \sum_{j=1}^{3} \sum_{i=1}^{j} \left ( \tau_{ij,n}^{fDNS} - \tau_{ij,n} \right )^2,
    \label{eq:loss_func}
\end{equation}
where $\tau_{ij}^{fDNS}$ and $\tau_{ij}$ are the true SGS stress obtained from fDNS and the predicted SGS stress, respectively. $N_{batch}$ represents the size of the mini-batch and is set to 128~\citep{park2021}. 

\begin{table}
  \begin{center}
\def~{\hphantom{0}}
  \begin{tabular}{lcccc}
      Model  & S-106H & SL-106H & SL-286H & SL-106+286H \\[3pt]
      \\
      Inputs & $\left\{\overline{S}_{ij}\right\}$ & $\left\{\overline{S}_{ij}, L_{ij}\right\}$ & $\left\{\overline{S}_{ij}, L_{ij}\right\}$ & $\left\{\overline{S}_{ij}, L_{ij}\right\}$ \\
      Outputs & {$\tau_{ij}$} & {$\tau_{ij}$} & {$\tau_{ij}$} & {$\tau_{ij}$} \\
      $Re_{\lambda}$ of training data & 106 & 106 & 286 & 106, 286 \\
  \end{tabular}
  \caption{Input variables and Reynolds numbers of training dataset for different ANN-SGS models}
  \label{tab:ANN}
  \end{center}
\end{table}

The sets of input and output variables and Reynolds numbers of the training datasets for ANN-based SGS models (hereafter ANN-SGS models) are listed in table~\ref{tab:ANN}. Each ANN-SGS model is named such that the character and the number represent the input variables and the Reynolds number at which the ANN is trained, respectively. 
The character H at the end of the model names denotes homogeneous isotropic turbulence. For example, S-106H uses six components of $\overline{S}_{ij}$ as inputs, whereas the other models use $\overline{S}_{ij}$ and the resolved stress $L_{ij}$ as inputs. Effects of input variables on the characteristics and the accuracy of ANN-SGS models is discussed in Section~\ref{sec:FHIT}. The ANN-SGS models are trained with fDNS datasets of forced homogeneous isotropic turbulence at the given Reynolds number.

\begin{figure}
	\centering
	\begin{subfigure}[b]{0.49\textwidth}
		\includegraphics[width=\textwidth]{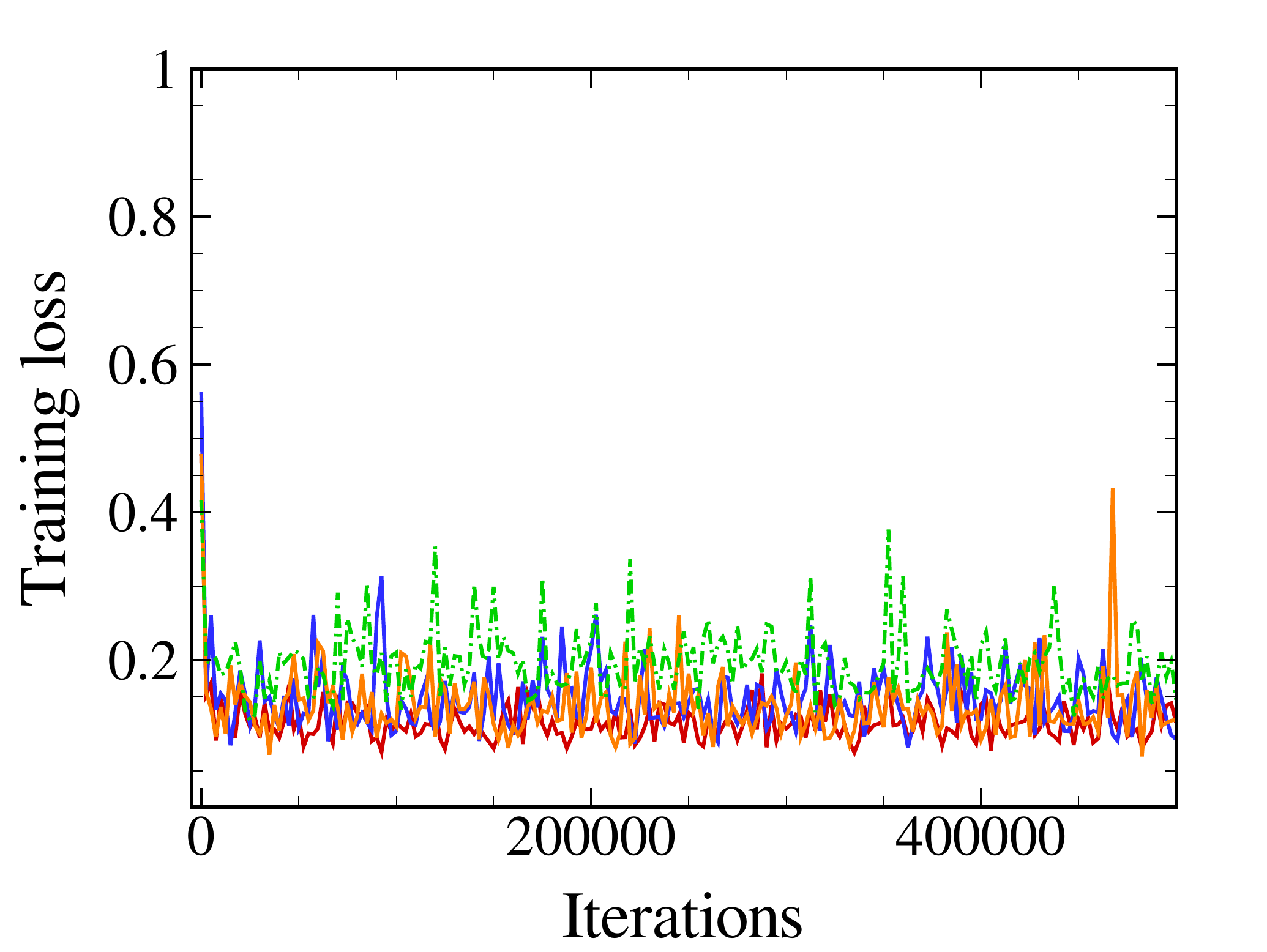}
		\caption{}
		\label{}
	\end{subfigure}
	\begin{subfigure}[b]{0.49\textwidth}
		\includegraphics[width=\textwidth]{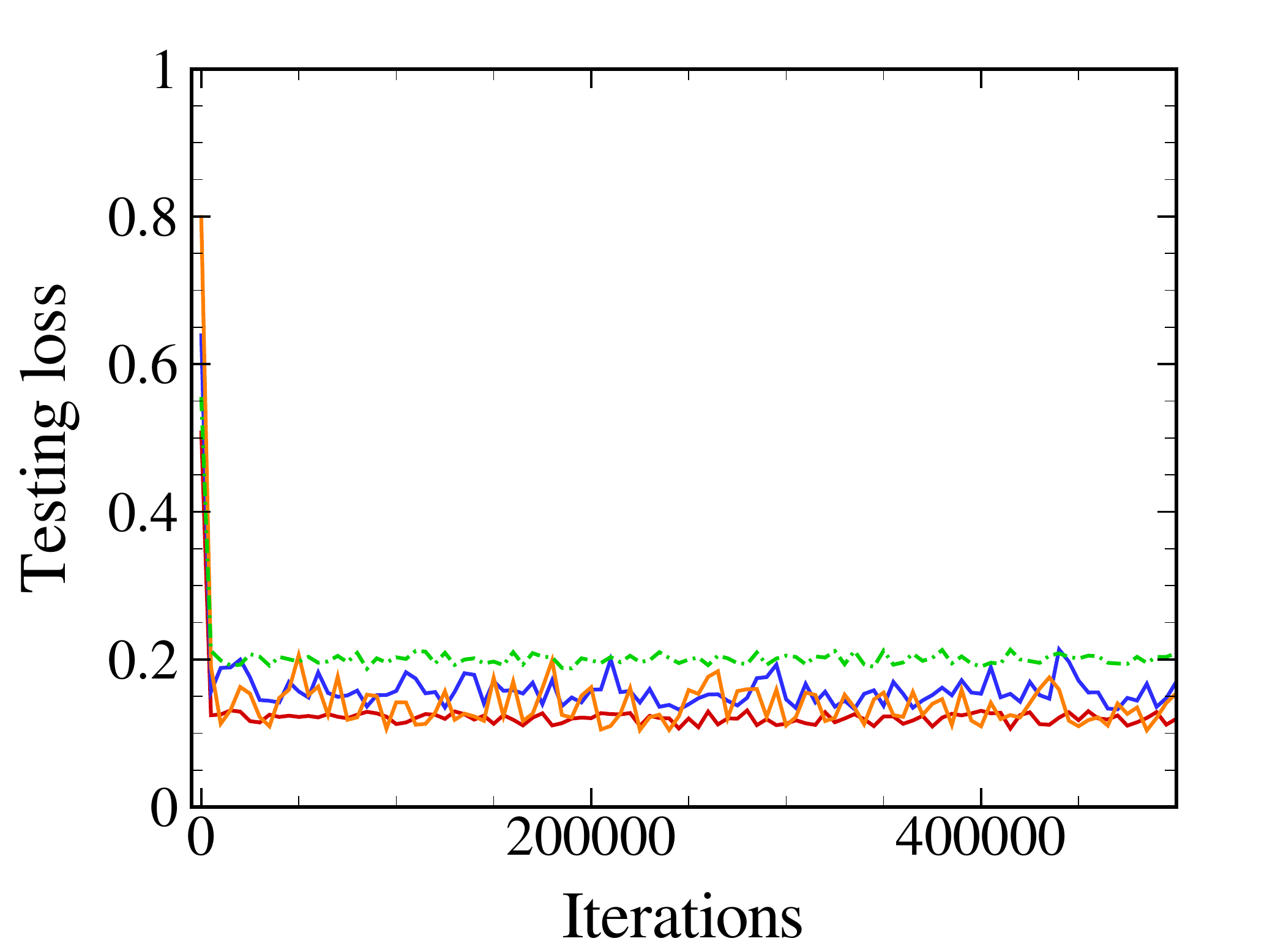}
		\caption{}
		\label{}
	\end{subfigure}
	\caption{Learning curve of ANN-SGS models. (a) Training loss and (b) testing loss. {\color{red}\sampleline{}}, SL-106H; {\color{blue}\sampleline{}}, SL-286H; {\color{YellowOrange}\sampleline{}}, SL-106+286H; {\color{green}\sampleline{dash pattern=on .7em off .2em on .2em off .2em}}, S-106H.}
    \label{fig:learning_curve}
\end{figure}

The Adam optimiser with a learning rate of $10^{-4}$ is utilised to optimise the trainable parameters. ANN-SGS models listed in table~\ref{tab:ANN} are trained for $5 \times 10^5$ iterations using Python library PyTorch~\citep{paszke2017}. A total of $2\times10^{8}$ data points sampled from 30 snapshots of fDNS results are used for the training, whereas the rest are used for testing (80$\%$ of the snapshots for training and 20$\%$ for testing). Figure~\ref{fig:learning_curve} shows the learning curve of ANN-SGS models. The training loss and the testing loss show similar values and converge to a stationary value, indicating that all ANN-SGS models are trained without overfitting.

The performance of an ANN is highly affected by the normalisation of the input and output variables~\citep{sola1997}. In the present study, the input and the output tensors are normalised as $\overline{S}_{ij}^* = \overline{S}_{ij}/\left< \left| \overline{S} \right|\right>, L_{ij}^* = L_{ij}/\left< \left| L \right|\right>$, and  $\tau_{ij}^* = \tau_{ij}/\left< \left| \tau \right|\right>$, where $\left< \cdot \right>$ and $\left| \cdot \right|$ denote a volume average and an $L_2$ norm of the tensor, respectively. However, since the denominator $\left< \left| \tau \right|\right>$ is unknown in actual LES, the normalisation factor of the SGS stress is estimated using the gradient model term $(\tau_{ij}^{grad} = \frac{1}{12}\mybar{\Delta}^2\frac{\partial \overline{u}_i}{\partial x_k}\frac{\partial \overline{u}_j}{\partial x_k})$; so that the normalisation of the SGS stress is performed as $\tau_{ij}^* = \tau_{ij}/\left< \left| \tau^{grad}\right|\right>$. Detailed discussion regarding the normalisation factor of the SGS stress, particularly in the context of generalisation to untrained conditions, is presented in Section~\ref{sec:DHIT}.

\section{Results and discussion}\label{sec:Results}

In Section~\ref{sec:FHIT}, the performance of S-106H is compared with that of DSM from \emph{a priori} and \emph{a posteriori} tests of forced isotropic turbulence at $Re_{\lambda} = 106$. Based on the above results, S-106H is improved by providing an additional input $L_{ij}$ (SL-106H). Accordingly, S-106H and SL-106H are compared in terms of the correlation coefficients, PDF of the SGS dissipation, PDF of the SGS stress, and the energy spectrum. 
In Section~\ref{sec:DHIT}, special attention is given to the normalisation of the input and the output tensors for generalisation of ANN-SGS models. Consequently, the generalisability of ANN-SGS models to transient flow is investigated from LESs of decaying isotropic turbulence at various untrained conditions (\emph{i.e.}, initial Reynolds numbers and grid resolution). 
In Sections~\ref{sec:DHIT_mixed} and~\ref{sec:DHIT_cost}, the performance and the computational cost of the developed ANN-SGS models are compared with those of conventional algebraic dynamic mixed models.
Finally, the application of the developed models to LES of turbulent channel flow is presented in Section~\ref{sec:channel}.

\subsection{Effects of input variables: a priori and a posteriori tests}\label{sec:FHIT}

Before conducting an actual LES, effects of input variables on ANN-SGS models are investigated in an \emph{a priori} test of forced homogeneous isotropic turbulence at $Re_{\lambda} = 106$. While the use of $\overline{S}_{ij}$ is expected to be preferable for homogeneous isotropic turbulence in terms of the accuracy and stability~\citep{park2021}, it is of interest to investigate how differently the predicted SGS stress aligns with respect to $\overline{S}_{ij}$. To examine the alignment, the correlation coefficients between $\overline{S}_{ij}$ and the predicted SGS stress ($\tau_{ij}^{ANN}$) are calculated and listed in table~\ref{tab:corr_tau_Sij}. The correlation coefficient between the components of arbitrary second-order tensors $\alpha_{ij}$ and $\beta_{ij}$ is defined as
\begin{equation}
    Corr(\alpha_{ij},\beta_{ij}) = \frac{\left< \alpha_{ij} \beta_{ij} \right>}{\left< \alpha_{ij}^2 \right>^{1/2} \left< \beta_{ij}^2 \right>^{1/2}},
    \label{eq:correlation}
\end{equation}
where $\left< \cdot \right>$ denotes the ensemble averaging. 
As shown in table~\ref{tab:corr_tau_Sij}, S-106H has much higher correlation coefficients between $\overline{S}_{ij}$ and $\tau_{ij}^{ANN}$ than fDNS, with the value excedding $-0.8$, which indicates that the predicted SGS stress is aligned closer to the strain-rate tensor rather than the true SGS stress. 

\begin{table}
  \begin{center}
\def~{\hphantom{0}}
  \begin{tabular}{lccccccc}
      $ $  & $\tau_{11}$ & $\tau_{22}$ & $\tau_{33}$ & $\tau_{12}$ & $\tau_{13}$ & $\tau_{23}$ \\[3pt]
       S-106H   & 0.8476 & 0.8465 & 0.8392 & 0.4175 & 0.3805 & 0.3652 \\
       SL-106H   & 0.9051 & 0.9103 & 0.9028 & 0.7195 & 0.7167 & 0.6941 \\
  \end{tabular}
  \caption{Correlation coefficients ($Corr(\tau_{ij}^{fDNS}, \tau_{ij}^{ANN})$) between the true SGS stress ($\tau_{ij}^{fDNS}$) and the predicted SGS stress by ANN-SGS models ($\tau_{ij}^{ANN}$) from \emph{a priori} test of forced homogeneous isotropic turbulence at $Re_{\lambda} = 106$.}
  \label{tab:corr_tau}
  \end{center}
\end{table}

\begin{table}
  \begin{center}
\def~{\hphantom{0}}
  \begin{tabular}{lccc}
      $ $  & $Corr(\overline{S}_{12}, \tau_{12})$ & $Corr(\overline{S}_{13}, \tau_{13})$ & $Corr(\overline{S}_{23}, \tau_{23})$ \\[3pt]
       fDNS  & $-0.2997$ & $-0.3229$ & $-0.2544$\\
       S-106H  & $-0.8202$ & $-0.8501$ & $-0.8045$ \\
       SL-106H  & $-0.3745$ & $-0.4473$ & $ -0.3258$\\
  \end{tabular}
  \caption{Correlation coefficients ($Corr(\overline{S}_{ij}, \tau_{ij})$) between $\overline{S}_{ij}$ and the predicted SGS stress by ANN-SGS models from \emph{a priori} test of forced homogeneous isotropic turbulence at $Re_{\lambda} = 106$.}
  \label{tab:corr_tau_Sij}
  \end{center}
\end{table}

\begin{figure}
	\centering
	\begin{subfigure}[b]{0.49\textwidth}
		\includegraphics[width=\textwidth]{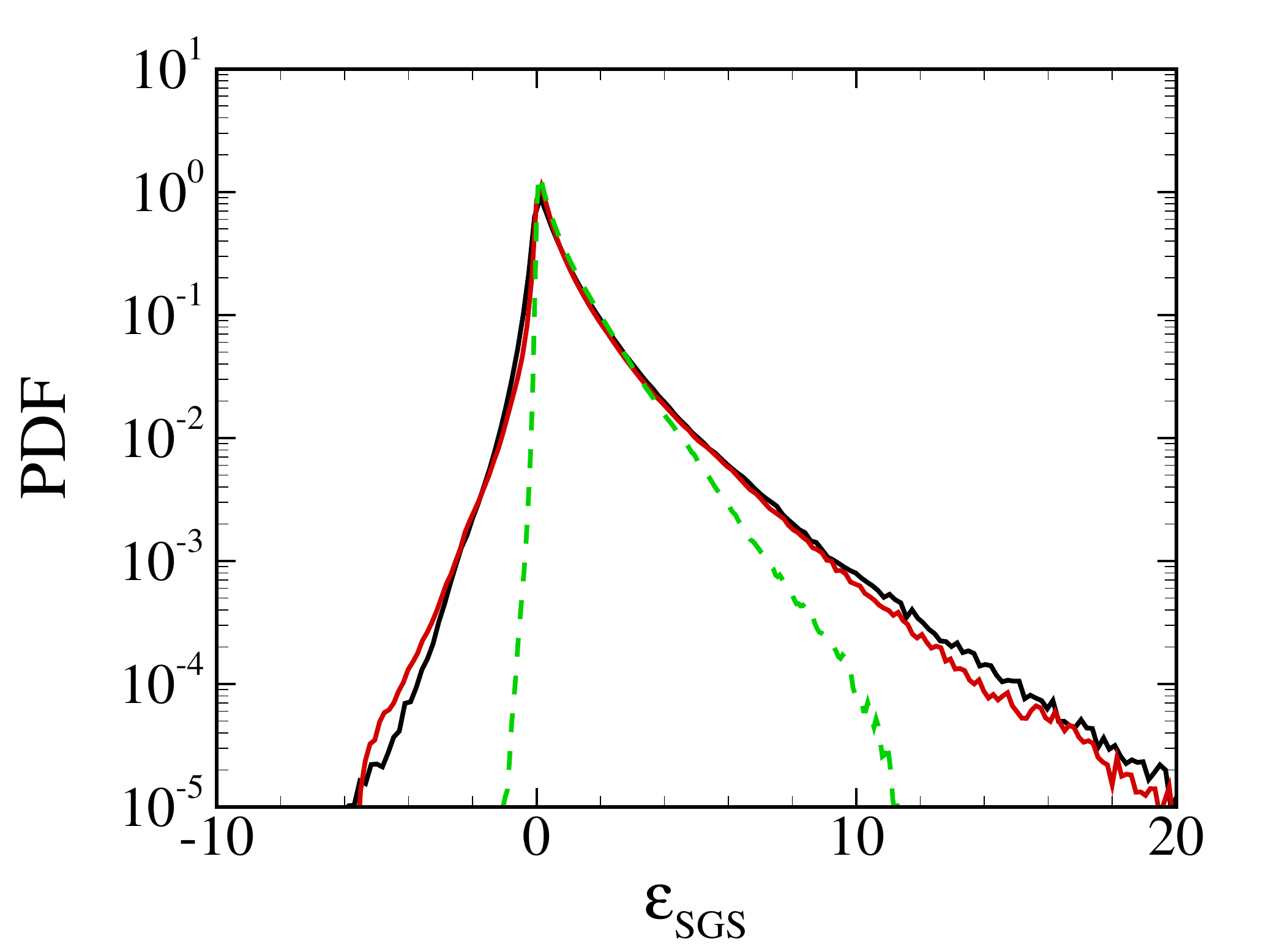}
		\caption{}
		\label{}
	\end{subfigure}
	\begin{subfigure}[b]{0.49\textwidth}
		\includegraphics[width=\textwidth]{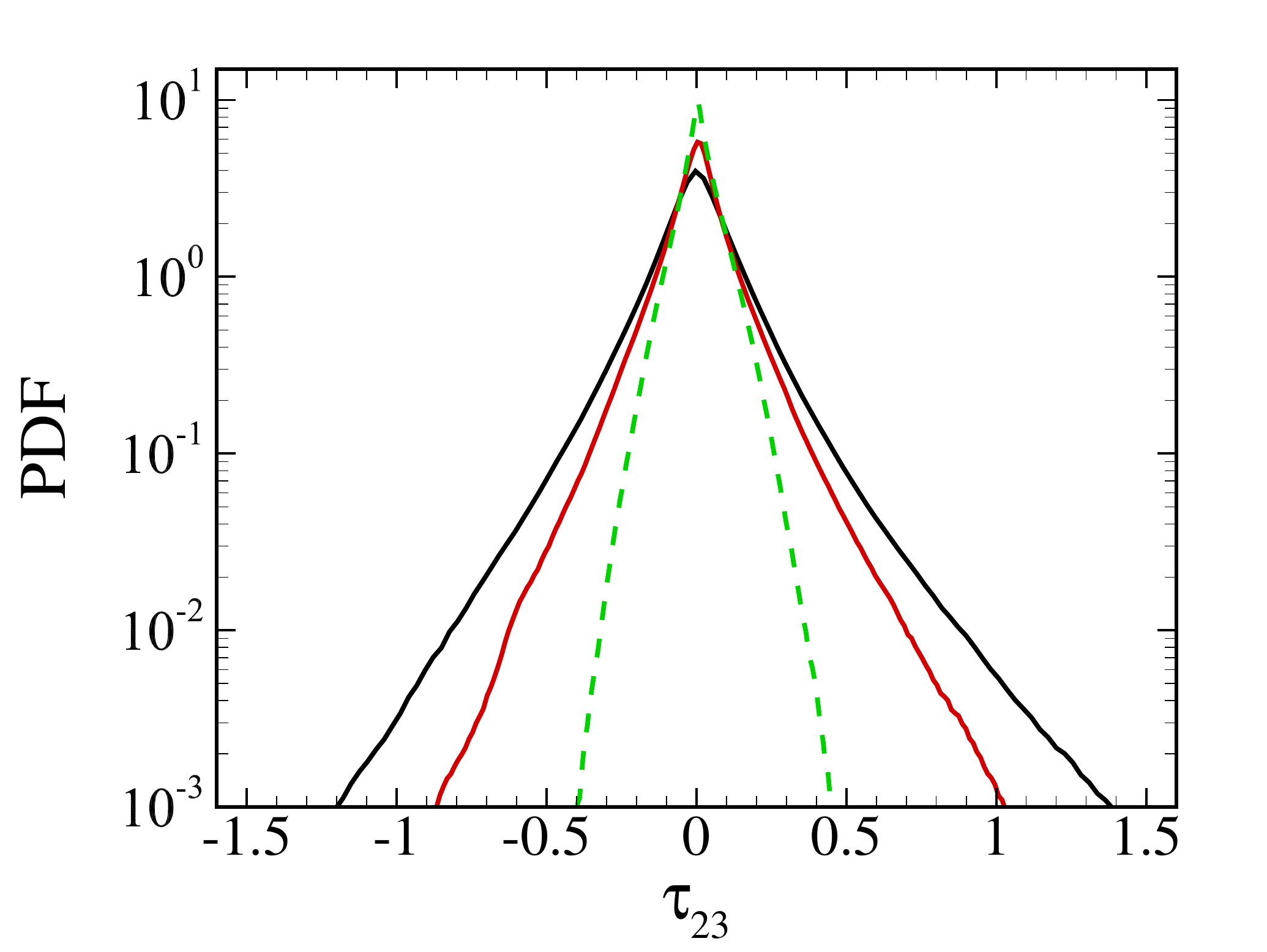}
		\caption{}
		\label{}
	\end{subfigure}
	\caption{Results from an \emph{a-priori} test of forced homogeneous isotropic turbulence at $Re_\lambda = 106$. (a) PDF of the SGS dissipation $\varepsilon_{SGS}$ ($= -\tau_{ij}\overline{S}_{ij}$); (b) PDF of $\tau_{23}$. {\color{black}\sampleline{}}, fDNS; {\color{red}\sampleline{}}, SL-106H; {\color{green}\sampleline{dash pattern=on 6pt off 2pt}}, S-106H.}
   \label{fig:a-priori_PDF}
\end{figure}

Additionally, the PDF of the SGS dissipation $\varepsilon_{SGS}$ ($ = -\tau_{ij}\overline{S}_{ij}$) predicted by S-106H is shown in figure~\ref{fig:a-priori_PDF}(a). Negative SGS dissipation corresponds to the backscatter, which is the kinetic energy transfer from the subgrid-scale to the resolved scale. Interestingly, S-106H  shows the similar characteristic to that of the eddy-viscosity models which are not capable of predicting backscatter. It can be considered that an ANN-SGS model which uses the resolved strain-rate tensor alone as the input only re-scales the given input $\overline{S}_{ij}$ to predict the SGS stress while maintaining its alignment.

Based on this observation, the performance of S-106H in actual LES is compared with that of DSM by conducting \emph{a posteriori} tests of forced isotropic turbulence at $Re_\lambda = 106$. The energy spectrum and PDFs of the SGS dissipation and the SGS stress from S-106H are compared with those from DSM and fDNS. The SGS stress of DSM~\citep{germano91, lilly92} is given as follows:
\begin{equation}
    \tau_{ij} - \frac{1}{3}\delta_{ij}\tau_{kk} = -2C\mybar{\Delta}^2|\mybar{S}|\mybar{S}_{ij},
    \label{eq:DSM}
\end{equation}
where $|\mybar{S}|=\sqrt{2\mybar{S}_{ij} \mybar{S}_{ij}}$, $\mybar{S}_{ij}=\frac{1}{2}(\partial \mybar{u}_i/\partial x_j+\partial \mybar{u}_j/\partial x_i)$, $C=\left< L_{ij}M_{ij}\right>/\left< M_{ij}M_{ij}\right>$, $L_{ij}=\myhat{\mybar{u}_i\mybar{u}_j}-\myhat{\mybar{u}}_i\myhat{\mybar{u}}_j$, $M_{ij}=-2\myhat{\Delta}^2|\myhat{\mybar{S}}|\myhat{\mybar{S}}_{ij}+2\mybar{\Delta}^2|\widehat{\mybar{S}|\mybar{S}_{ij}}$, $\mybar{\Delta}$ and $\myhat{\Delta} (= \sqrt{5}\mybar{\Delta})$ are the grid-filter and the test-filter scales, respectively. $\overline{\left(\,\cdot\,\right)}$ denotes the grid-level filter at $\mybar{\Delta}$ scale, $\widehat{\left(\,\cdot\,\right)}$ denotes the test-filter at $\myhat{\Delta}$ scale, and $\left< \cdot \right>$ denotes averaging over homogeneous directions (volume averaging for homogeneous isotropic turbulence). LESs are performed using a pseudospectral code \codename{HIT3D}~\citep{chumakov2007,chumakov2008} with the dealiasing method of the $2/3$ rule. The second-order Adams--Bashforth scheme is used for time integration and the time step size is set so that the CFL number of LES is the same as that of DNS~\citep{wang2018}.  The present ANN-SGS models predict the SGS stress tensor at each local grid point using resolved flow variables at the corresponding grid point as inputs. LESs with ANN-SGS models are performed without any \emph{ad-hoc} stabilisation procedures.

\begin{figure}
\centering \
\includegraphics[width=0.60\textwidth]{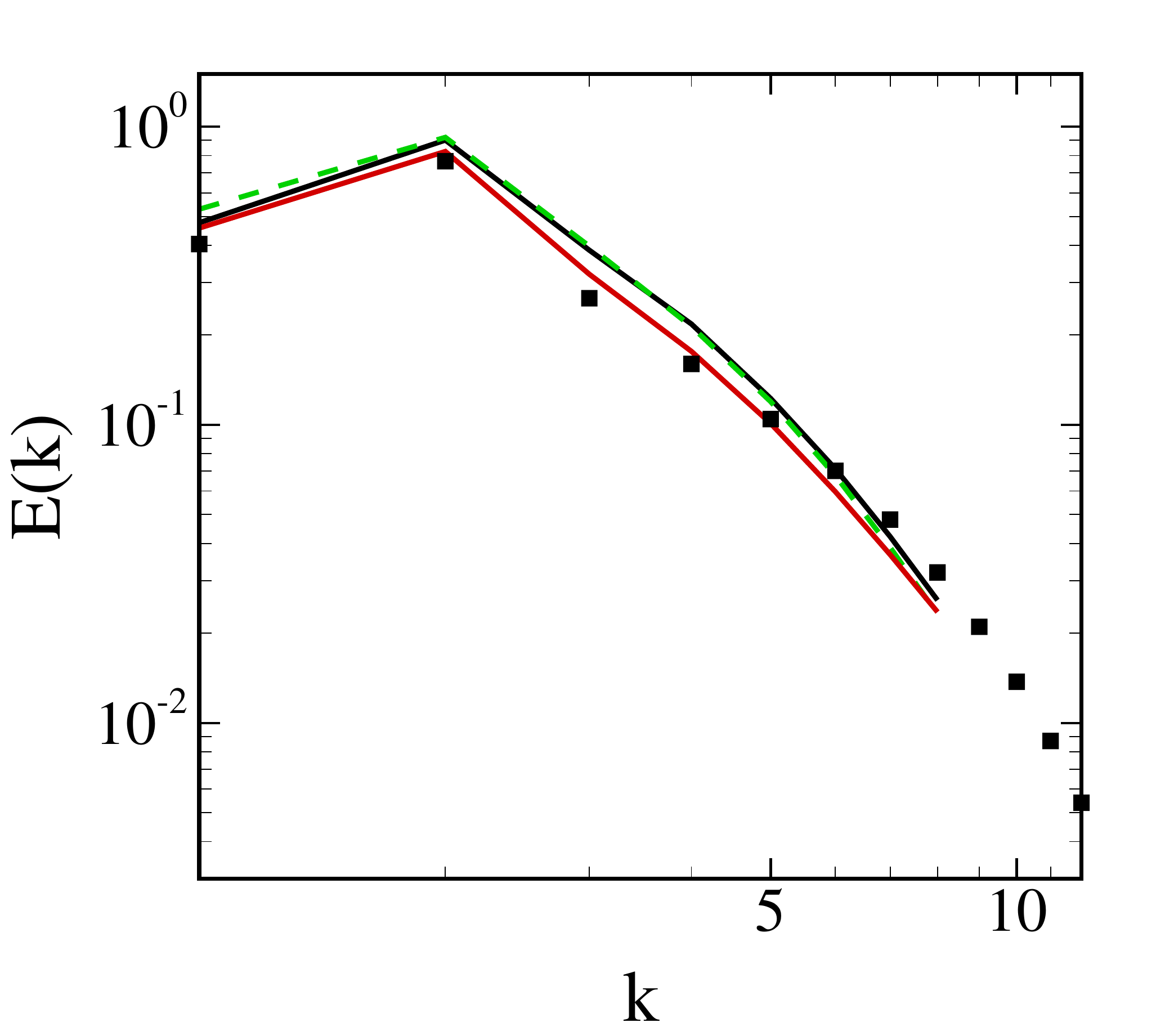}
\caption{Energy spectra from fDNS and LESs of forced homogeneous isotropic turbulence at $Re_\lambda = 106$ with grid resolution of $48^3$. $\blacksquare$, fDNS; {\color{black}\sampleline{}}, DSM; {\color{red}\sampleline{}}, SL-106H; {\color{green}\sampleline{dash pattern=on 6pt off 2pt}}, S-106H.}
\label{fig:FHIT106_ir_SL106}
\end{figure}

\begin{figure}
	\centering
	\begin{subfigure}[b]{0.49\textwidth}
		\includegraphics[width=\textwidth]{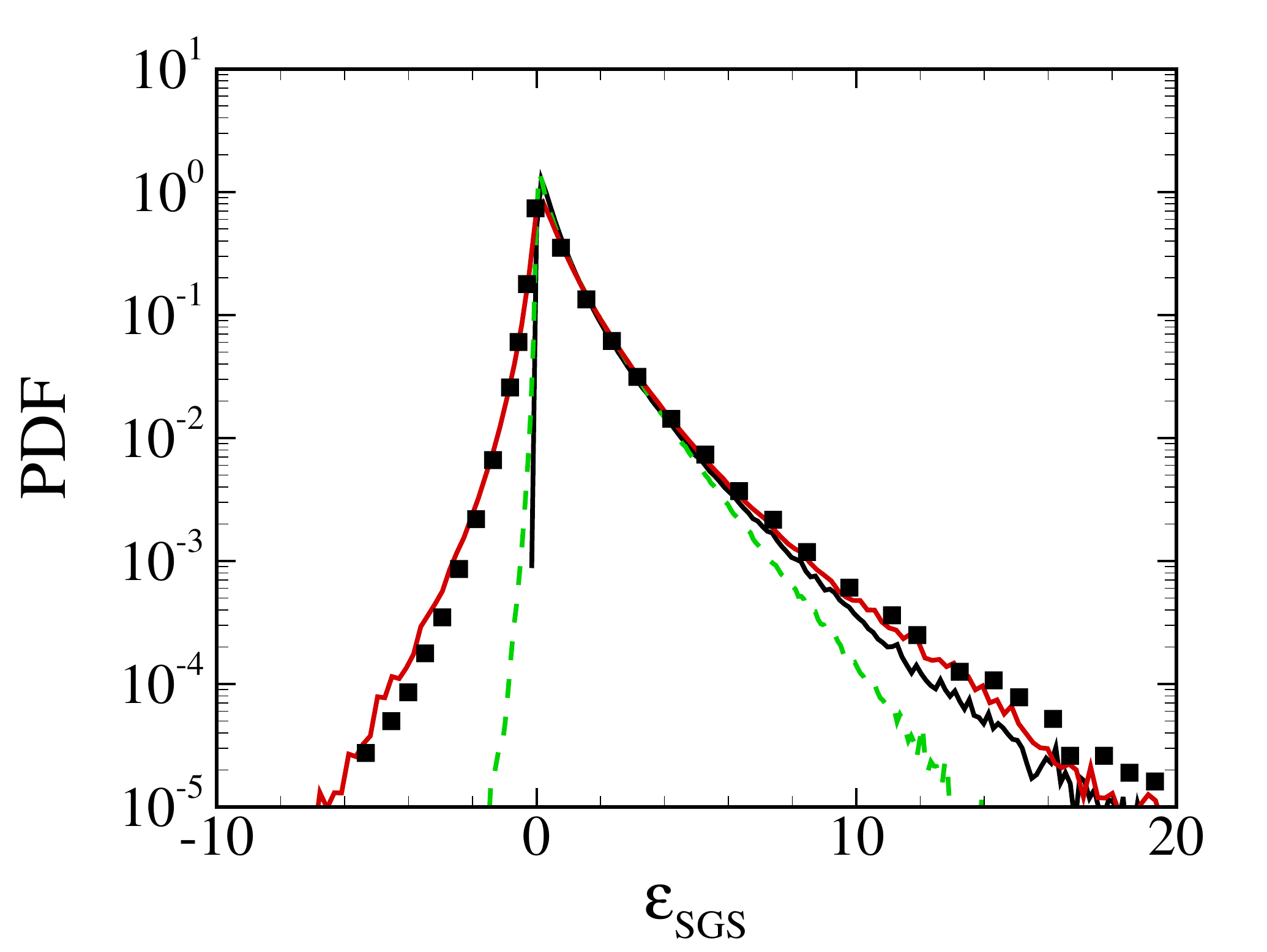}
		\caption{}
		\label{}
	\end{subfigure}
	\begin{subfigure}[b]{0.49\textwidth}
		\includegraphics[width=\textwidth]{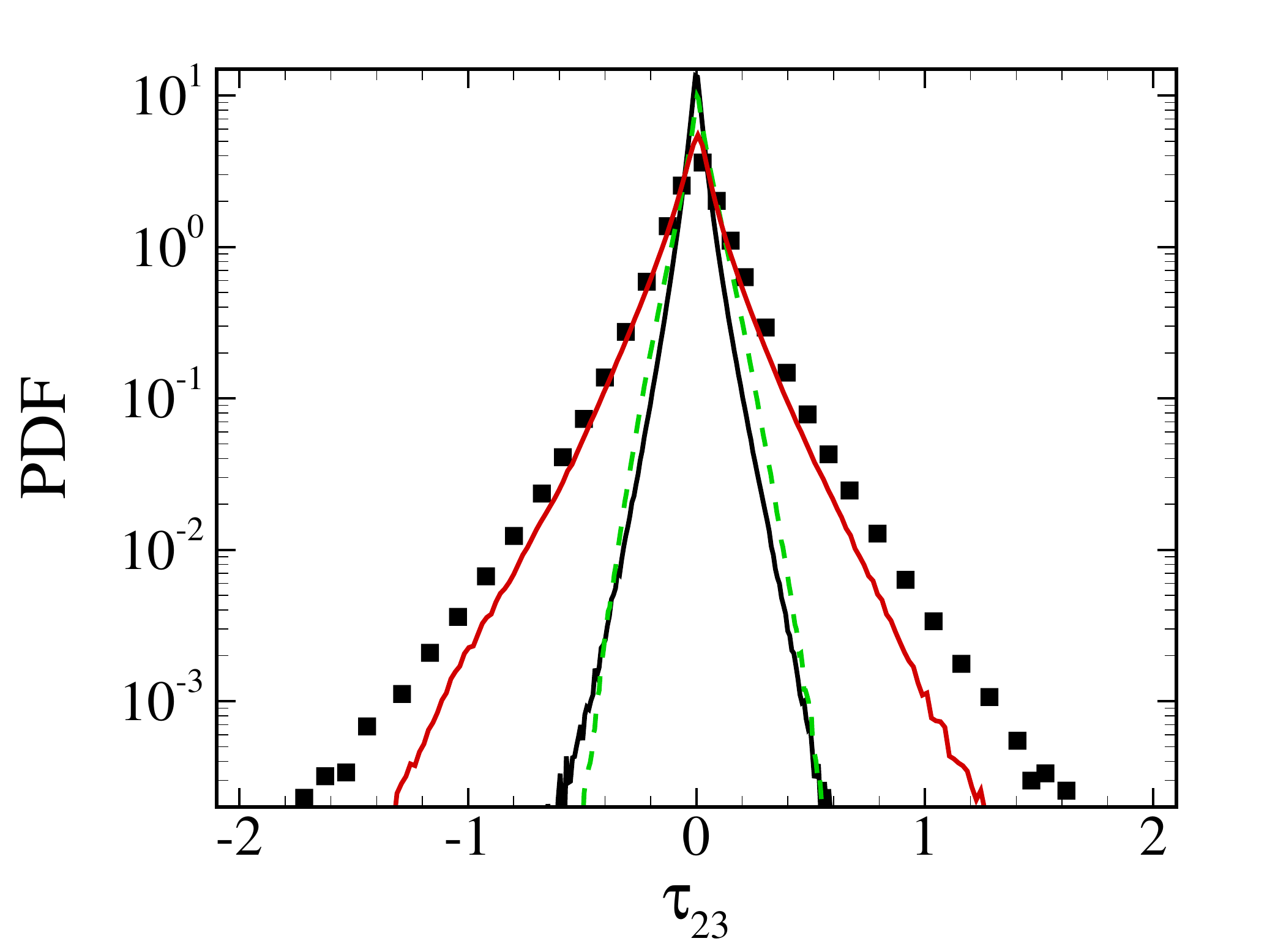}
		\caption{}
		\label{}
	\end{subfigure}
	\caption{Results from an \emph{a-posteriori} test of forced homogeneous isotropic turbulence at $Re_\lambda = 106$ with grid resolution of $48^3$. (a) PDF of the SGS dissipation $\varepsilon_{SGS}$ ($= -\tau_{ij}\overline{S}_{ij}$); (b) PDF of $\tau_{23}$. $\blacksquare$, fDNS; {\color{black}\sampleline{}}, DSM; {\color{red}\sampleline{}}, SL-106H; {\color{green}\sampleline{dash pattern=on 6pt off 2pt}}, S-106H.}
    \label{fig:FHIT106_ir_pdf_SL106}
\end{figure}

Figure~\ref{fig:FHIT106_ir_SL106} shows energy spectra from LESs of forced isotropic turbulence at $Re_\lambda = 106$ with grid resolution of $48^3$, which is the same as that of the training data. DSM is found to overestimate the energy spectrum in the range of $k \leq 5$. Interestingly, S-106H shows nearly identical performance to that of DSM in predicting the energy spectrum. In addition, from the PDF of the SGS dissipation shown in figure~\ref{fig:FHIT106_ir_pdf_SL106}(a), S-106H and DSM have the same characteristic that they are not capable of predicting backscatter. Note that the PDF of the SGS stress ($\tau_{23}$) predicted by S-106H is almost identical to that of DSM as both PDFs are narrower than that of fDNS (figure~\ref{fig:FHIT106_ir_pdf_SL106}(b)). This indicates that S-106H produces the SGS stress similar to that of DSM.

From the above observation, it can be concluded that the ANN-SGS model with $\overline{S}_{ij}$ as an input predicts the SGS stress which is closely aligned with the given input tensor, and performs similarly to DSM. Consequently, it is expected that S-106H can be improved by the similar concept employed in dynamic mixed models~\citep{liu1994, anderson1999}. It is reported that a linear combination of the eddy-viscosity model with the resolved stress $L_{ij} (=\myhat{\mybar{u}_i\mybar{u}_j}-\myhat{\mybar{u}}_i\myhat{\mybar{u}}_j)$ improves the accuracy of the predicted SGS stress with better alignment to the true SGS stress. 

Similarly to the algebraic dynamic mixed models, it is expected that an ANN-SGS model can achieve closer alignment to the true SGS stress and more accurate prediction of the magnitude of the SGS stress if the resolved stress tensor $L_{ij}$ is considered as an input in addition to the strain-rate tensor $\overline{S}_{ij}$ (\emph{i.e.}, ANN-SGS mixed model). To investigate effect of the additional input variable, results of S-106H and SL-106H are compared in \emph{a-priori} and \emph{a-posteriori} tests of forced homogeneous isotropic turbulence at $Re_\lambda = 106$. Results of SL-106H are also compared with those of the algebraic dynamic mixed models in Section~\ref{sec:DHIT_mixed}.

The performance of SL-106H is compared with that of S-106H in an \emph{a-priori} test. The correlation coefficients between the true and the predicted SGS stress tensors are calculated following the definition of (\ref{eq:correlation}) and shown in table~\ref{tab:corr_tau}. The correlation coefficients of SL-106H are significantly improved compared with those of S-106H, which indicates that SL-106H reconstructs the instantaneous SGS stress closer to the true SGS stress. In table~\ref{tab:corr_tau_Sij}, SL-106H more accurately predicts the correlation coefficients between $\overline{S}_{ij}$ and the predicted SGS stress $\tau_{ij}$ ($Corr(\overline{S}_{ij}, \tau_{ij})$) than S-106H, indicating that SL-106H is capable of aligning the principal axes of the predicted SGS stress closer to the true SGS stress. In addition, SL-106H provides excellent prediction of the PDF of the SGS dissipation (figure~\ref{fig:a-priori_PDF}(a)), which almost overlaps with that of fDNS. At the same time, SL-106H predicts the PDF of the SGS stress ($\tau_{23}$) closer to that of fDNS, whereas S-106H predicts a narrower PDF as shown in figure~\ref{fig:a-priori_PDF}(b).

An \emph{a-posteriori} test of forced isotropic turbulence at $Re_\lambda = 106$ is also conducted using SL-106H. As shown in figure~\ref{fig:FHIT106_ir_SL106}, SL-106H predicts the energy spectrum closer to that of fDNS than S-106H and DSM in the range of $k \leq  5$. Furthermore, SL-106H noticeably better predicts the PDF of the SGS dissipation than S-106H (figure~\ref{fig:FHIT106_ir_pdf_SL106}(a)). It is worth noting that SL-106H is capable of accurately predicting the backscatter even in the \emph{a-posteriori} test; consequently, LES becomes stable without any \emph{ad-hoc} procedures, which is a clear advantage over the algebraic dynamic SGS models. SL-106H is found to predict the PDF of the SGS stress ($\tau_{23}$) more accurately as shown in figure~\ref{fig:FHIT106_ir_pdf_SL106}(b).

Results of \emph{a-priori} and \emph{a-posteriori} tests using SL-106H clearly indicate that considering the resolved stress $L_{ij}$ as an input in addition to $\overline{S}_{ij}$ improves the performance of the ANN-SGS model. Unlike the SGS stress predicted by S-106H which is closely aligned with the given input $\overline{S}_{ij}$, the SGS stress predicted by SL-106H is closely aligned with the true SGS stress. Although the performance of S-106H is almost identical to that of DSM, both S-106H and SL-106H have advantages over DSM as they are free from the requirement of a stabilisation process such as averaging over statistically homogeneous directions or clipping of the negative model coefficients.

\subsection{Generalisation to untrained flow: LES of decaying homogeneous isotropic turbulence}\label{sec:DHIT}

\begin{figure}
	\centering
	\begin{subfigure}[b]{0.49\textwidth}
		\includegraphics[width=\textwidth]{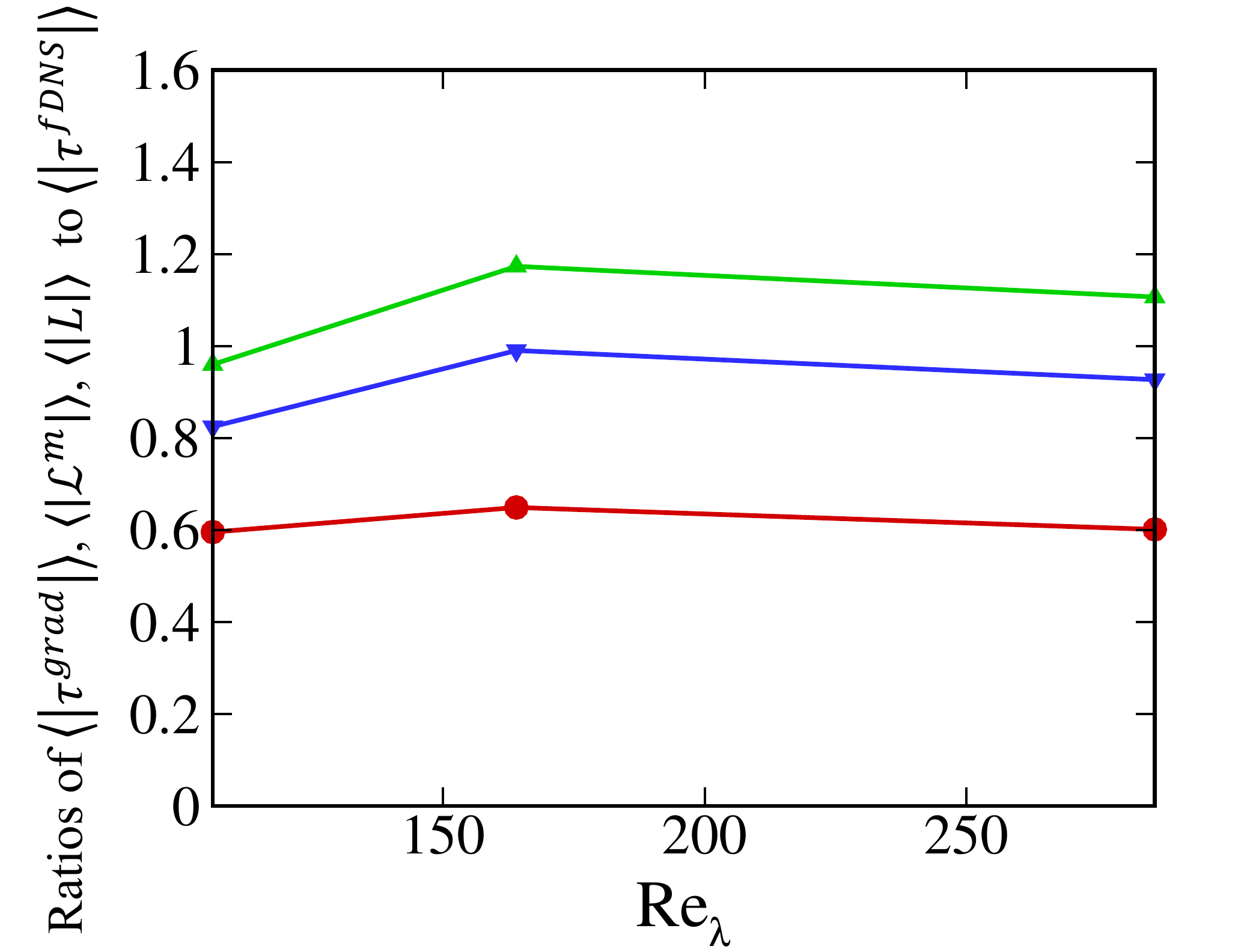}
		\caption{}
		\label{}
	\end{subfigure}
	\begin{subfigure}[b]{0.49\textwidth}
		\includegraphics[width=\textwidth]{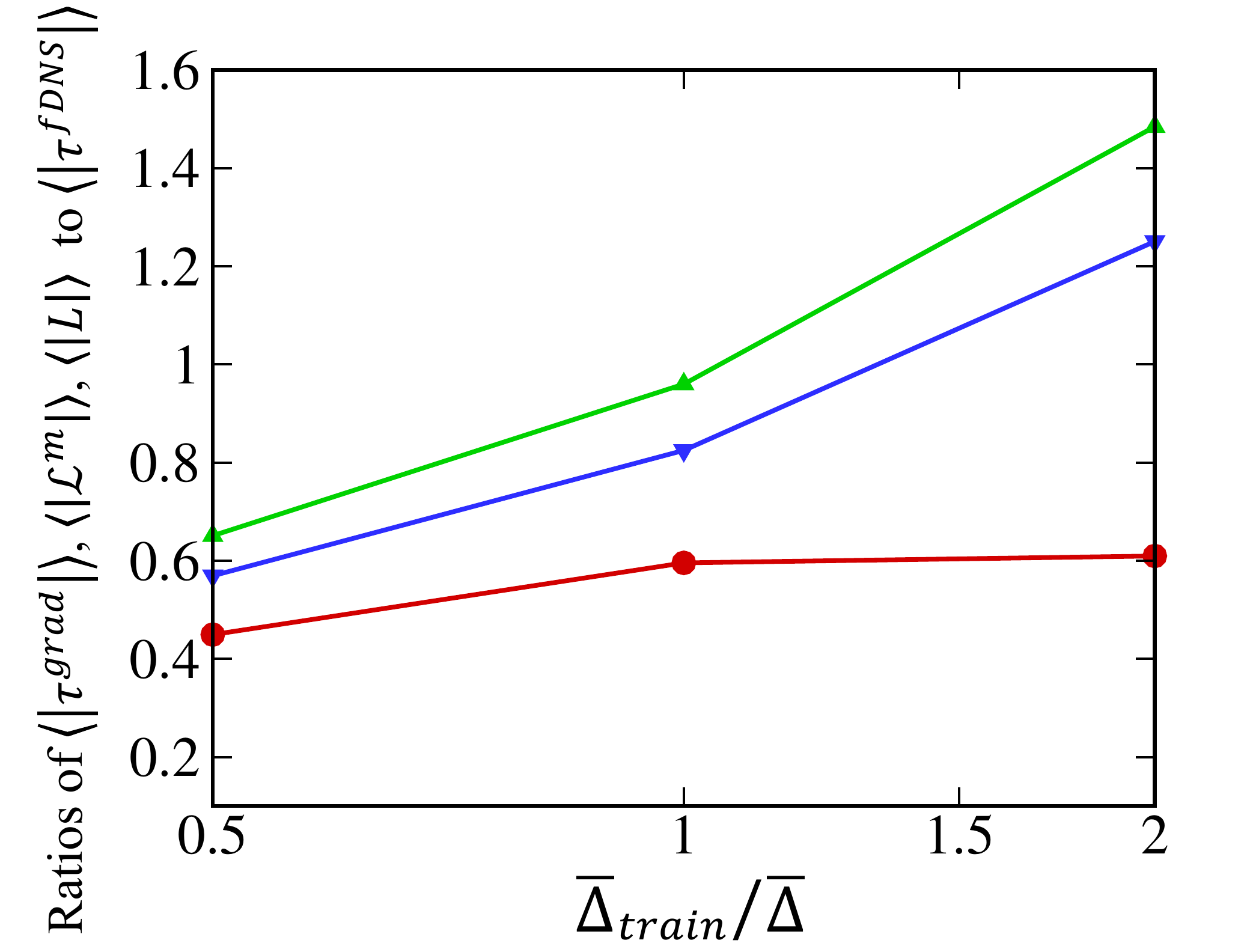}
		\caption{}
		\label{}
	\end{subfigure}
	\caption{Ratios of $\left< \left| \tau^{grad}\right|\right>$, $\left< \left| \mathcal{L}^m \right|\right>$, and $\left< \left| L \right|\right>$ to $\left< \left| \tau^{fDNS}\right|\right>$ at (a) $Re_{\lambda}$=106, 168, and 286 for $\mybar{\Delta}_{train}/\mybar{\Delta}$=1, and for (b) $\mybar{\Delta}_{train}/\mybar{\Delta}$=1/2, 1, 2 at $Re_{\lambda}$=106. {\color{red}\sampleline{}}, $\left< \left| \tau^{grad}\right|\right>/\left< \left| \tau^{fDNS}\right|\right>$; {\color{blue}\sampleline{}}, $\left< \left| \mathcal{L}^m \right|\right>/\left< \left| \tau^{fDNS}\right|\right>$; {\color{green}\sampleline{}}, $\left< \left| L \right|\right>/\left< \left| \tau^{fDNS}\right|\right>$.}
    \label{fig:normalization}
\end{figure}

\begin{figure}
	\centering
	\begin{subfigure}[b]{0.49\textwidth}
		\includegraphics[width=\textwidth]{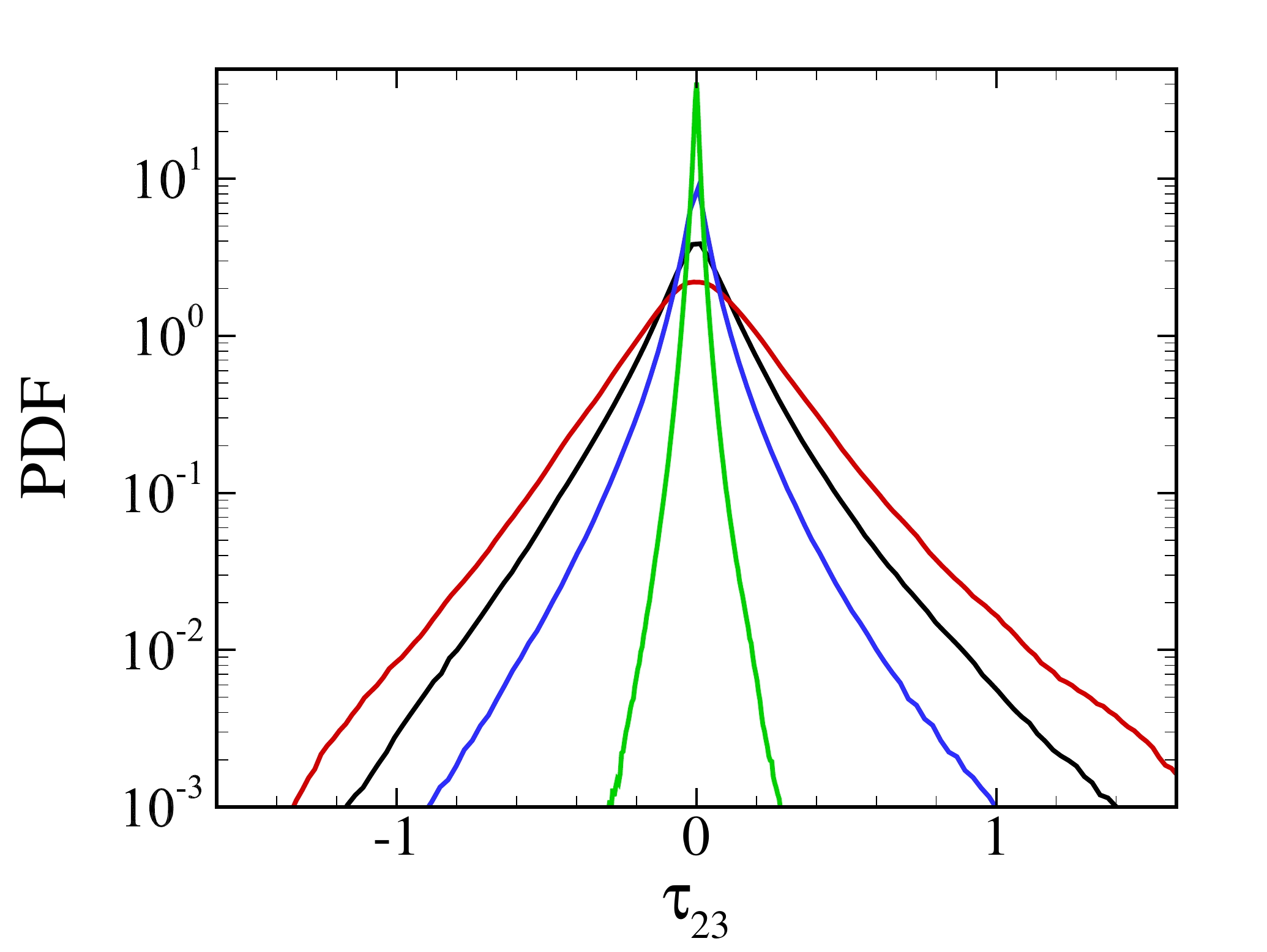}
		\caption{}
		\label{}
	\end{subfigure}
	\begin{subfigure}[b]{0.49\textwidth}
		\includegraphics[width=\textwidth]{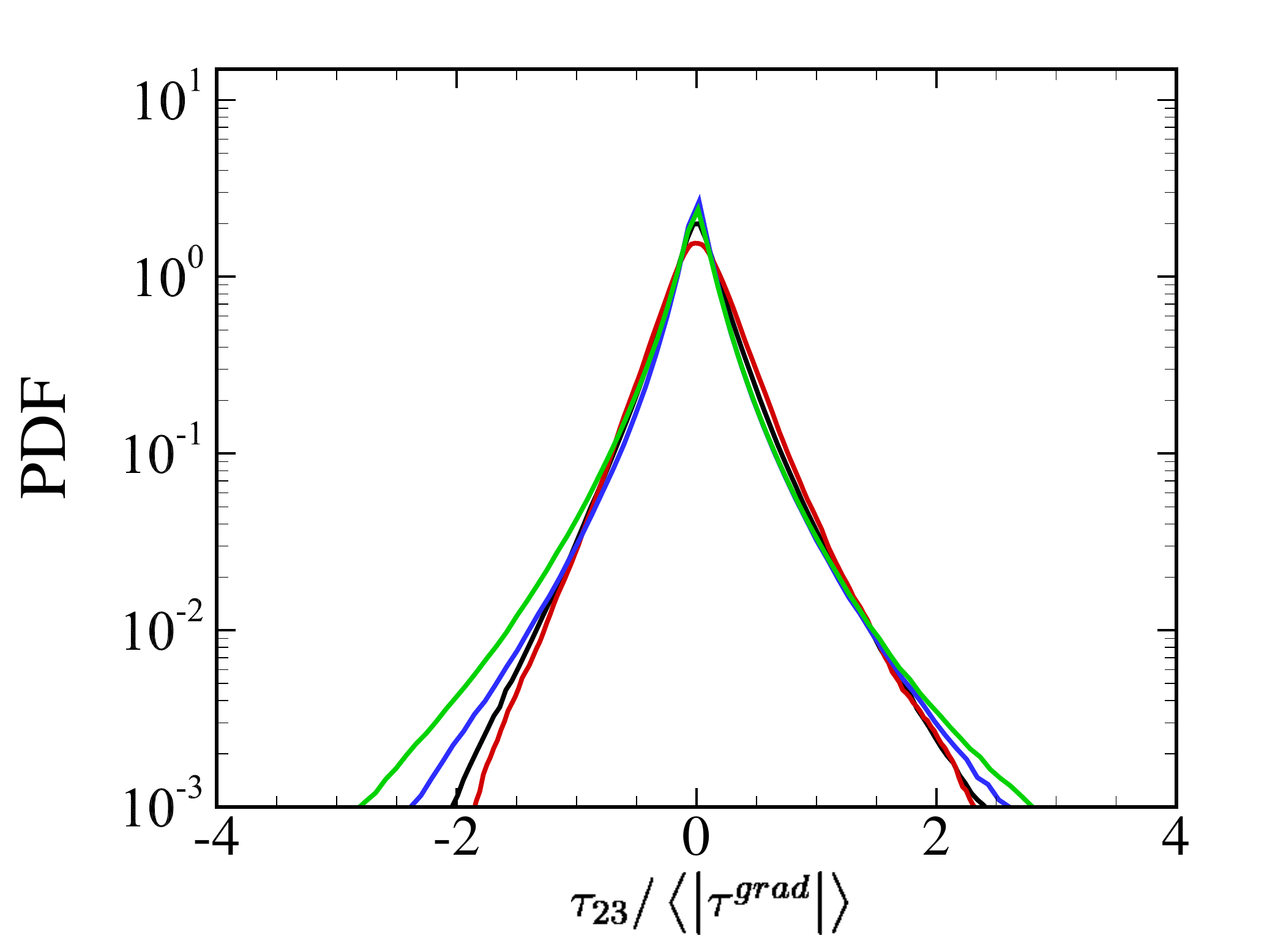}
		\caption{}
		\label{}
	\end{subfigure}
	\caption{PDFs of the SGS stress $\tau_{23}$ from fDNS of forced isotropic turbulence with various Reynolds numbers and grid resolution. (a) PDF of $\tau_{23}$ without normalisation; (b) PDF of normalised $\tau_{23}$ with $\left<\left| \tau^{grad}\right|\right>$. {\color{black}\sampleline{}}, $Re_{\lambda}$=106, $\mybar{\Delta}_{train}/\mybar{\Delta}$=1; {\color{red}\sampleline{}}, $Re_{\lambda}$=106, $\mybar{\Delta}_{train}/\mybar{\Delta}$=1/2; {\color{blue}\sampleline{}}, $Re_{\lambda}$=106, $\mybar{\Delta}_{train}/\mybar{\Delta}$=2; {\color{green}\sampleline{}}, $Re_{\lambda}$=286, $\mybar{\Delta}_{train}/\mybar{\Delta}$=1.}
    \label{fig:norm_tau}
\end{figure}

In this section, the application of ANN-SGS models trained with only forced homogeneous isotropic turbulence data to untrained decaying isotropic turbulence is discussed. For successful generalisation to decaying isotropic turbulence, ANN-SGS models have to provide accurate predictions of the SGS stress for various Reynolds numbers and grid resolution.

In this regard, normalisation of variables is important, as it plays a critical role for the consistent performance of an ANN under various conditions. As explained in Section~\ref{sec:Artificial neural network for subgrid-scale modelling}, the input and the output tensors are normalised as $\overline{S}_{ij}^* = \overline{S}_{ij}/\left< \left| \overline{S} \right|\right>, L_{ij}^* = L_{ij}/\left< \left| L \right|\right>$, and $\tau_{ij}^* = \tau_{ij}/\left< \left| \tau \right|\right>$, where the denominator $\left< \left| \tau \right|\right>$ requires an approximation in actual LESs. The input variables ($\overline{S}_{ij}$ and $L_{ij}$) are properly normalised by the magnitudes of the variables to have similar distributions for various conditions. However, normalising the SGS stress tensor to have similar distributions for various conditions is a challenging task because the approximation for $\left< \left| \tau \right|\right>$ is not accurate enough. In other words, an inaccurate approximation of the normalisation factor of the output SGS stress results in a significantly different distribution of the normalised output for different conditions. In this situation, an ANN-based model could suffer from the prior probability shift issue, which occurs when the output variable distributions are different at training and test conditions (\emph{i.e.}, $P_{train}(y) \neq P_{test}(y)$, where $P(y)$ is a probability distribution of an output variable $y$)~\citep{quinonero2008, moreno2012}. Dataset shift, including the prior probability shift, is not desirable for ANNs as it can cause significant changes in their performance during testing with untrained data~\citep{gawlikowski2021}. Consequently, the performance of an ANN-SGS model trained at a specific Reynolds number and on certain grid resolution could be different from that at other flow conditions.

\begin{figure}
\centering \
\includegraphics[width=0.60\textwidth]{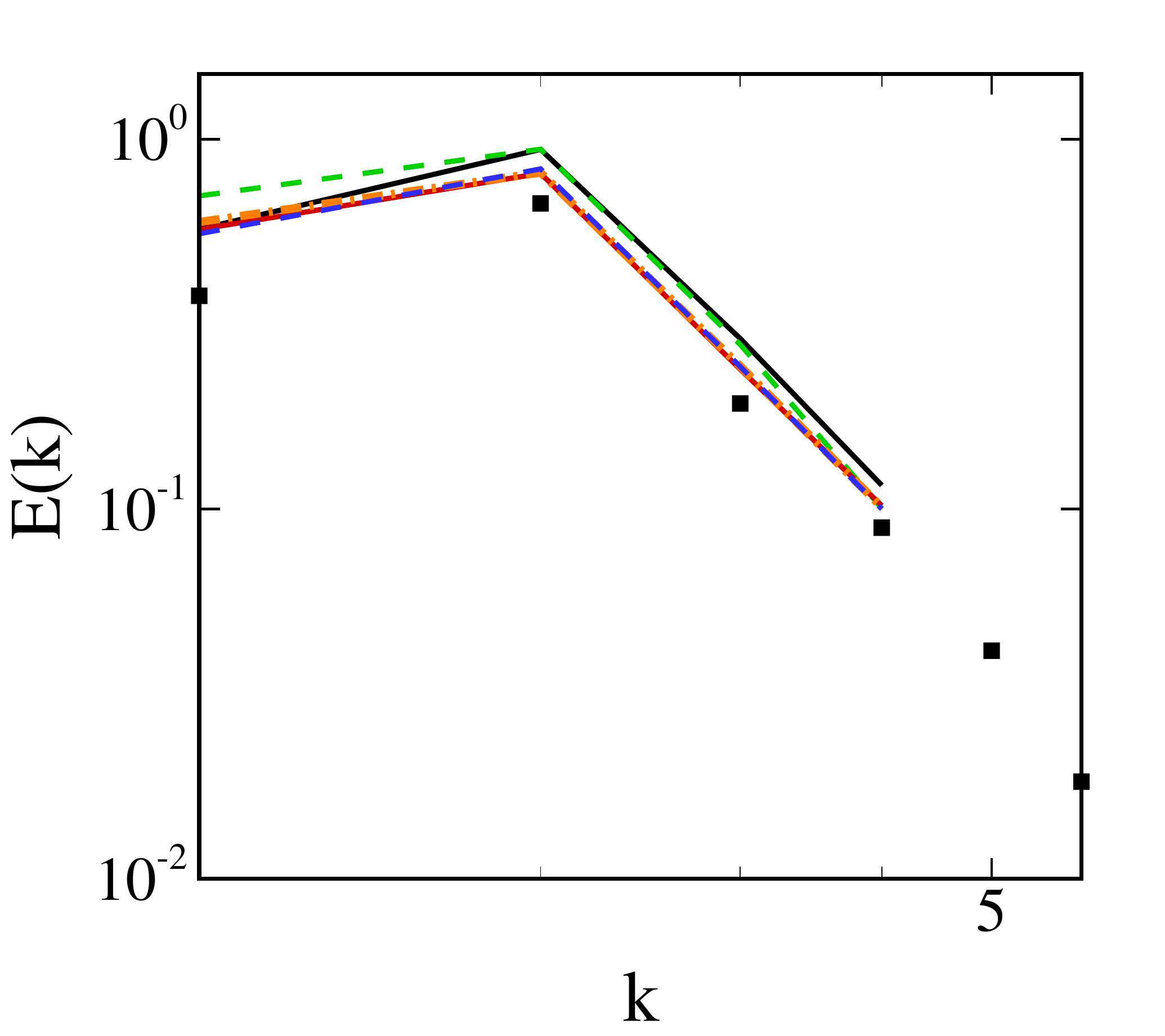}
\caption{Energy spectra from fDNS and LESs of forced homogeneous isotropic turbulence at $Re_\lambda=106$ with grid resolution of $24^3$ (two-times coarser resolution than that of training data). $\blacksquare$, fDNS; {\color{black}\sampleline{}}, DSM; {\color{red}\sampleline{}}, SL-106H; {\color{blue}\sampleline{dash pattern=on 6pt off 2pt}}, SL-286H; {\color{YellowOrange}\sampleline{dash pattern=on .7em off .2em on .2em off .2em}}, SL-106+286H; {\color{green}\sampleline{dash pattern=on 6pt off 2pt}}, S-106H.}
\label{fig:FHIT106_c}
\end{figure}

Therefore, it is suggested that a normalisation factor, which enables the distributions of the normalised inputs and outputs to remain unchanged for various Reynolds numbers and grid resolution, should be selected to avoid such performance inconsistency of an ANN-SGS model during the generalisation. The modified Leonard term $\mathcal{L}_{ij}^m$ ($= \overline{\overline{u}_i \overline{u}_j} - \overline{\overline{u}}_i \overline{\overline{u}}_j)$, the resolved stress $L_{ij}$ ($ = \widehat{\overline{u}_i \overline{u}_j} - \widehat{\overline{u}}_i \widehat{\overline{u}}_j)$, and the gradient model term $\tau_{ij}^{grad}$ ($ = \frac{1}{12}\mybar{\Delta}^2\frac{\partial \overline{u}_i}{\partial x_k}\frac{\partial \overline{u}_j}{\partial x_k})$ are considered to approximate the normalisation factor for the SGS stress. Among the three candidates, the term that has the most constant ratio to the true SGS stress $\left< \left| \tau \right|\right>$ for various Reynolds numbers and grid resolution should be selected. Figure~\ref{fig:normalization} shows ratios of $\left< \left| \tau^{grad}\right|\right>$, $\left< \left| \mathcal{L}^m \right|\right>$, and $\left< \left| L \right|\right>$ to $\left< \left| \tau^{fDNS}\right|\right>$, obtained at three different Reynolds numbers on a fixed grid resolution and three different grid resolution at $Re_\lambda=106$. The gradient model term $\left< \left| \tau^{grad}\right|\right>$ shows the most constant ratio to the true SGS stress $\left< \left| \tau^{fDNS}\right|\right>$ for various Reynolds numbers and grid resolution, compared with the other two terms. Therefore, in the present study, the averaged $L_2$ norm of the gradient model term is selected to normalise and rescale the output SGS stress tensor as $\tau_{ij}^* = \tau_{ij}/\left< \left| \tau^{grad}\right|\right>$. 

\begin{table}
	\begin{center}
		\def~{\hphantom{0}}
		\begin{tabular}{cp{0.75cm}cp{0.75cm}cp{0.75cm}c}
			LES case&&
			Initial $\Rey_{\lambda}$&&
			$N^3$&&
			SGS model\\[3pt]
			\hline
			 &&  &&  && SL-106H \\[3pt]
			 &&  &&  && SL-286H\\[3pt]
			DHIT106 && 106 && $48^3$ && SL-106+286H\\[3pt]
			 &&  &&  && S-106H\\[3pt]
			 &&  &&  && DSM\\
			\hline
			 &&  &&  && SL-106H \\[3pt]
			DHIT106c && 106 && $24^3$ && SL-286H\\[3pt]
			 &&  &&  && S-106H\\[3pt]
			 &&  &&  && DSM\\
			\hline
			 &&  &&  && SL-106H \\[3pt]
			DHIT106f && 106 && $96^3$ && SL-286H\\[3pt]
			 &&  &&  && S-106H\\[3pt]
			 &&  &&  && DSM\\
			\hline
			 &&  &&  && SL-106H \\[3pt]
			DHIT286 && 286 && $128^3$ && SL-286H\\[3pt]
			 &&  &&  && S-106H\\[3pt]
			 &&  &&  && DSM\\
		\end{tabular}
		\caption{Test LES cases for decaying homogeneous isotropic turbulence with ANN-SGS models and DSM. $N$ is the number of grid points in each direction. Effects of grid resolution and the initial Reynolds number on the performance of ANN-SGS models are considered.}
		\label{tab:DHIT_case}
	\end{center}
\end{table}

Figure~\ref{fig:norm_tau} shows the PDF of the SGS stress from fDNS of forced isotropic turbulence with and without normalisation. Distributions of the SGS stress without normalisation vary significantly depending on the Reynolds number and grid resolution, while the normalised SGS stress tensors show similar distributions for various Reynolds numbers and grid resolution, which enables an ANN-SGS model trained at a single condition to be successfully generalised to other conditions. In figure~\ref{fig:FHIT106_c}, effectiveness of the normalisation is verified in LES of forced homogeneous isotropic turbulence at $Re_\lambda=106$ with grid resolution of $24^3$, which is two-times coarser than that of the training data. As shown in figure~\ref{fig:FHIT106_c}, SL-106H, SL-286H, and SL-106+286H predict the energy spectra more accurately than DSM and S-106H at all wavenumbers. This indicates that the selection of a proper normalisation factor for the output SGS stress enables the present ANN-SGS mixed models to perform accurately at untrained coarser grid resolution. 

\begin{figure}
	\centering
	\begin{subfigure}[b]{0.49\textwidth}
		\includegraphics[width=\textwidth]{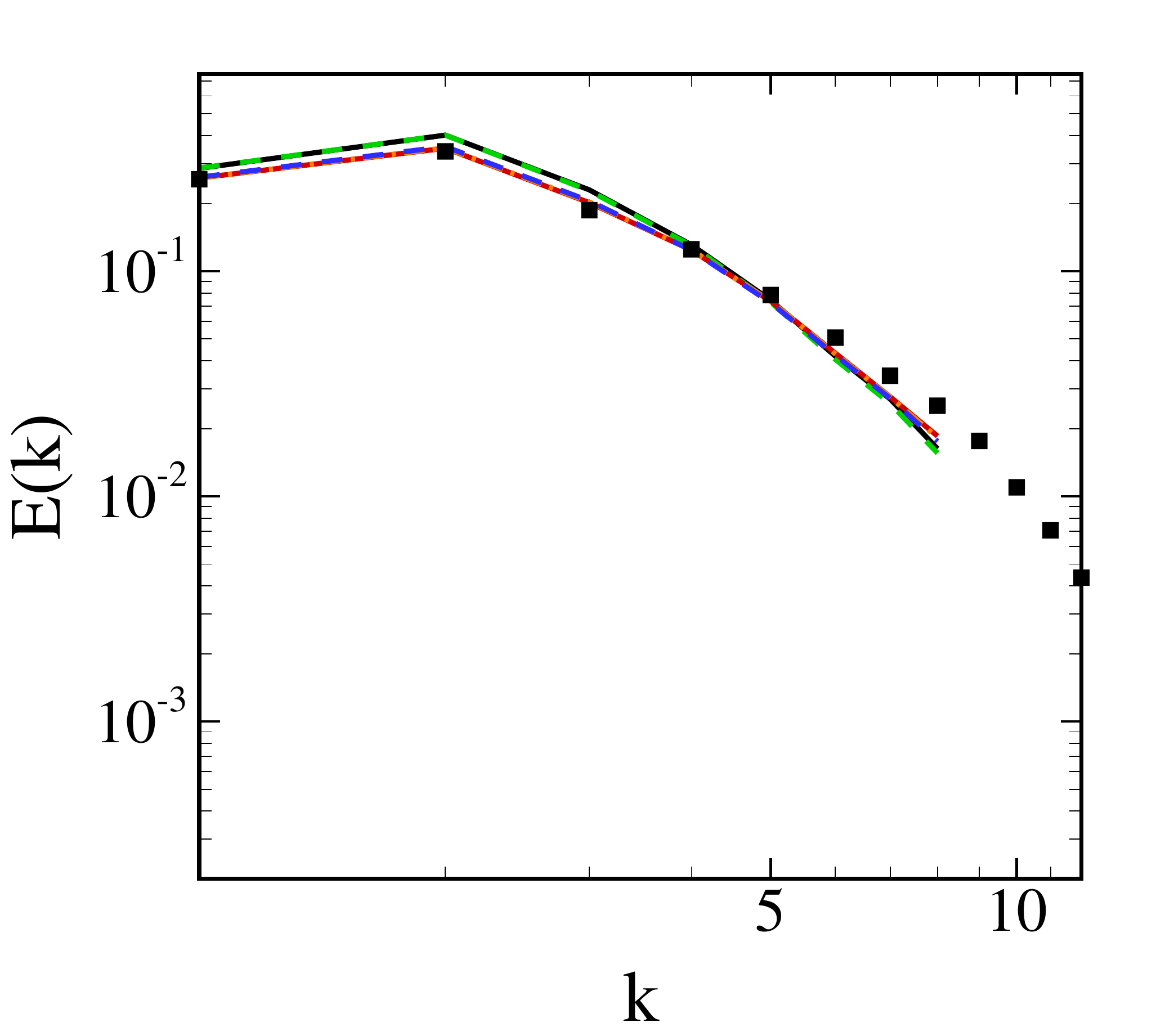}
		\caption{}
		\label{}
	\end{subfigure}
	\begin{subfigure}[b]{0.49\textwidth}
		\includegraphics[width=\textwidth]{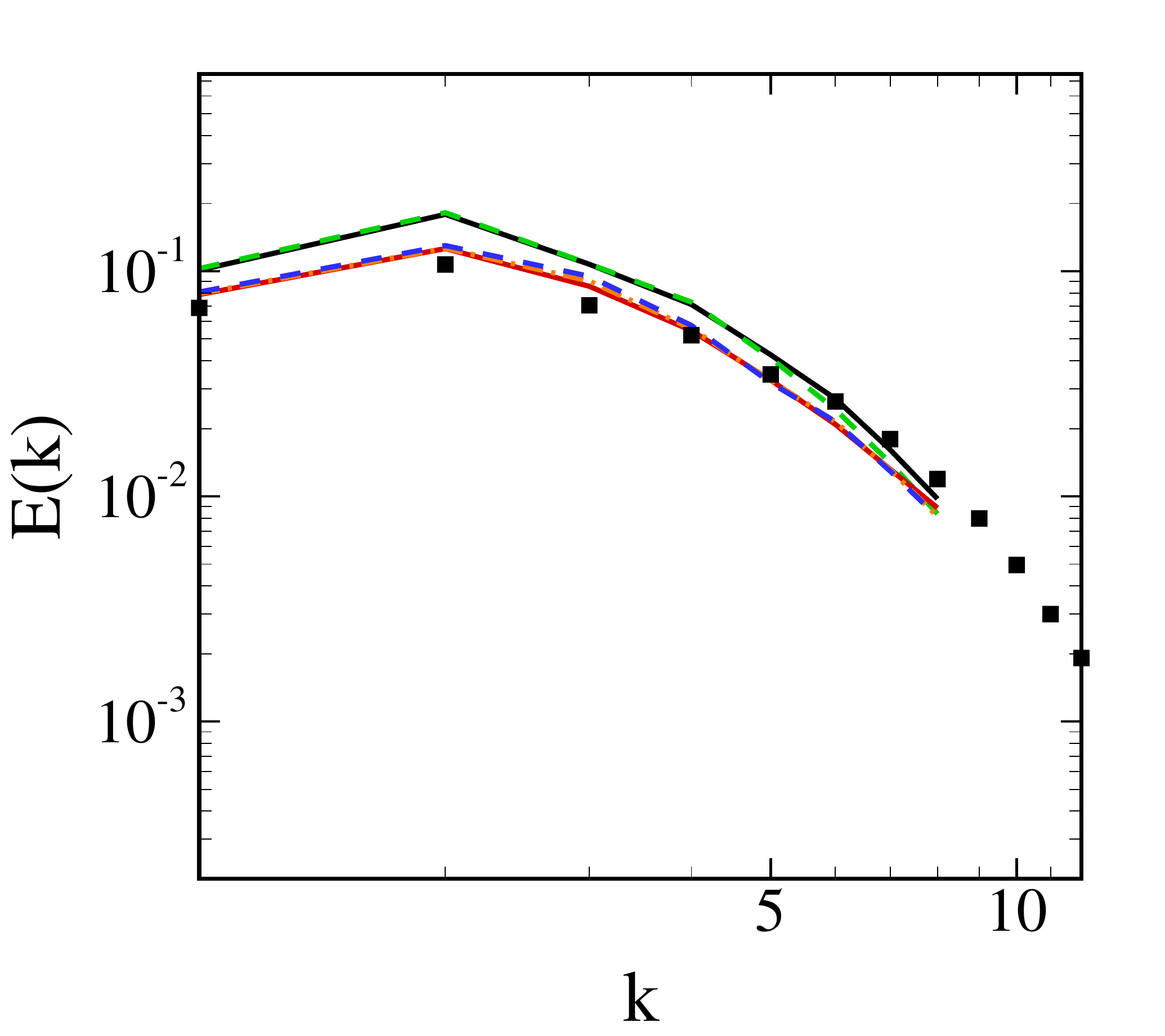}
		\caption{}
		\label{}
	\end{subfigure}
	\begin{subfigure}[b]{0.49\textwidth}
		\includegraphics[width=\textwidth]{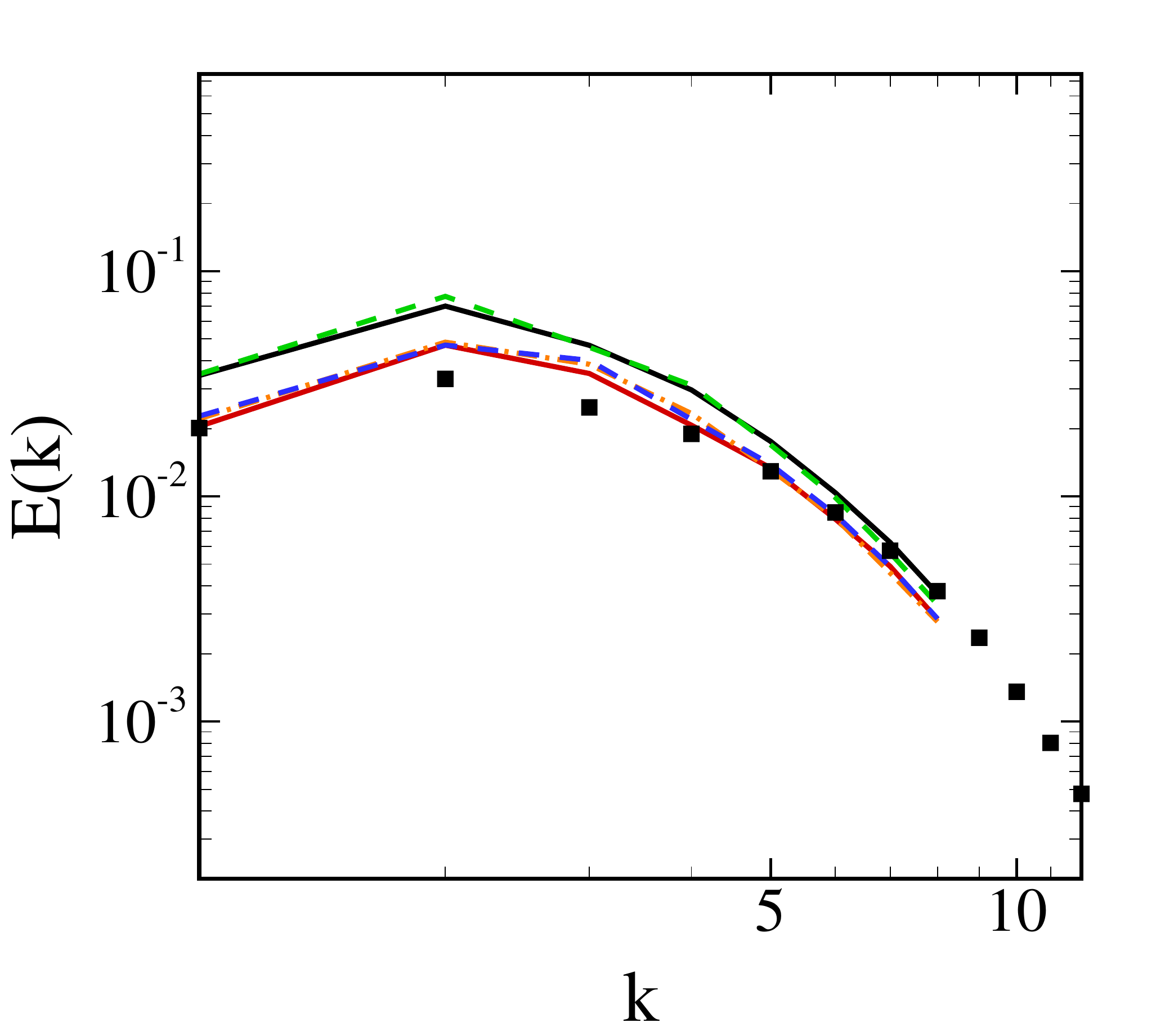}
		\caption{}
		\label{}
	\end{subfigure}
	\caption{Energy spectra from fDNS and LESs of decaying homogeneous isotropic turbulence at the initial Reynolds number $Re_\lambda$ of 106 with grid resolution of $48^3$ (DHIT106 case). (a) $t/T_{e,0} = 1.1$; (b) $t/T_{e,0} = 3.3$; (c) $t/T_{e,0} = 6.6$. $\blacksquare$, fDNS; {\color{black}\sampleline{}}, DSM; {\color{red}\sampleline{}}, SL-106H; {\color{blue}\sampleline{dash pattern=on 6pt off 2pt}}, SL-286H; {\color{YellowOrange}\sampleline{dash pattern=on .7em off .2em on .2em off .2em}}, SL-106+286H; {\color{green}\sampleline{dash pattern=on 6pt off 2pt}}, S-106H.}
    \label{fig:DHIT106_ir}
\end{figure}

\begin{figure}
	\centering
	\begin{subfigure}[b]{0.49\textwidth}
		\includegraphics[width=\textwidth]{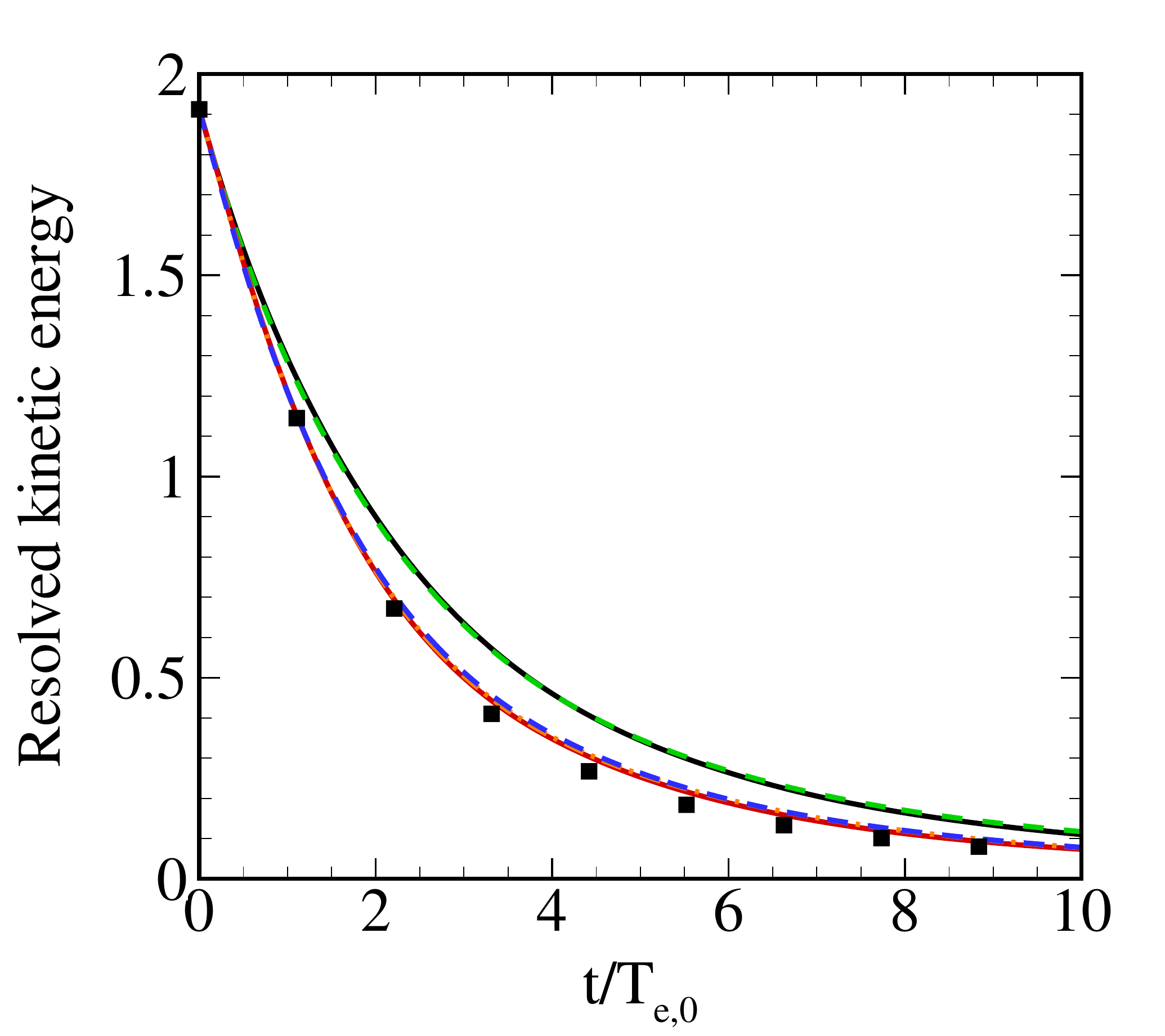}
		\caption{}
		\label{}
	\end{subfigure}
	\begin{subfigure}[b]{0.49\textwidth}
		\includegraphics[width=\textwidth]{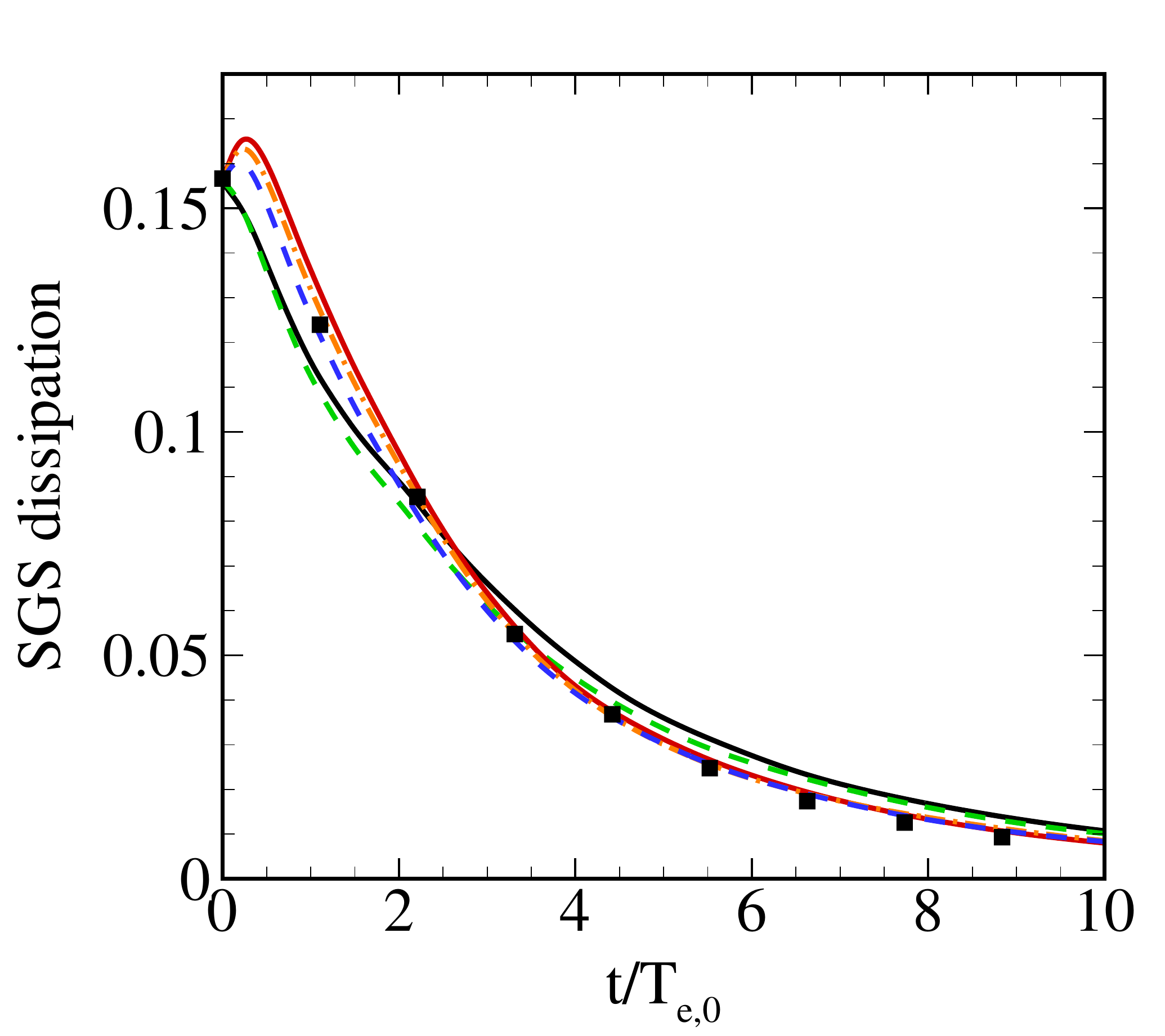}
		\caption{}
		\label{}
	\end{subfigure}
	\caption{Results from fDNS and LESs of decaying isotropic turbulence at the initial Reynolds number $Re_\lambda$ of 106 with grid resolution of $48^3$ (DHIT106 case). Temporal evolution of (a) the resolved kinetic energy and (b) the mean SGS dissipation  $\left<\varepsilon_{SGS}\right>$ ($= \left<-\tau_{ij}\overline{S}_{ij}\right>$) are shown. $\blacksquare$, fDNS; {\color{black}\sampleline{}}, DSM; {\color{red}\sampleline{}}, SL-106H; {\color{blue}\sampleline{dash pattern=on 6pt off 2pt}}, SL-286H; {\color{YellowOrange}\sampleline{dash pattern=on .7em off .2em on .2em off .2em}}, SL-106+286H; {\color{green}\sampleline{dash pattern=on 6pt off 2pt}}, S-106H.}
    \label{fig:DHIT106_ir_temporal_evol}
\end{figure}

Using $\left< \left| \tau^{grad}\right|\right>$ as the normalisation factor, ANN-SGS models trained with only forced homogeneous isotropic turbulence data are applied to LES for decaying homogeneous isotropic turbulence to assess the generalisability to untrained transient flow. Table~\ref{tab:DHIT_case} shows the initial Reynolds numbers and grid resolution of test LES cases for decaying isotropic turbulence. The numerical methods are the same as those used for LESs of forced isotropic turbulence. A fully converged instantaneous flow field of fDNS of forced homogeneous isotropic turbulence is selected as the initial field of decaying homogeneous isotropic turbulence. 

Figure~\ref{fig:DHIT106_ir} shows energy spectra from LESs of decaying homogeneous isotropic turbulence at the initial Reynolds number $Re_\lambda$ of 106 with grid resolution of $48^3$. Energy spectra are obtained at $t/T_{e,0} = 1.1, 3.3$, and $6.6$, where time $t$ is normalised by the initial eddy turnover time $T_{e,0}$. Energy spectra from S-106H and DSM are found to be almost identical. This is consistent with the \emph{a posteriori} test of forced isotropic turbulence where S-106H performs similar to DSM. S-106H and DSM overestimate the energy spectra in the range of $k < 5$, and errors increase as  flow evolves. In contrast, energy spectra from SL-106H, SL-286H, and SL-106+286H show better agreement with those of fDNS than DSM at all time steps.

Additionally, temporal evolution of the resolved kinetic energy and the mean SGS dissipation are investigated from the same simulations and presented in figure~\ref{fig:DHIT106_ir_temporal_evol}. DSM and S-106H predict smaller SGS dissipation than fDNS, which leads to a slower decaying rate of the resolved kinetic energy. Compared to DSM, SL-106H, SL-286H, and SL-106+286H significantly better predict temporal evolutions of the resolved kinetic energy and the SGS dissipation, which are in good agreement with those of fDNS.

Similarly to the results of forced isotropic turbulence, SL-106H, SL-286H, and SL-106+286H show significantly better performance than DSM for LES of decaying isotropic turbulence, which clearly indicates that the additional input variable $L_{ij}$ is beneficial. In addition, SL-106H, SL-286H, and SL-106+286H provide stable solutions without any \emph{ad-hoc} stabilisation procedures unlike DSM during the temporal evolution of the flow.

Note that SL-106H, SL-286H, and SL-106+286H show almost identical performance, although they are trained at different Reynolds numbers. This is because the present normalisation of the input and the output variables enables the mapping of an ANN between the input resolved variables and the output SGS stress to remain unchanged for various flow conditions. Therefore, in the present study, training an ANN-SGS model at a single Reynolds number and on a single grid resolution is found to be sufficient for generalisation to untrained flow conditions.

\begin{figure}
	\centering
	\begin{subfigure}[b]{0.49\textwidth}
		\includegraphics[width=\textwidth]{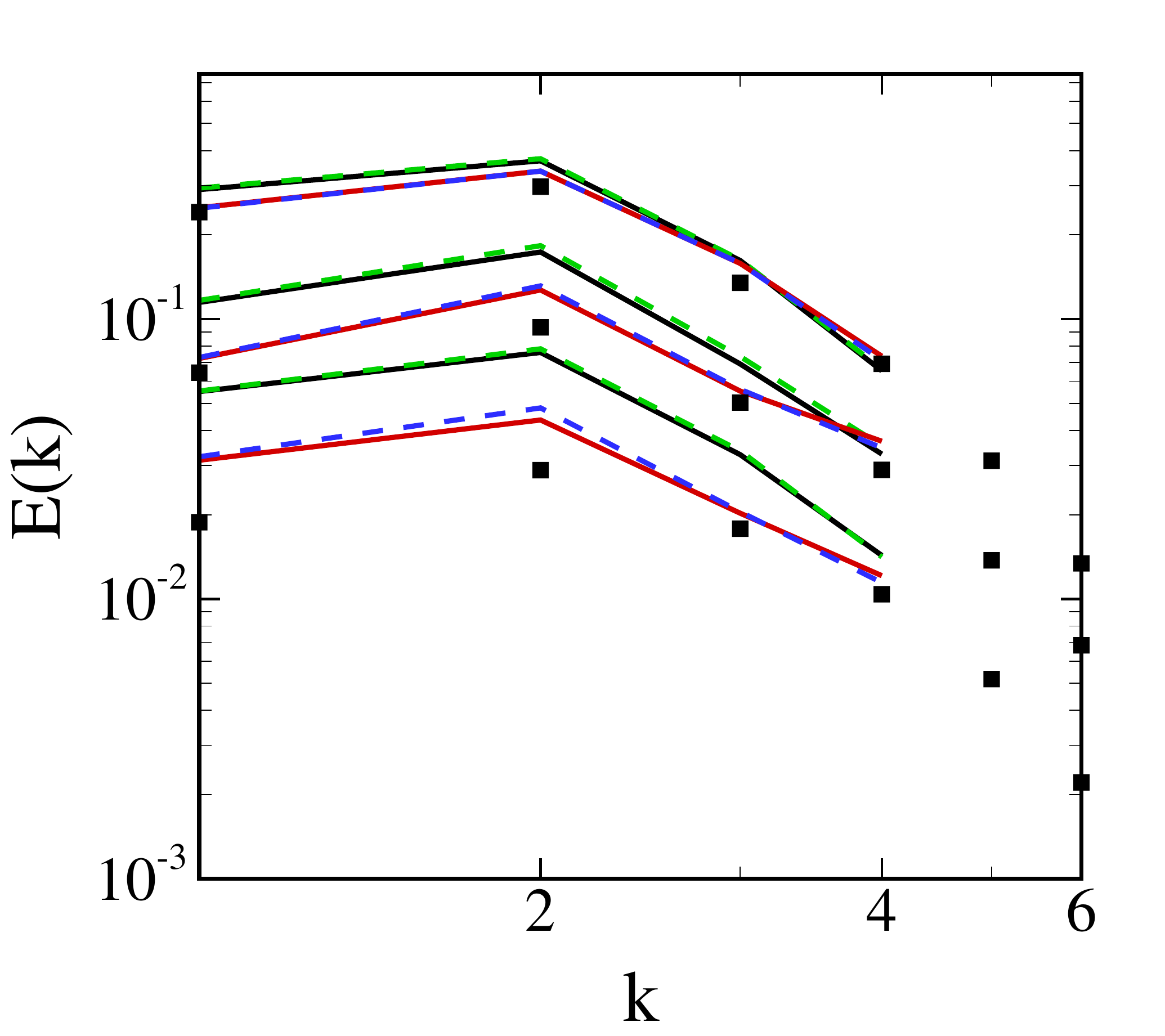}
		\caption{}
		\label{}
	\end{subfigure}
	\begin{subfigure}[b]{0.49\textwidth}
		\includegraphics[width=\textwidth]{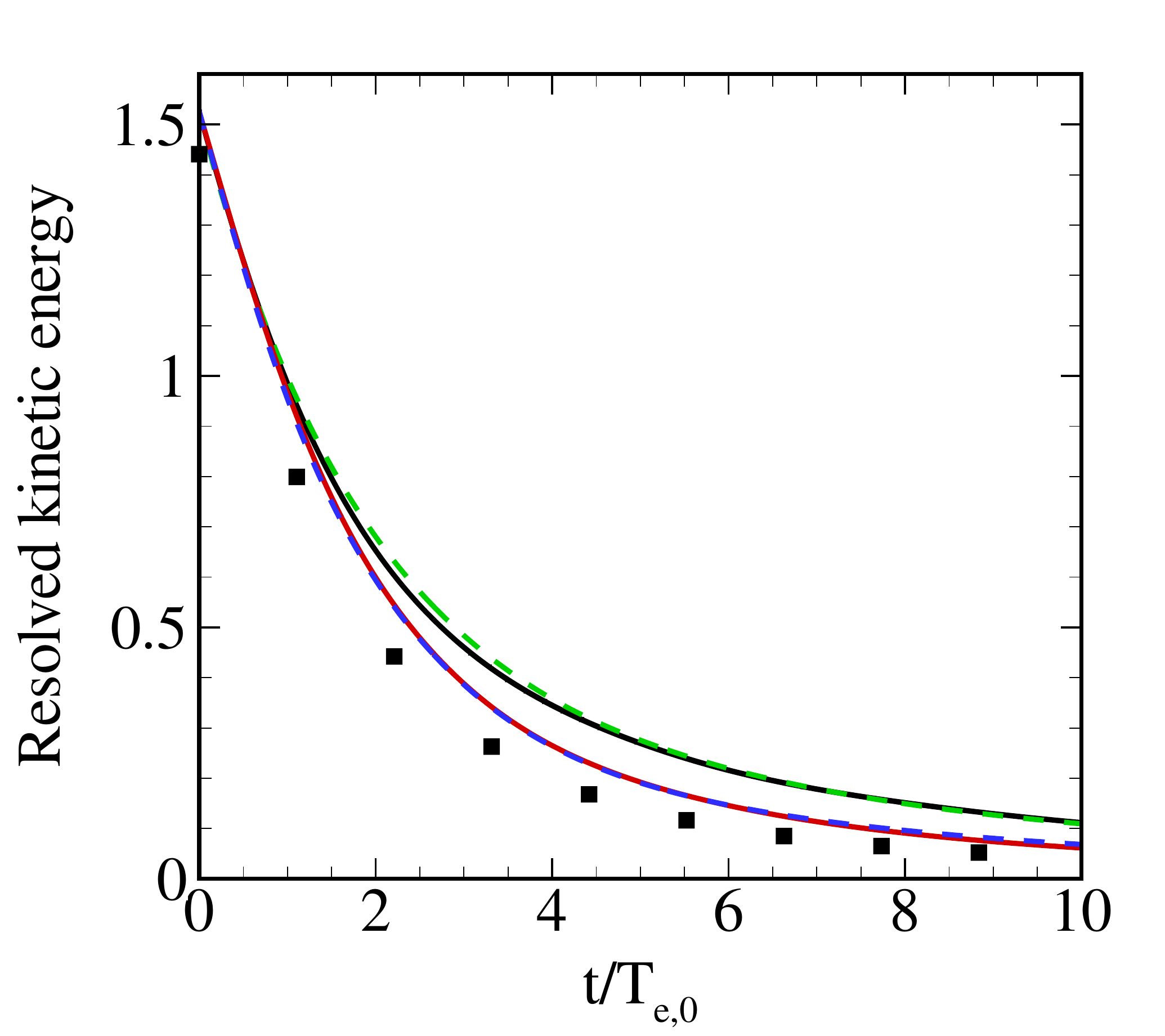}
		\caption{}
		\label{}
	\end{subfigure}
	\caption{Results from fDNS and LESs of decaying isotropic turbulence at the initial Reynolds number $Re_\lambda$ of 106 with grid resolution of $24^3$ (DHIT106c case). (a) Energy spectra at $t/T_{e,0} = 1.1$, $3.3$ and $6.6$; (b) temporal evolution of the resolved kinetic energy. $\blacksquare$, fDNS; {\color{black}\sampleline{}}, DSM; {\color{red}\sampleline{}}, SL-106H; {\color{blue}\sampleline{dash pattern=on 6pt off 2pt}}, SL-286H; {\color{green}\sampleline{dash pattern=on 6pt off 2pt}}, S-106H.}
    \label{fig:DHIT106_c}
\end{figure}

\begin{figure}
	\centering
	\begin{subfigure}[b]{0.49\textwidth}
		\includegraphics[width=\textwidth]{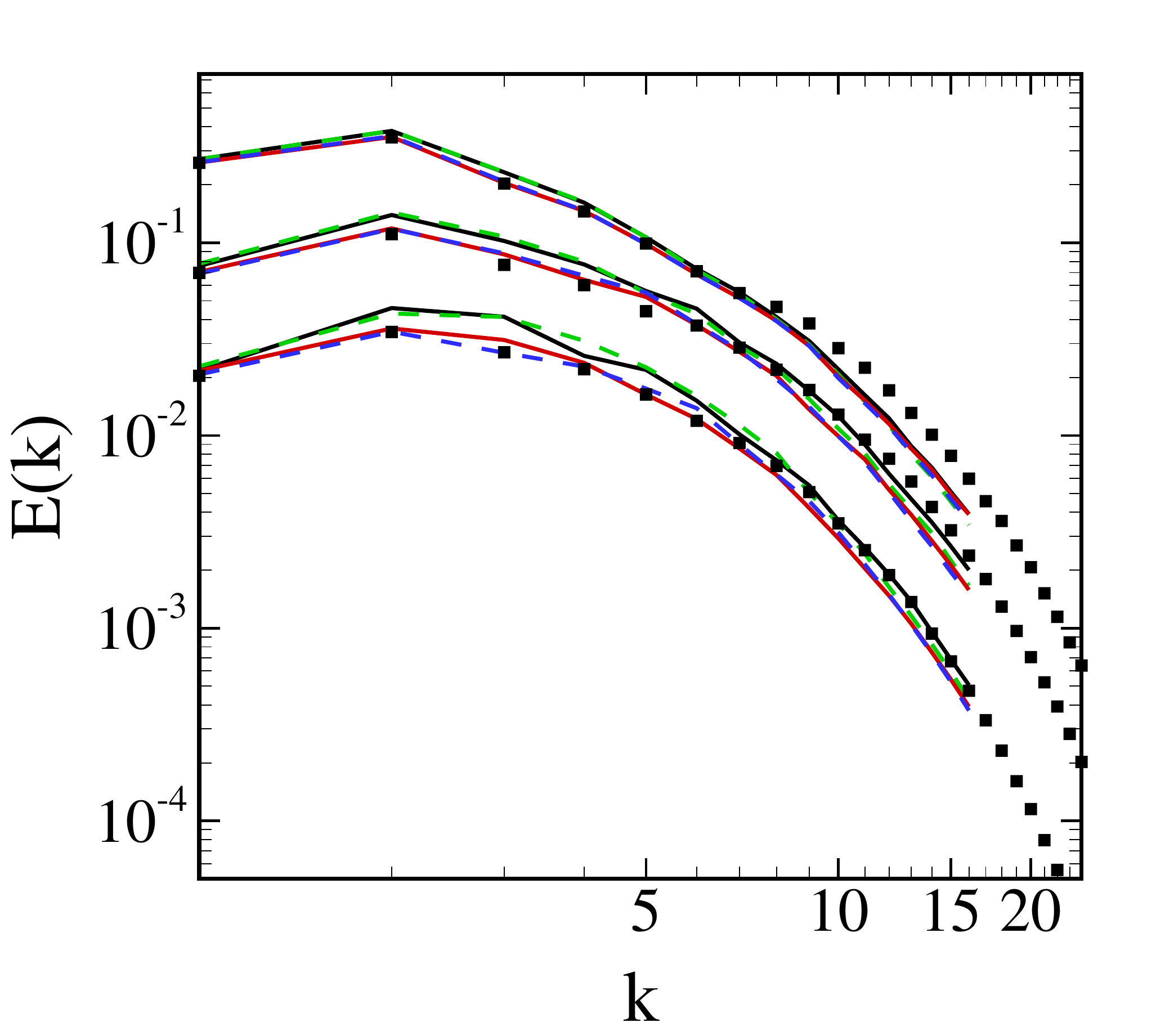}
		\caption{}
		\label{}
	\end{subfigure}
	\begin{subfigure}[b]{0.49\textwidth}
		\includegraphics[width=\textwidth]{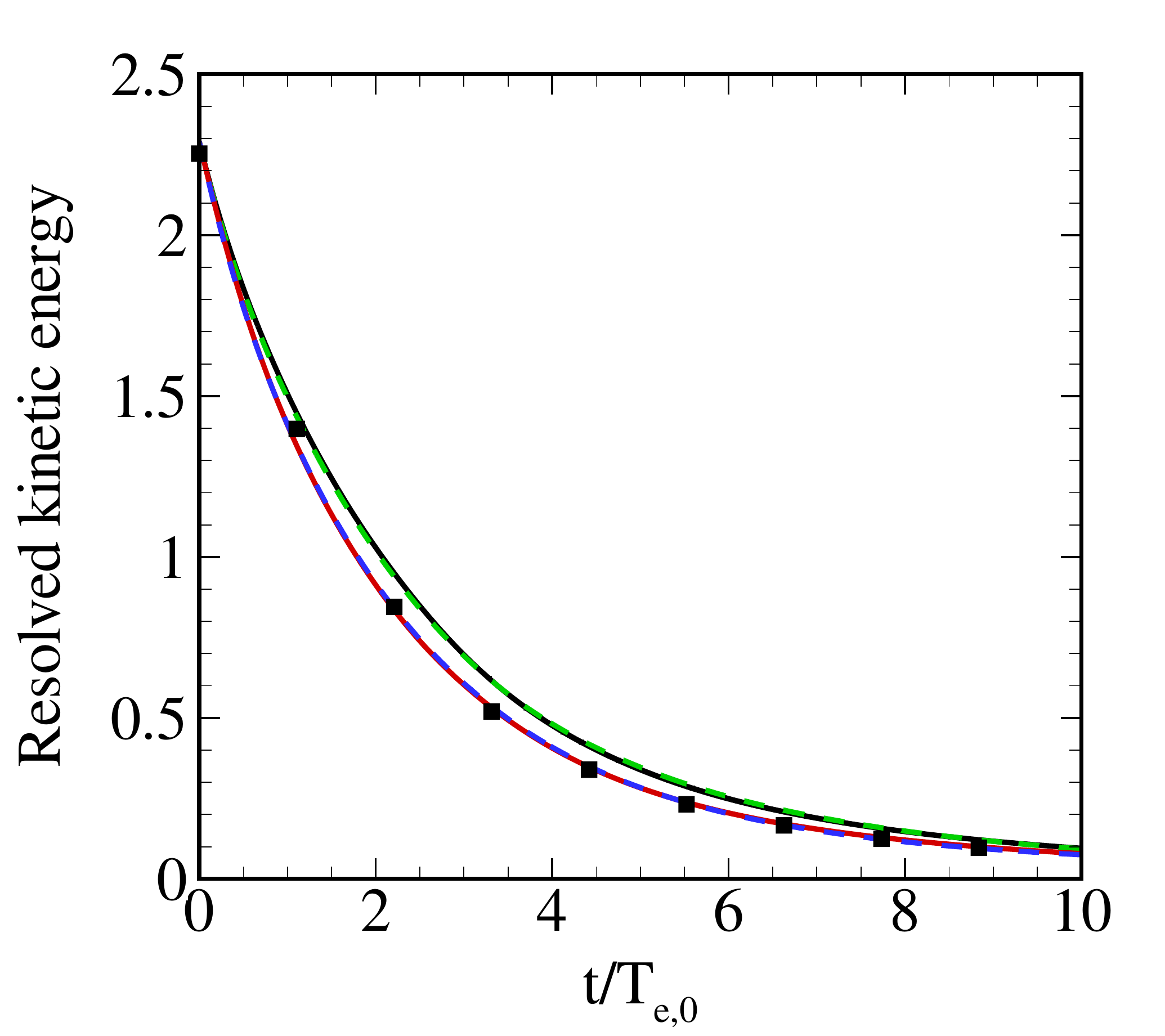}
		\caption{}
		\label{}
	\end{subfigure}
	\caption{Results from fDNS and LESs of decaying isotropic turbulence at the initial Reynolds number $Re_\lambda$ of 106 with grid resolution of $96^3$ (DHIT106f case). (a) Energy spectra at $t/T_{e,0} = 1.1$, $3.3$ and $6.6$; (b) temporal evolution of the resolved kinetic energy. $\blacksquare$, fDNS; {\color{black}\sampleline{}}, DSM; {\color{red}\sampleline{}}, SL-106H; {\color{blue}\sampleline{dash pattern=on 6pt off 2pt}}, SL-286H; {\color{green}\sampleline{dash pattern=on 6pt off 2pt}}, S-106H.}
    \label{fig:DHIT106_f}
\end{figure}

The performance of ANN-SGS models is further investigated through LESs of decaying isotropic turbulence with different grid resolution. Two times coarser ($24^3$) and finer ($96^3$) grid resolution than that of the training dataset ($48^3$) are considered. These are challenging cases, as the magnitudes of the SGS stress and the resolved kinetic energy differ from those of the training data as the turbulence decays. Nevertheless, SL-106H and SL-286H consistently provide better results than DSM.

LES of decaying homogeneous isotropic turbulence at the initial Reynolds number $Re_\lambda$ of 106 with grid resolution of $24^3$ is performed and results are shown in figure~\ref{fig:DHIT106_c}. DSM and S-106H show similar performances; they overestimate the energy spectra at all wavenumbers, and the resolved kinetic energy decays slower than that of fDNS. On the other hand, SL-106H and SL-286H provide more accurate predictions of the energy spectra than DSM, especially with smaller errors at $k \leq 2$. Furthermore, SL-106H and SL-286H predict the temporal evolution of the resolved kinetic energy more accurately than DSM and S-106H. 

Figure~\ref{fig:DHIT106_f} shows results from LES of decaying homogeneous isotropic turbulence at the initial Reynolds number $Re_\lambda$ of 106 with grid resolution of $96^3$. As the grid resolution becomes finer, the energy spectra and the temporal evolution of the resolved kinetic energy obtained by SL-106H and SL-286H become more accurate and almost overlapped with those of fDNS. However, DSM and S-106H overestimate energy spectra in the range of $k \leq 6$ and predict slower decaying rates of the resolved kinetic energy.

\begin{figure}
	\centering
	\begin{subfigure}[b]{0.49\textwidth}
		\includegraphics[width=\textwidth]{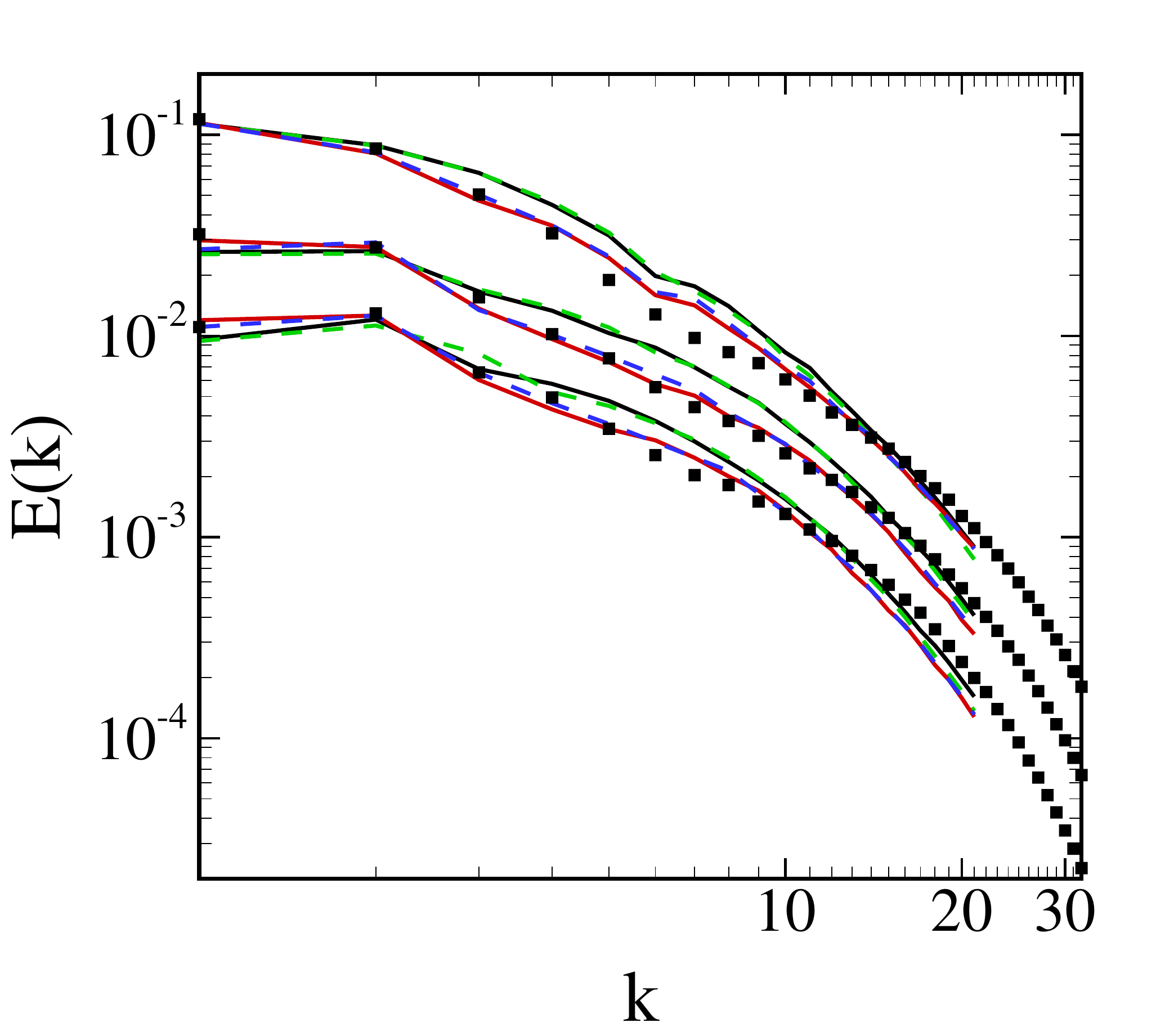}
		\caption{}
		\label{}
	\end{subfigure}
	\begin{subfigure}[b]{0.49\textwidth}
		\includegraphics[width=\textwidth]{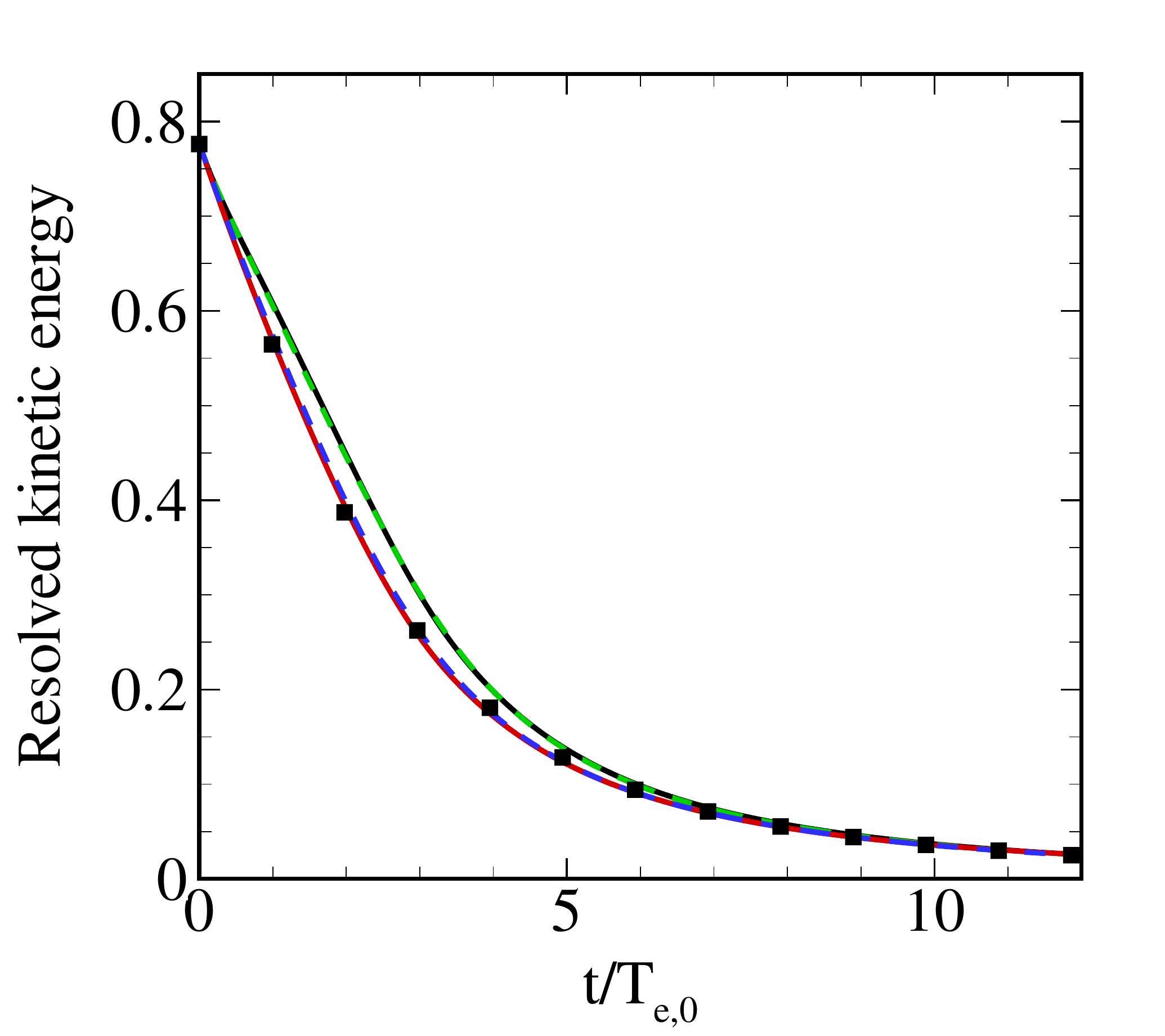}
		\caption{}
		\label{}
	\end{subfigure}
	\caption{Results from fDNS and LESs of decaying isotropic turbulence at the initial Reynolds number $Re_\lambda$ of 286 with grid resolution of $128^3$ (DHIT286 case). (a) Energy spectra at $t/T_{e,0} = 2.0, 4.9, 7.9$; (b) temporal evolution of the resolved kinetic energy. $\blacksquare$, fDNS; {\color{black}\sampleline{}}, DSM; {\color{red}\sampleline{}}, SL-106H; {\color{blue}\sampleline{dash pattern=on 6pt off 2pt}}, SL-286H; {\color{green}\sampleline{dash pattern=on 6pt off 2pt}}, S-106H.}
    \label{fig:DHIT286_ir}
\end{figure}

The performance of ANN-SGS models is tested through LES at a higher initial Reynolds number $Re_\lambda$ of 286. Figure~\ref{fig:DHIT286_ir} shows results from LES of decaying homogeneous isotropic turbulence with grid resolution of $128^3$, in which the cut-off wavenumber is located in the inertial range, same as in the training data. Notably, SL-106H is successfully generalised to the higher Reynolds number, and it predicts the energy spectra and the temporal evolution of the resolved kinetic energy more accurately than DSM. In figure~\ref{fig:DHIT286_ir}(a), SL-106H and SL-286H accurately predict the energy spectra, which are in good agreement with those of fDNS. In contrast, DSM and S-106H slightly underestimate the energy at low wavenumbers ($k \leq 2$), while they overestimated the energy at $4 \leq k \leq  10$. In figure~\ref{fig:DHIT286_ir}(b), temporal evolutions of the resolved kinetic energy of SL-106H and SL-286H are well matched with that of fDNS at all time steps, whereas DSM and S-106H show slower decaying rates of the resolved kinetic energy until $t/T_{e,0} \approx 5$.

Despite the fact that the present ANN-SGS models are not trained for the transient characteristics of decaying isotropic turbulence, they can be successfully utilised for LES of transient decaying isotropic turbulence with various initial Reynolds numbers and grid resolution. This is again possible due to the selection of proper normalisation factors for the input and the output variables such that the normalisation enables the mapping of an ANN between the input resolved variables and the output SGS stress to remain unchanged during the temporal decay of turbulence. 

\subsection{Comparison with algebraic dynamic mixed models}\label{sec:DHIT_mixed}

In Section~\ref{sec:FHIT}, the predictive performance of SL-106H is found to meet the expectation that the use of both $\overline{S}_{ij}$ and $L_{ij}$ as inputs could perform better than DSM due to better prediction of the SGS stress in terms of the magnitude and alignment to the true SGS stress. Consequently, it is of interest to investigate how well the ANN-SGS mixed model performs compared with the algebraic dynamic mixed models. Therefore, the performance of the developed ANN-SGS mixed model is compared with one-parameter~\citep{zang1993, vreman1994} and two-parameter~\citep{anderson1999} algebraic dynamic mixed models through LESs of both forced and decaying isotropic turbulence. 


The one-parameter dynamic mixed model that combines the eddy-viscosity model with the modified Leonard term $(\mathcal{L}_{ij}^m = \overline{\overline{u}_i \overline{u}_j} - \overline{\overline{u}}_i \overline{\overline{u}}_j)$, which was formulated by \citet{zang1993} and later modified by \citet{vreman1994} is denoted as DMM. The model coefficient of DMM is dynamically determined to close the SGS stress as follows:
\begin{equation}
    \tau_{ij} - \frac{1}{3}\delta_{ij}\tau_{kk} = -2C\mybar{\Delta}^2|\mybar{S}|\mybar{S}_{ij} + \mathcal{L}_{ij}^m - \frac{1}{3}\delta_{ij}\mathcal{L}_{kk}^m,
    \label{eq:DMM}
\end{equation}
where $C=\left< M_{ij}(L_{ij}-H_{ij})\right>/\left< M_{ij}M_{ij}\right>$,
$M_{ij}=-2\myhat{\Delta}^2|\myhat{\mybar{S}}|\myhat{\mybar{S}}_{ij}+2\mybar{\Delta}^2|\widehat{\mybar{S}|\mybar{S}_{ij}}$, $L_{ij}=\myhat{\mybar{u}_i\mybar{u}_j}-\myhat{\mybar{u}}_i\myhat{\mybar{u}}_j$, $H_{ij}=\myhat{\mybar{\myhat{\mybar{u}}_i\myhat{\mybar{u}}_j}}-\myhat{\mybar{\myhat{\mybar{u}}_i}}\myhat{\mybar{\myhat{\mybar{u}}_j}}-(\myhat{\mybar{\mybar{u}_i\mybar{u}_j}}-\myhat{\mybar{\mybar{u}}_i\mybar{\mybar{u}}_j})$, $\mathcal{L}_{ij}^m = \overline{\overline{u}_i \overline{u}_j} - \overline{\overline{u}}_i \overline{\overline{u}}_j$.

\citet{anderson1999} formulated the dynamic two-parameter mixed model ($DTM_{sim}$), which combines the eddy-viscosity model with the resolved stress and calculates two model coefficients for eddy-viscosity and mixed terms using the dynamic procedure as follows:
\begin{equation}
    \tau_{ij} = -2C_{1}\mybar{\Delta}^2|\mybar{S}|\mybar{S}_{ij} + C_{2}L_{ij},
    \label{eq:DTM_sim}
\end{equation}

\begin{equation}
    C_{1} = \frac{\left<L_{ij}M_{ij}\right>\left<N_{ij}N_{ij}\right> - \left<L_{ij}N_{ij}\right>\left<M_{ij}N_{ij}\right>}{\left<M_{ij}M_{ij}\right>\left<N_{ij}N_{ij}\right> - \left<N_{ij}M_{ij}\right>^2},
    \label{eq:DTM_c1}
\end{equation}

\begin{equation}
    C_{2} = \frac{\left<L_{ij}N_{ij}\right>\left<M_{ij}M_{ij}\right> - \left<L_{ij}M_{ij}\right>\left<M_{ij}N_{ij}\right>}{\left<M_{ij}M_{ij}\right>\left<N_{ij}N_{ij}\right> - \left<N_{ij}M_{ij}\right>^2},
    \label{eq:DTM_c2}
\end{equation}
where $L_{ij}=\myhat{\mybar{u}_i\mybar{u}_j}-\myhat{\mybar{u}}_i\myhat{\mybar{u}}_j$, $M_{ij}=-2\myhat{\Delta}^2|\myhat{\mybar{S}}|\myhat{\mybar{S}}_{ij}+2\mybar{\Delta}^2|\widehat{\mybar{S}|\mybar{S}_{ij}}$, $N_{ij}=(\widetilde{\myhat{\mybar{u}}_i\myhat{\mybar{u}}_j}-\widetilde{\myhat{\mybar{u}}_i}\widetilde{\myhat{\mybar{u}}_j}) - (\myhat{\myhat{\mybar{u}_i\mybar{u}_j}}-\myhat{\myhat{\mybar{u}}_i\myhat{\mybar{u}}_j})$, 
and $\widetilde{\left(\,\cdot\,\right)}$ denotes the filtering at the $5\mybar{\Delta}$ scale.

\citet{anderson1999} proposed another dynamic two-parameter mixed model ($DTM_{nl}$), which combines the eddy-viscosity model with the Clark model as follows:
\begin{equation}
    \tau_{ij} = -2C_{1}\mybar{\Delta}^2|\mybar{S}|\mybar{S}_{ij} + C_{2}\mybar{\Delta}^2\frac{\partial \overline{u}_i}{\partial x_k}\frac{\partial \overline{u}_j}{\partial x_k},
    \label{eq:DTM_nl}
\end{equation}
where $C_{1}$ and $C_{2}$ are determined by (\ref{eq:DTM_c1}) and (\ref{eq:DTM_c2}), respectively, with $L_{ij}=\myhat{\mybar{u}_i\mybar{u}_j}-\myhat{\mybar{u}}_i\myhat{\mybar{u}}_j$, $M_{ij}=-2\myhat{\Delta}^2|\myhat{\mybar{S}}|\myhat{\mybar{S}}_{ij}+2\mybar{\Delta}^2|\widehat{\mybar{S}|\mybar{S}_{ij}}$,
$N_{ij}=\myhat{\Delta}^2\frac{\partial \myhat{\mybar{u}}_i}{\partial x_k}\frac{\partial \myhat{\mybar{u}}_j}{\partial x_k} - \mybar{\Delta}^2\myhat{\frac{\partial \mybar{u}_i}{\partial x_k}\frac{\partial \mybar{u}_j}{\partial x_k}}$.

\begin{figure}
\centering \
\includegraphics[width=0.60\textwidth]{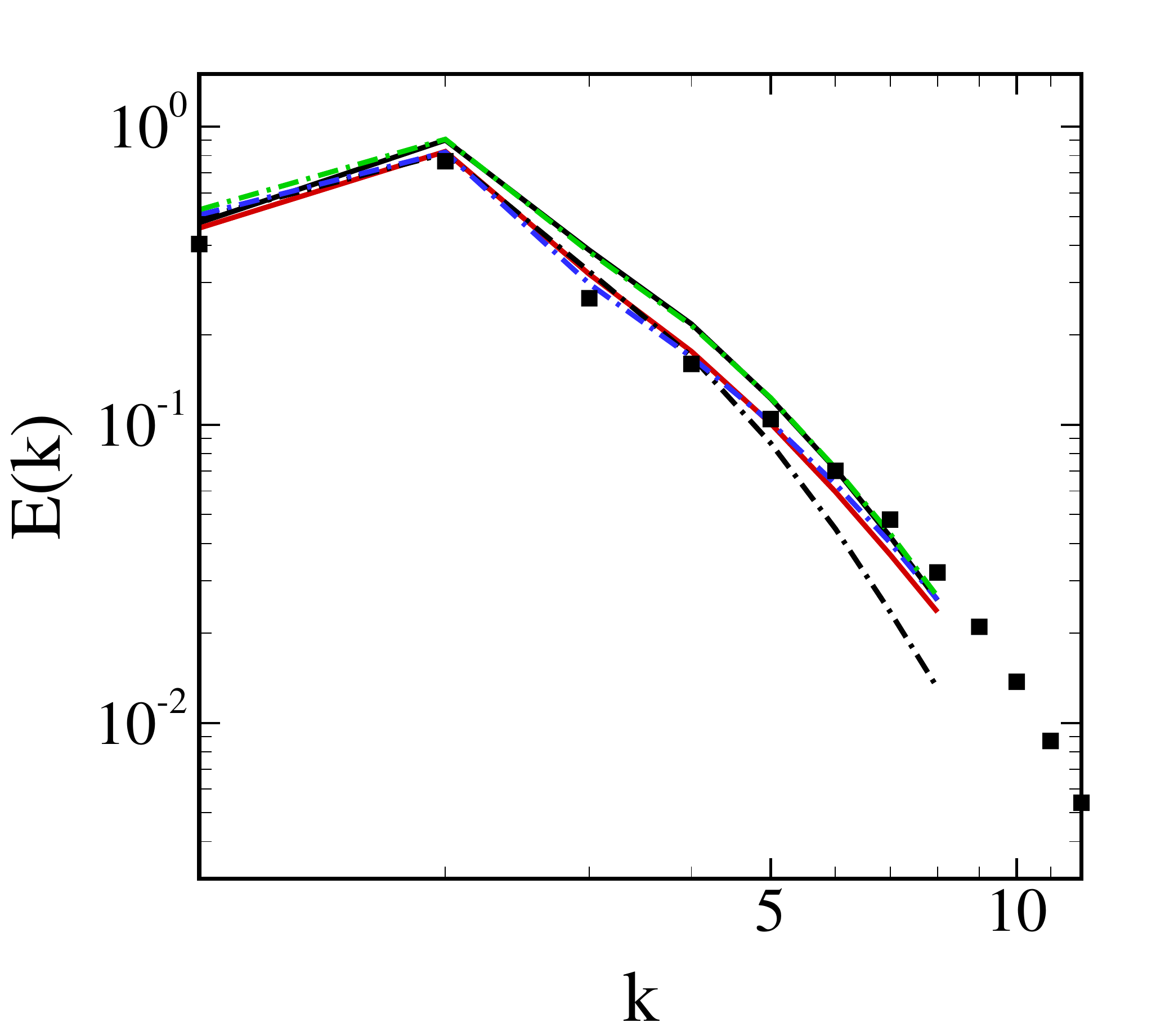}
\caption{Energy spectra from fDNS and LESs of forced homogeneous isotropic turbulence at $Re_\lambda=106$ with grid resolution of $48^3$. $\blacksquare$, fDNS; {\color{black}\sampleline{}}, DSM; {\color{black}\sampleline{dash pattern=on .7em off .2em on .2em off .2em}}, DMM; {\color{blue}\sampleline{dash pattern=on .7em off .2em on .2em off .2em}}, $DTM_{nl}$; {\color{green}\sampleline{dash pattern=on .7em off .2em on .2em off .2em}}, $DTM_{sim}$; {\color{red}\sampleline{}}, SL-106H.}
\label{fig:FHIT106_ir_mixed_es}
\end{figure}

\begin{figure}
	\centering
	\begin{subfigure}[b]{0.49\textwidth}
		\includegraphics[width=\textwidth]{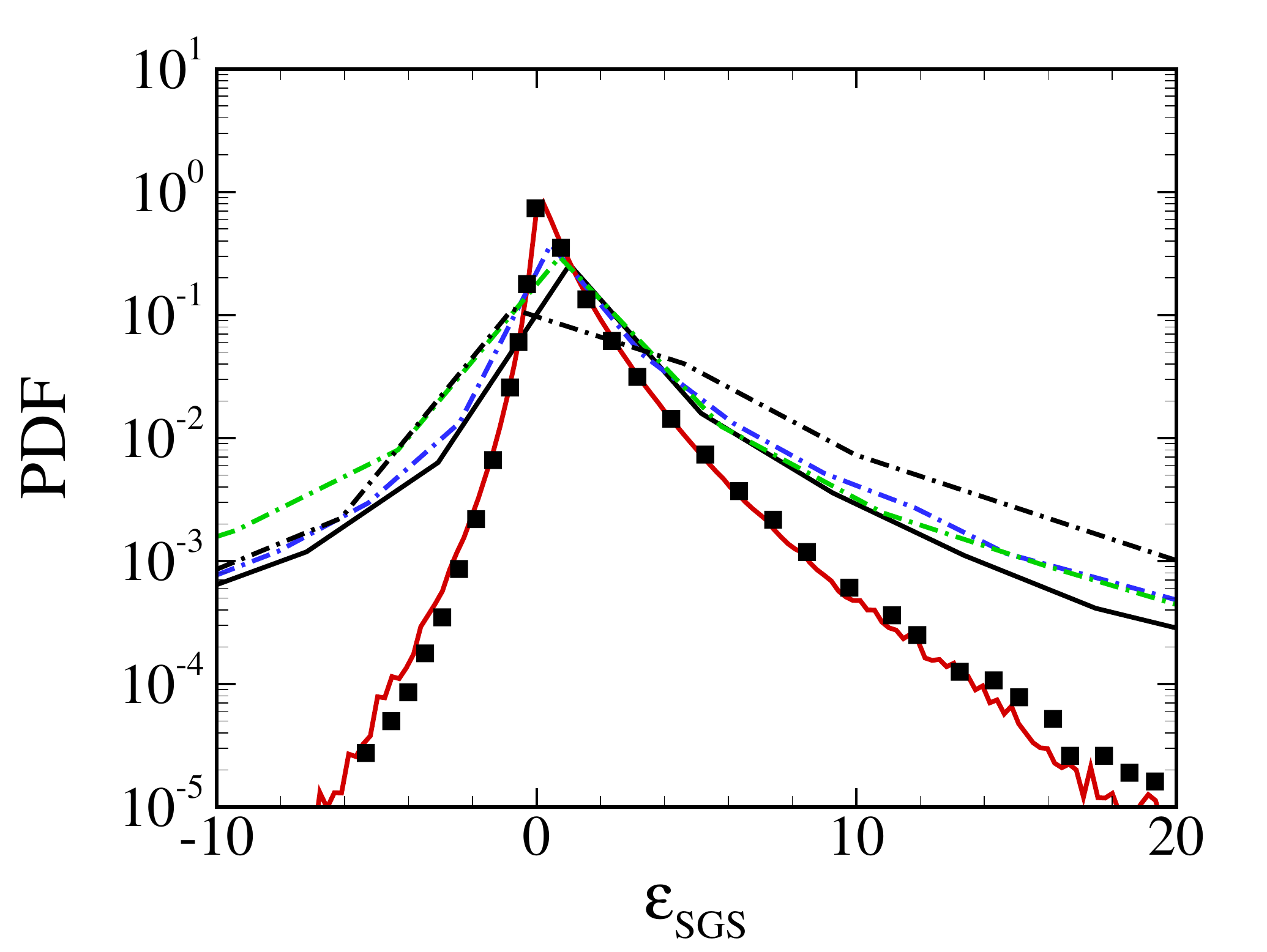}
		\caption{}
		\label{}
	\end{subfigure}
	\begin{subfigure}[b]{0.49\textwidth}
		\includegraphics[width=\textwidth]{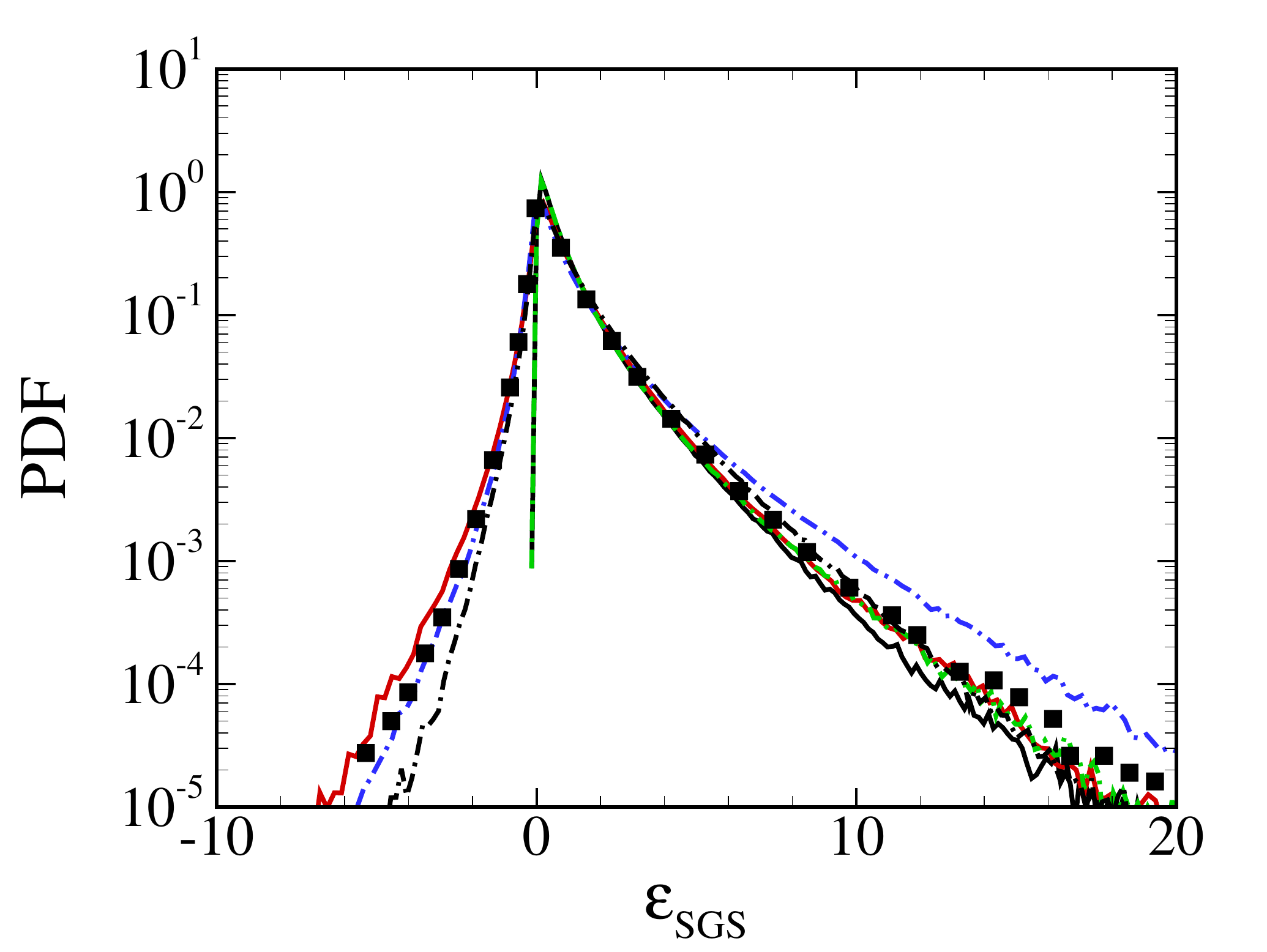}
		\caption{}
		\label{}
	\end{subfigure}
	\caption{PDF of the SGS dissipation $\varepsilon_{SGS}$ from LES of forced homogeneous isotropic turbulence at $Re_\lambda=106$ with grid resolution of $48^3$. (a) Without \emph{ad-hoc} stabilisation; (b) with \emph{ad-hoc} stabilisation including averaging of the model coefficients in statistically homogeneous directions and clipping of the negative model coefficients. $\blacksquare$, fDNS; {\color{black}\sampleline{}}, DSM; {\color{black}\sampleline{dash pattern=on .7em off .2em on .2em off .2em}}, DMM; {\color{blue}\sampleline{dash pattern=on .7em off .2em on .2em off .2em}}, $DTM_{nl}$; {\color{green}\sampleline{dash pattern=on .7em off .2em on .2em off .2em}}, $DTM_{sim}$; {\color{red}\sampleline{}}, SL-106H.}
    \label{fig:FHIT106_ir_mixed_pdf}
\end{figure}

Figure~\ref{fig:FHIT106_ir_mixed_es} shows energy spectra from LESs of forced isotropic turbulence at $Re_\lambda=106$ with grid resolution of $48^3$, including those from the algebraic dynamic mixed models. Energy spectra from DMM and $DTM_{nl}$ show good agreement with that of fDNS, especially in the range of $k < 5$. However, the energy spectrum of DMM deviates from that of fDNS in the range of $k \geq 5$, showing large errors near the cut-off wavenumber.  $DTM_{nl}$ accurately predicts the energy spectrum at all wavenumbers, whereas the energy spectrum predicted by $DTM_{sim}$ is almost identical to that of DSM. Thus, $DTM_{nl}$ is the most accurate among the tested dynamic SGS models. In comparison with $DTM_{nl}$, SL-106H produces a relatively accurate prediction of the energy spectrum at $k=1$ and nearly identical at other wavenumbers in LES of forced isotropic turbulence. Notably, SL-106H shows the best prediction of the SGS dissipation, whereas DMM, DSM, and $DTM_{sim}$ underestimate the backscatter and $DTM_{nl}$ overestimates the forward-scatter, as shown in figure~\ref{fig:FHIT106_ir_mixed_pdf}(b).

\begin{figure}
	\centering
	\begin{subfigure}[b]{0.49\textwidth}
		\includegraphics[width=\textwidth]{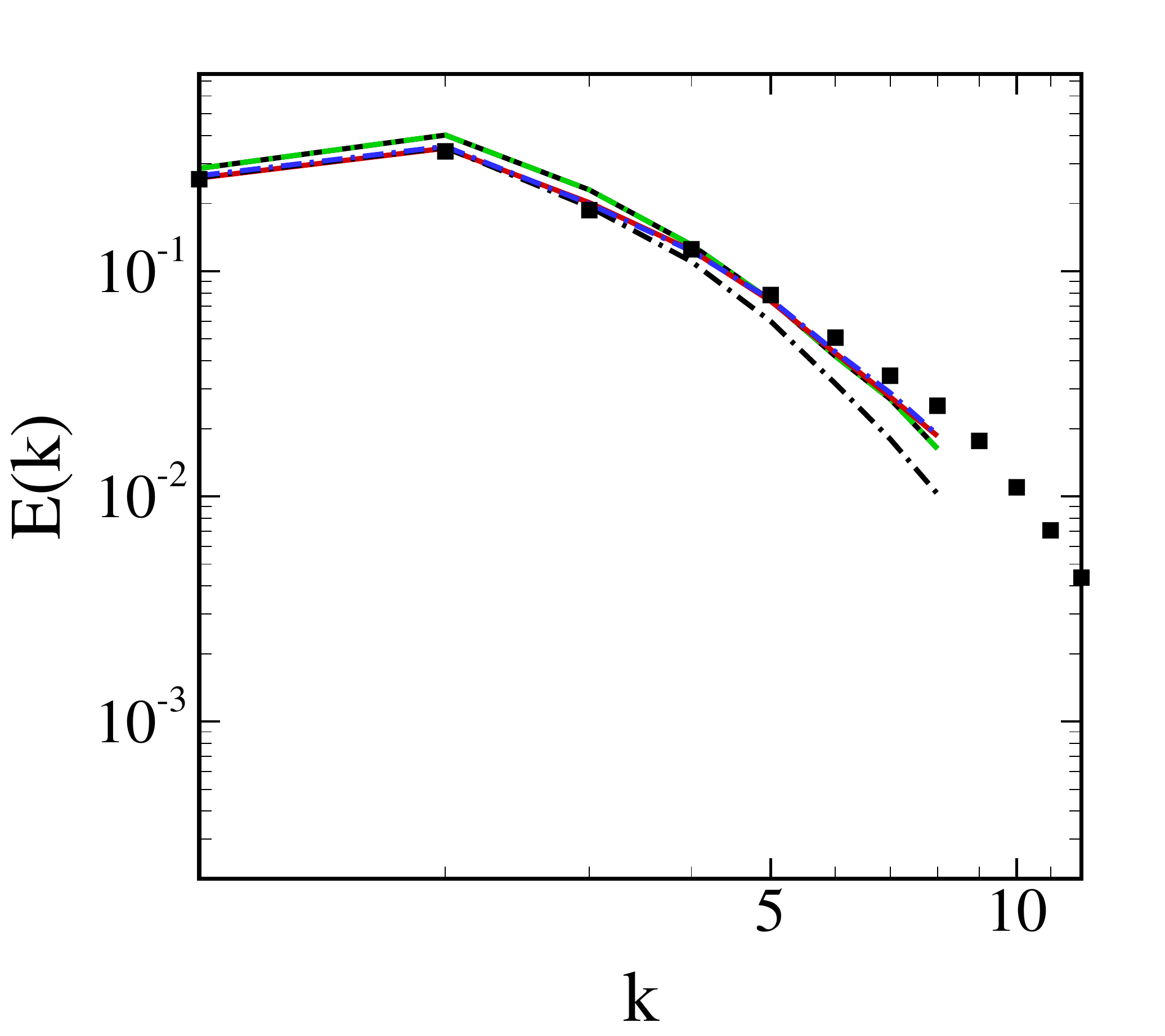}
		\caption{}
		\label{}
	\end{subfigure}
	\begin{subfigure}[b]{0.49\textwidth}
		\includegraphics[width=\textwidth]{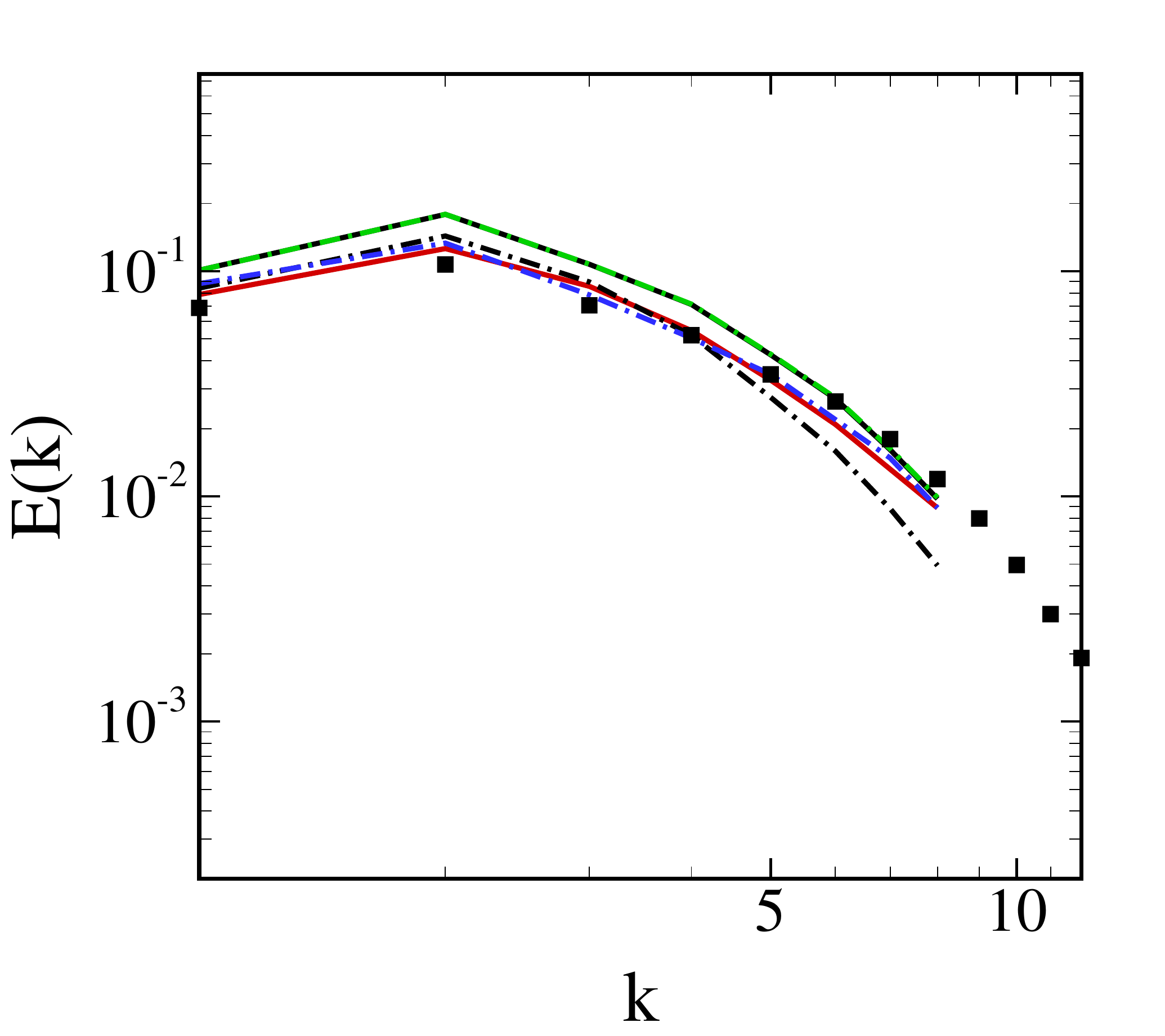}
		\caption{}
		\label{}
	\end{subfigure}
	\begin{subfigure}[b]{0.49\textwidth}
		\includegraphics[width=\textwidth]{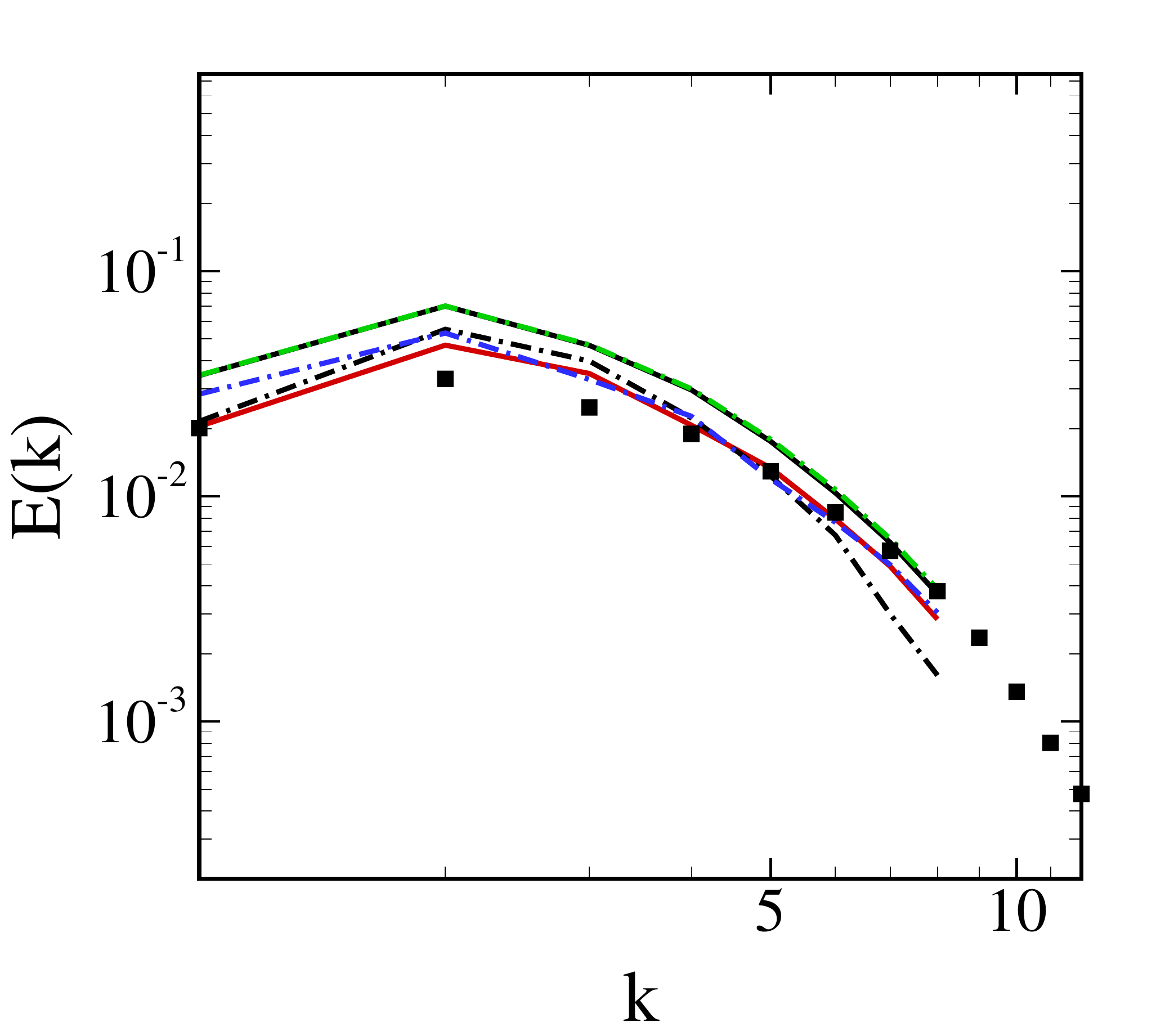}
		\caption{}
		\label{}
	\end{subfigure}
	\caption{Energy spectra from fDNS and LESs of decaying isotropic turbulence at the initial Reynolds number $Re_\lambda=106$ with grid resolution of $48^3$ (DHIT106 case). (a) $t/T_{e,0} = 1.1$; (b) $t/T_{e,0} = 3.3$; (c) $t/T_{e,0} = 6.6$. $\blacksquare$, fDNS; {\color{black}\sampleline{}}, DSM; {\color{black}\sampleline{dash pattern=on .7em off .2em on .2em off .2em}}, DMM; {\color{blue}\sampleline{dash pattern=on .7em off .2em on .2em off .2em}}, $DTM_{nl}$; {\color{green}\sampleline{dash pattern=on .7em off .2em on .2em off .2em}}, $DTM_{sim}$; {\color{red}\sampleline{}}, SL-106H.}
    \label{fig:DHIT106_ir_mixed}
\end{figure}

\begin{figure}
	\centering
	\begin{subfigure}[b]{0.49\textwidth}
		\includegraphics[width=\textwidth]{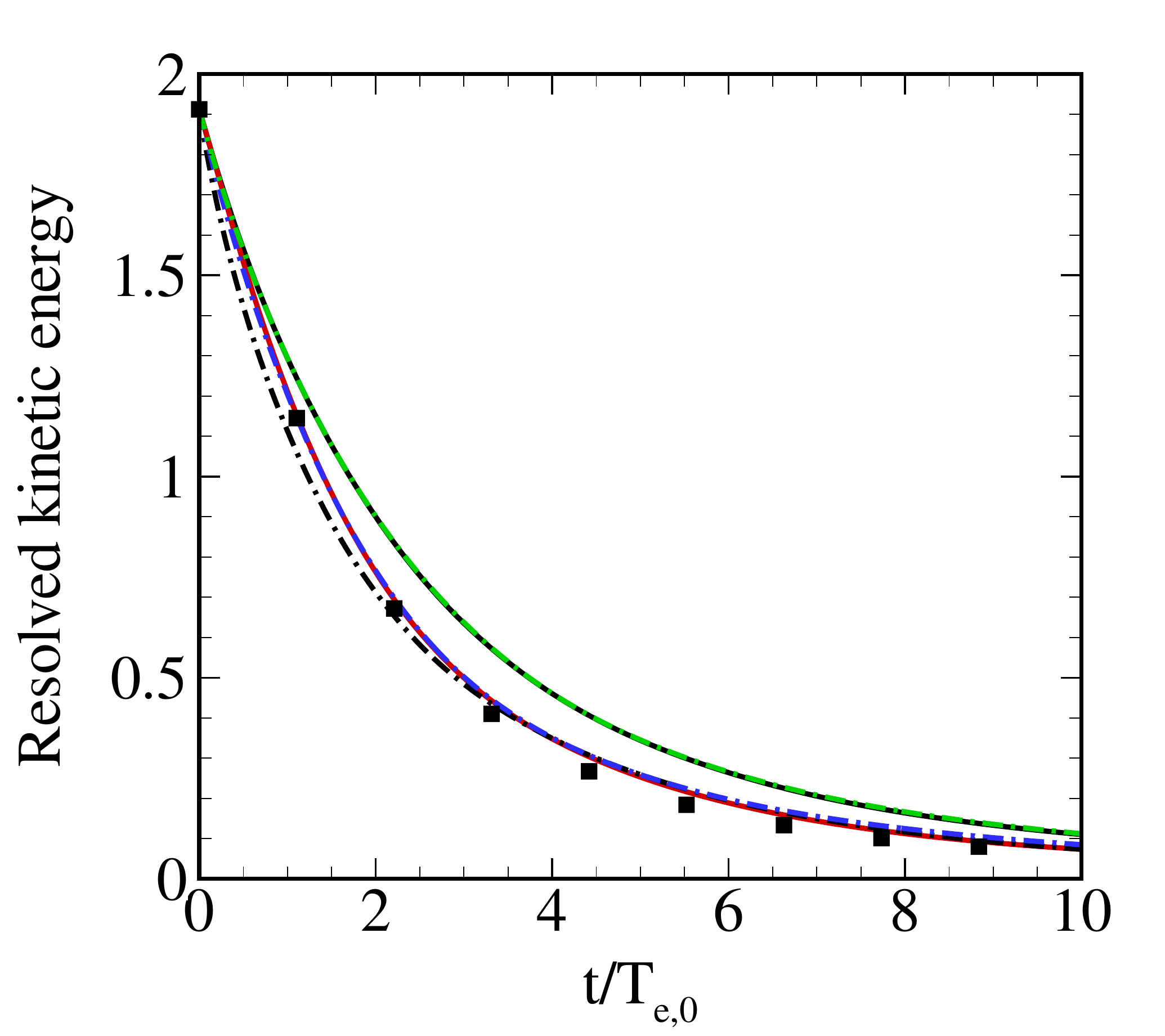}
		\caption{}
		\label{}
	\end{subfigure}
	\begin{subfigure}[b]{0.49\textwidth}
		\includegraphics[width=\textwidth]{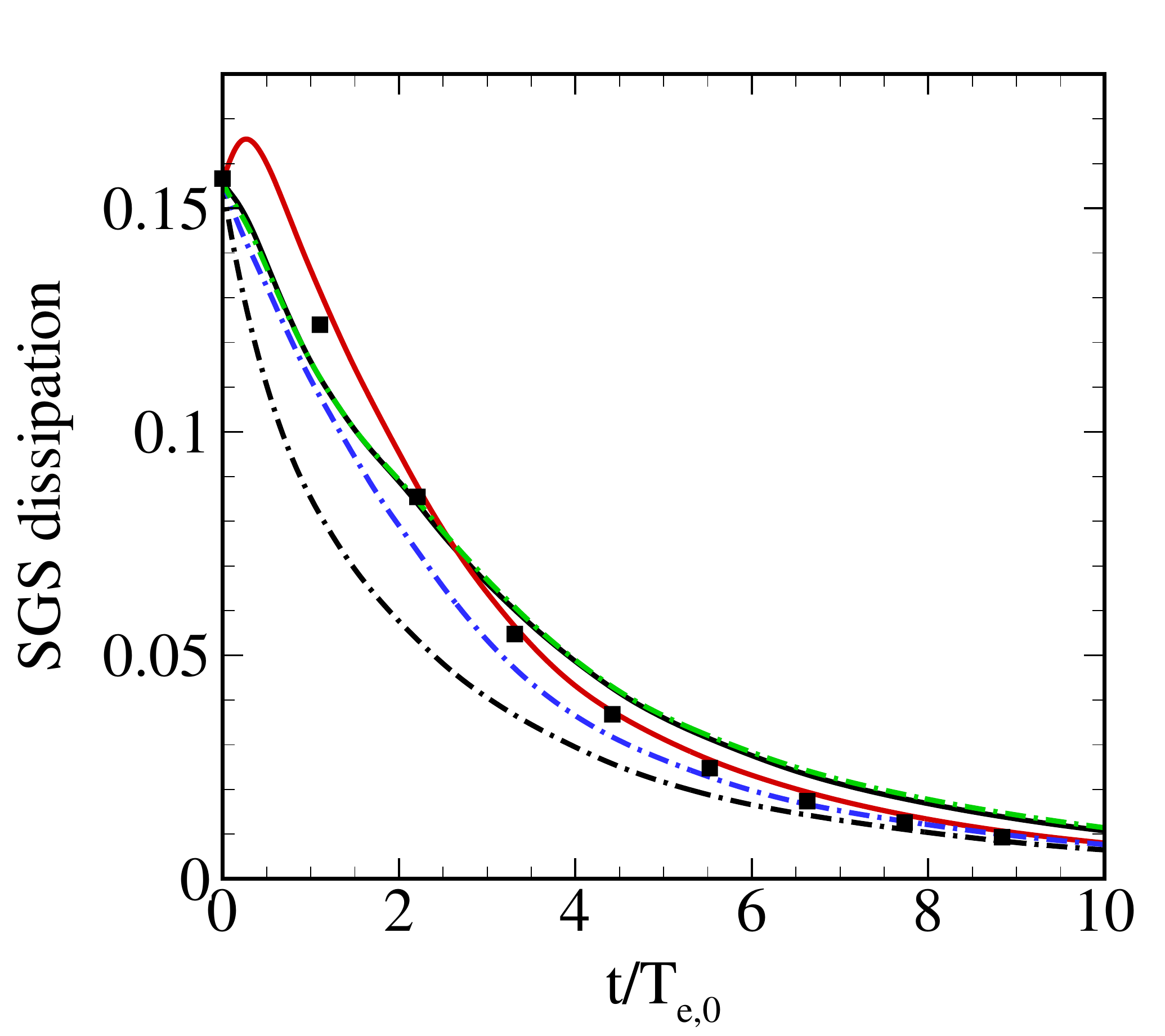}
		\caption{}
		\label{}
	\end{subfigure}
	\caption{Results from fDNS and LESs of decaying isotropic turbulence at the initial Reynolds number $Re_\lambda$ of 106 with grid resolution of $48^3$ (DHIT106 case). Temporal evolution of (a) the resolved kinetic energy and (b) the mean SGS dissipation  $\left<\varepsilon_{SGS}\right>$ ($= \left<-\tau_{ij}\overline{S}_{ij}\right>$) are shown. $\blacksquare$, fDNS; {\color{black}\sampleline{}}, DSM; {\color{black}\sampleline{dash pattern=on .7em off .2em on .2em off .2em}}, DMM; {\color{blue}\sampleline{dash pattern=on .7em off .2em on .2em off .2em}}, $DTM_{nl}$; {\color{green}\sampleline{dash pattern=on .7em off .2em on .2em off .2em}}, $DTM_{sim}$; {\color{red}\sampleline{}}, SL-106H.}
    \label{fig:DHIT106_ir_temporal_evol_mixed}
\end{figure}

Note that the algebraic dynamic SGS models tested in the present study are stable only after applying \emph{ad-hoc} procedures (\emph{i.e.}, averaging of the model coefficients in statistically homogeneous directions and clipping of the negative model coefficients). A possible reason for this problem can be found from the PDF of the SGS dissipation without \emph{ad-hoc} procedures, as shown in figure~\ref{fig:FHIT106_ir_mixed_pdf}(a). DSM and algebraic dynamic mixed models significantly overestimate both backscatter and forward-scatter when the \emph{ad-hoc} procedures are not applied. The excessive backscatter causes LES to diverge within tens of time steps. In contrast, LES with SL-106H is stable without any \emph{ad-hoc} procedures, since SL-106H accurately predicts the PDF of the SGS dissipation, which is a notable advantage of the present ANN-SGS mixed model. 

Additionally, LES of decaying isotropic turbulence at the initial Reynolds number $Re_\lambda$ of  106 with grid resolution of $48^3$ is performed, and the results are shown in figures~\ref{fig:DHIT106_ir_mixed} and~\ref{fig:DHIT106_ir_temporal_evol_mixed}. Similarly to the results from LES of forced isotropic turbulence, DMM accurately predicts energy spectra at $k < 5$ but underpredicts the energy at $k \geq 5$. On the other hand, energy spectra of $DTM_{nl}$ are accurate at $k \geq 5$ but the energy of large scale at $k \leq 2$ is slightly overestimated at $t/T_{e,0} = 3.3$ and $6.6$. Compared to the algebraic dynamic mixed models, SL-106H predicts most accurately the energy spectra at all time steps; especially, it shows the smallest error in the range of $k < 5$. In addition, SL-106H most accurately predicts the temporal evolution of the resolved kinetic energy and the mean SGS dissipation, as shown in figure~\ref{fig:DHIT106_ir_temporal_evol_mixed}. This indicates that the present ANN-SGS mixed model with input tensors of $\overline{S}_{ij}$ and $L_{ij}$ is able to predict the decay of isotropic turbulence more accurately than the algebraic dynamic mixed models.

\subsection{Computational cost of ANN-based SGS models}\label{sec:DHIT_cost}

\begin{figure}
	\centering
	\begin{subfigure}[b]{0.49\textwidth}
		\includegraphics[width=\textwidth]{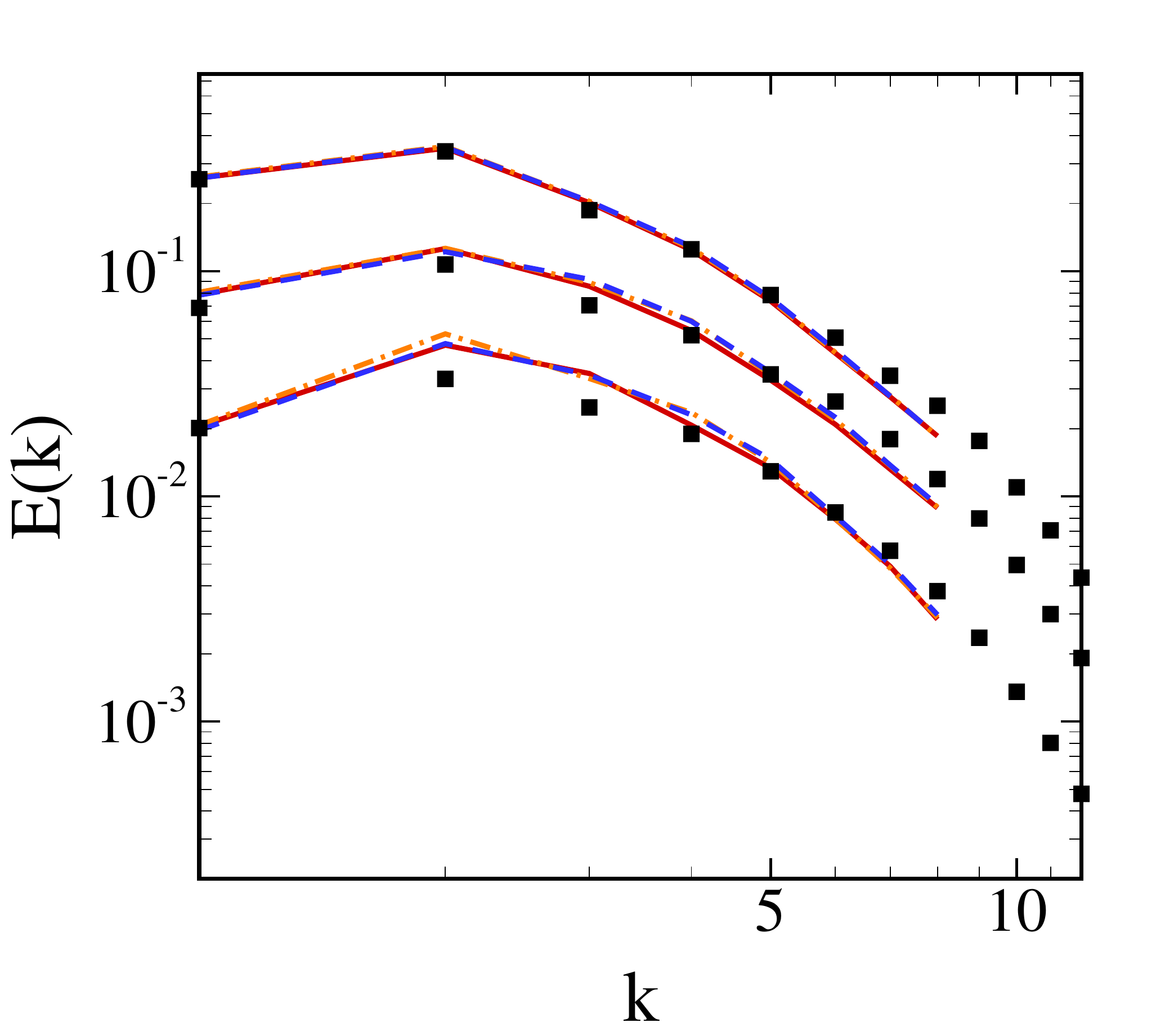}
		\caption{}
		\label{}
	\end{subfigure}
	\begin{subfigure}[b]{0.49\textwidth}
		\includegraphics[width=\textwidth]{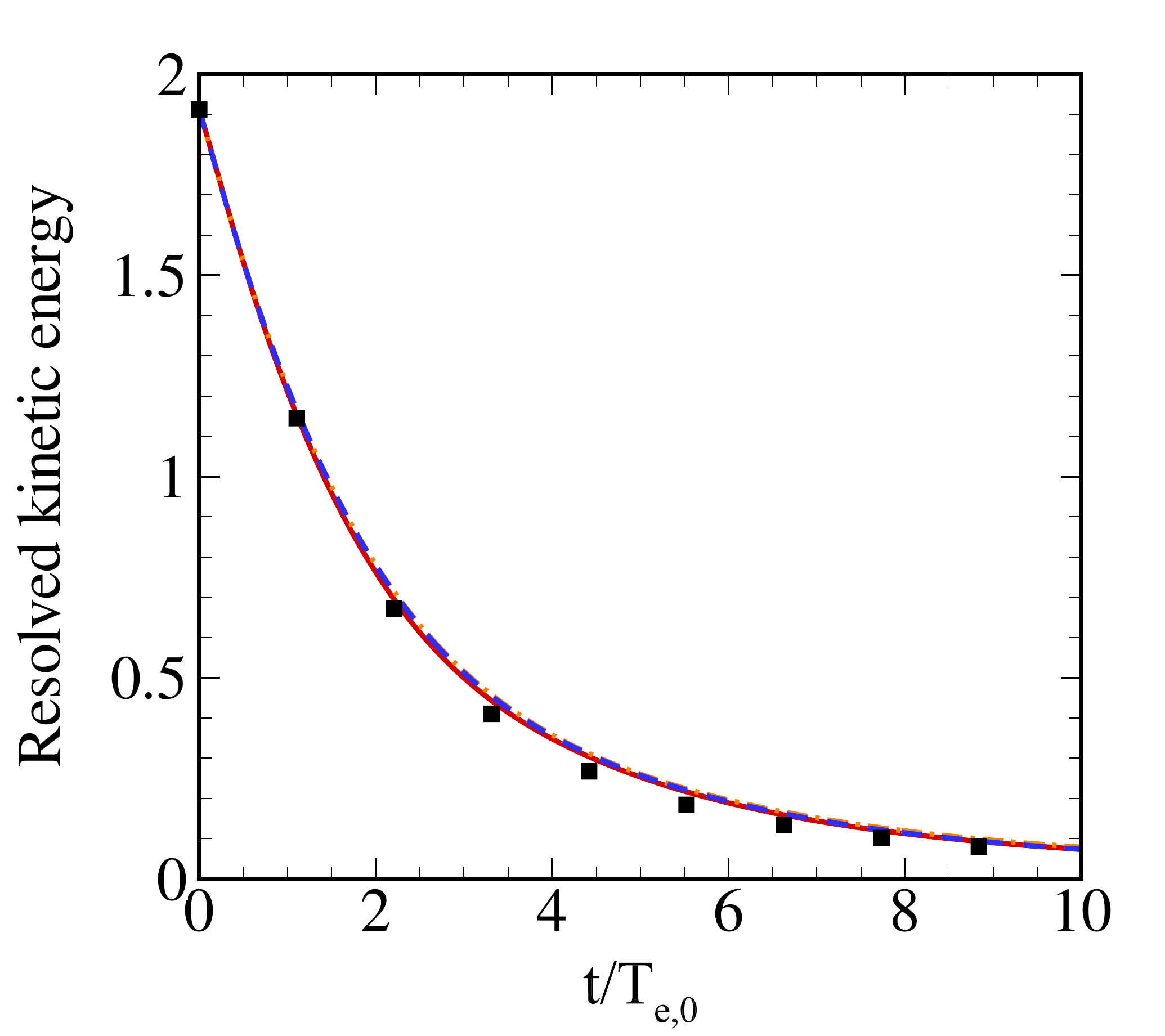}
		\caption{}
		\label{}
	\end{subfigure}
	\caption{Results from fDNS and LESs of decaying isotropic turbulence at the initial Reynolds number $Re_\lambda$ of 106 with grid resolution of $48^3$. (a) Energy spectra at $t/T_{e,0} = 1.1$, $3.3$ and $6.6$; (b) temporal evolution of the resolved kinetic energy. $\blacksquare$, fDNS; {\color{red}\sampleline{}}, ANN-SGS mixed model with 12 neurons per layer; {\color{YellowOrange}\sampleline{dash pattern=on .7em off .2em on .2em off .2em}}, ANN-SGS mixed model with 24 neurons per layer; {\color{blue}\sampleline{dash pattern=on 6pt off 2pt}}, ANN-SGS mixed model with 48 neurons per layer.}
    \label{fig:NN_node}
\end{figure}

Computational cost is an important factor of ANN-based SGS models, as it is one of the main obstacles to the practical application of the models. The previous studies~\citep{park2021, wang2018, yuan2020, xie2019comp, xie2020spatial} reported that ANN-based SGS models are computationally more expansive than DSM, which can be considered a major disadvantage. Therefore, a parameter study of the present ANN-SGS mixed model with the number of neurons per hidden layer is conducted to find an optimal size of the ANN, as the computational time for forward propagation of an ANN depends on the size of the ANN. 

Three different ANNs with 12, 24, and 48 neurons at each of two hidden layer are considered.  ANN-SGS mixed models with the three different ANNs are trained with the fDNS dataset of forced isotropic turbulence at $Re_\lambda = 106$. Figure~\ref{fig:NN_node} shows results from LES of decaying homogeneous isotropic turbulence with grid resolution of $48^3$ using ANN-SGS mixed models with ANNs of three different sizes. It is found that 12 neurons per hidden layer are sufficient, and additional neurons do not improve the performance of the ANN-SGS mixed model. However, the computational time to evaluate the SGS stress increases with the number of neurons as ratios of DSM : $12n : 24n : 48n = 1 : 0.71 : 0.89 : 1.42$, where $n$ denotes the number of neurons per hidden layer. Therefore, in the present study, an ANN with 12 neurons at each of the two hidden layer is selected.

\begin{figure}
\centering \
\includegraphics[width=0.6\textwidth]{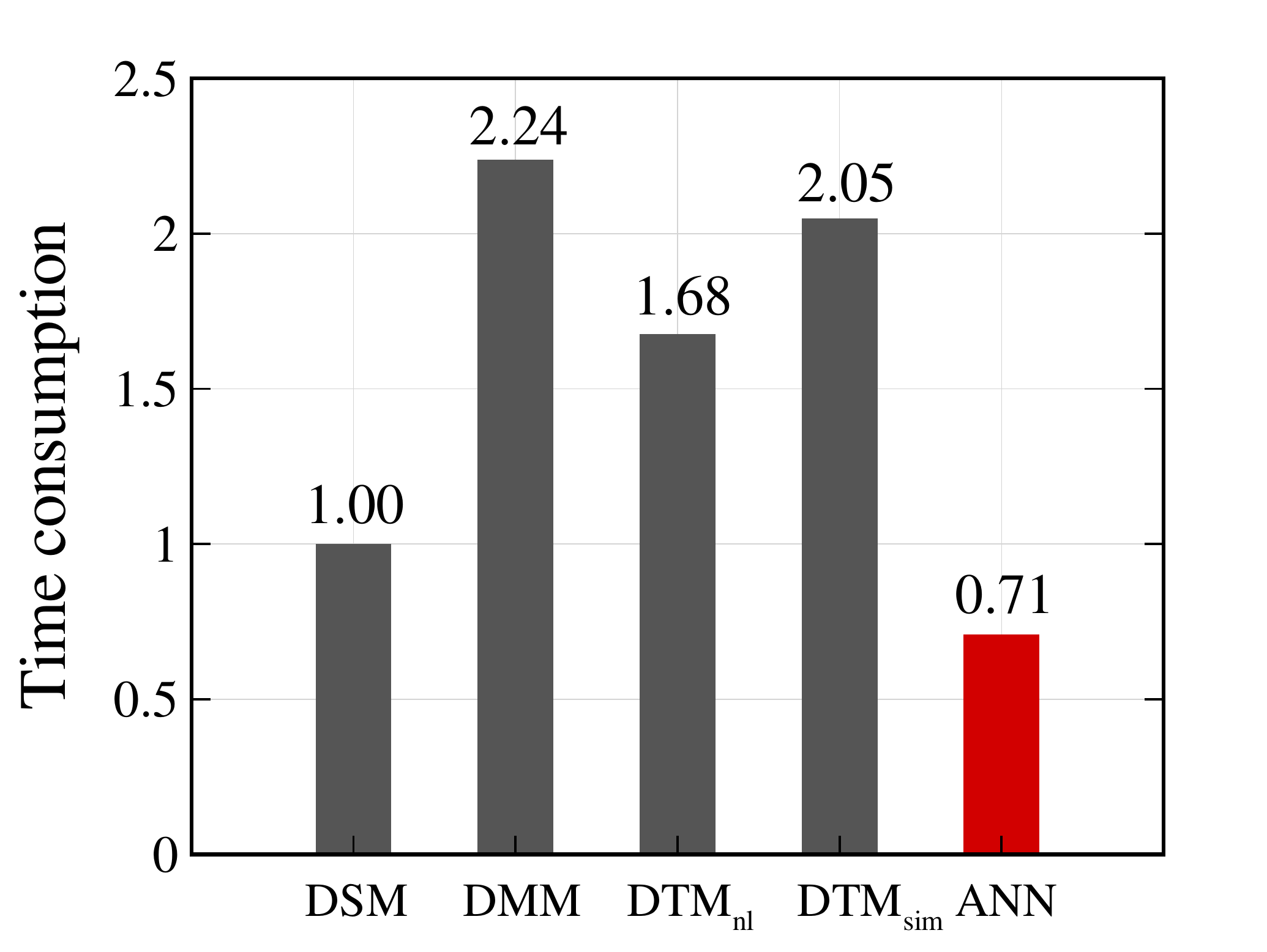}
\caption{Ratio of computational time of tested SGS models for evaluating the SGS stress. Computational time is normalised to that of DSM.}
\label{fig:cost}
\end{figure}

The computational cost of the developed ANN-SGS mixed model is compared with those of the algebraic dynamic SGS models. For a fair comparison, the matrix operations of the ANN are implemented on the LES solver. The weight matrices and the bias vectors of the trained ANN are saved and transferred to the solver to perform a forward propagation of the ANN. All simulations are conducted using 16 CPU cores of Intel(R) Xeon(R) E5-2650 v2, and the time consumption for evaluating the SGS stress tensors during the same 100 time steps are compared. 
Figure~\ref{fig:cost} shows ratios of the computational time of the tested SGS models for evaluating the SGS stress. The algebraic dynamic mixed models require 1.6 to 2.2 times greater computational time than DSM to calculate the SGS stress. It is noteworthy that the developed ANN-SGS mixed model is computationally cheaper than the algebraic dynamic mixed models and even than DSM. 

\subsection{Application to turbulent channel flow}\label{sec:channel}

\begin{table}
	\begin{center}
		\def~{\hphantom{0}}
		\begin{tabular}{cp{0.2cm}cp{0.2cm}cp{0.2cm}cp{0.2cm}cp{0.2cm}cp{0.2cm}cp{0.2cm}cp{0.2cm}c}
			$\Rey_{\tau}$&&
			$L_x$&&
			$L_z$&&
			$N_x\times N_y \times N_z$&&
			$\Delta x^+$&&
			$\Delta z^+$&&
			$\Delta y_{min}^+$&&
			$\Delta y_c^+$\\[3pt]
			180 && $4\pi \delta$ && $\frac{4}{3}\pi \delta$ && $192\times192\times192$ && $11.8$ && $3.9$ && $0.1$ && $4.6$\\[3pt]
			395 && $2\pi \delta$ && $\pi \delta$ && $192\times192\times192$ && $13.0$ && $6.5$ && $0.2$ && $10.0$\\
		\end{tabular}
		\caption{Parameters for DNS of turbulent channel flow. $L_x$ and $L_z$ are the streamwise and the spanwise domain sizes, respectively.  $\delta$ is the channel half width, and $N_x$, $N_y$ and $N_z$ are numbers of grid points in the streamwise, the wall-normal, and the spanwise directions, respectively. $\Delta x^+$ and $\Delta z^+$ are the streamwise and the spanwise grid sizes in wall units, respectively. $\Delta y_{min}^+$ and $\Delta y_{c}^+$ are the wall-normal grid sizes at the wall and the centerline, respectively.}
		\label{tab:DNS_channel}
	\end{center}
\end{table}

In this section, the application of developed ANN-SGS mixed model to wall-bounded flow is discussed. Turbulent channel flow at $Re_{\tau}=180$ and 395 is considered. The predictive capabilities of the ANN-SGS mixed model which uses both $\overline{S}_{ij}$ and $L_{ij}$ as inputs and the ANN-SGS model with $\overline{S}_{ij}$ as the only input are investigated through \emph{a posteriori} tests for turbulent channel flow with various untrained Reynolds number and grid resolution. Additionally, it is further investigated whether the ANN-SGS mixed model trained with both channel flow and homogeneous isotropic turbulence at a certain condition is capable of providing accurate solutions for both types of flow at untrained conditions. Lastly, results from LES of turbulent channel flow with the ANN-SGS mixed model which is trained only with homogeneous isotropic turbulence are discussed.

DNSs are performed with parameters summarized in table~\ref{tab:DNS_channel} to construct  training datasets and reference solutions for LESs. A fourth-order finite-difference code by \citet{bose2010} is utilised to conduct DNSs and LESs. The convection terms are discretized in a skew-symmetric form~\citep{morinish98}. A semi-implicit method is used for time marching. The viscous diffusion terms in the wall-normal direction are integrated implicitly with the Crank-Nicolson method, and the other terms are explicitly advanced with the third-order Runge-Kutta (RK3) method~\citep{morinish98, bose2010, kangphd}. 

The fDNS dataset of turbulent channel flow at $Re_{\tau} = 180$ is constructed by applying the grid- and test-filters with $\mybar{\Delta} = 4\Delta_{DNS}$ and $\myhat{\mybar{\Delta}} = 8\Delta_{DNS}$ to instantaneous DNS fields, where $\Delta_{DNS}$ is grid spacing of DNS. The filter width $\mybar{\Delta}$ of the fDNS dataset is the same as the grid size of LES with $N_x\times N_y \times N_z = 48\times48\times48$ cells at $Re_{\tau}=180$. Fourth-order commutative discrete filters by \citet{vasilyev1998} are utilised to construct the fDNS dataset and to conduct test-filtering of LES. A total of $3.5\times10^7$ data points are sampled from 5 instantaneous velocity fields of fDNS, when flow becomes fully developed.

\begin{table}
	\begin{center}
		\def~{\hphantom{0}}
		\resizebox{0.99\textwidth}{!}{%
		\begin{tabular}{cp{0.1cm}cp{0.1cm}cp{0.1cm}cp{0.1cm}cp{0.1cm}cp{0.1cm}cp{0.1cm}c}
			LES case&&
			$\Rey_{\tau}$&&
			$(L_x, L_z)$&&
			$(N_x, N_y, N_z)$&&
			$(\Delta x^+, \Delta z^+)$&&
			$(\Delta y_{min}^+, \Delta y_c^+)$&&
			model\\[3pt]
			\hline
			 &&  &&  &&  &&  &&  && SL-180C \\[3pt]
			 &&  &&  &&  &&  &&  && S-180C \\[3pt]
			LES180 && 180 && $(4\pi \delta, \frac{4}{3}\pi \delta)$ && $(48, 48, 48)$ && $(47.2, 15.7)$ && $(0.33, 18.2)$ && SL-106H180C \\[3pt]
			 &&  &&  &&  &&  &&  && SL-106H \\[3pt]
			 &&  &&  &&  &&  &&  && DSM \\[3pt]
			 &&  &&  &&  &&  &&  && no-SGS \\[3pt]
			\hline
			 &&  &&  &&  &&  &&  && SL-180C \\[3pt]
			LES180c && 180 && $(4\pi \delta, \frac{4}{3}\pi \delta)$ && $(32, 48, 32)$ && $(70.8, 23.6)$ && $(0.33, 18.2)$ && SL-106H180C \\[3pt]
			 &&  &&  &&  &&  &&  && DSM \\[3pt]
			 &&  &&  &&  &&  &&  && no-SGS \\[3pt]
			\hline
			 &&  &&  &&  &&  &&  && SL-180C \\[3pt]
			LES180f && 180 && $(4\pi \delta, \frac{4}{3}\pi \delta)$ && $(64, 64, 64)$ && $(35.4, 11.8)$ && $(0.24, 13.7)$ && SL-106H180C \\[3pt]
			 &&  &&  &&  &&  &&  && DSM \\[3pt]
			 &&  &&  &&  &&  &&  && no-SGS \\[3pt]
			\hline
			 &&  &&  &&  &&  &&  && SL-180C \\[3pt]
			 &&  &&  &&  &&  &&  && S-180C \\[3pt]
			LES395 && 395 && $(2\pi \delta, \pi \delta)$ && $(48, 48, 48)$ && $(52.0, 26.0)$ && $(0.72, 40.0)$ && SL-106H180C \\[3pt]
			 &&  &&  &&  &&  &&  && DSM \\[3pt]
			 &&  &&  &&  &&  &&  && no-SGS \\[3pt]
			\hline
			 &&  &&  &&  &&  &&  && SL-180C \\[3pt]
			 &&  &&  &&  &&  &&  && S-180C \\[3pt]
			LES395f && 395 && $(2\pi \delta, \pi \delta)$ && $(64, 64, 64)$ && $(39.0, 19.5)$ && $(0.53, 30.1)$ && SL-106H180C \\[3pt]
			 &&  &&  &&  &&  &&  && DSM \\[3pt]
			 &&  &&  &&  &&  &&  && no-SGS \\[3pt]
		\end{tabular}}
		\caption{Parameters for LESs of turbulent channel flow with ANN-based SGS models and DSM. The effects of grid resolution and the Reynolds number on the performance of ANN-based SGS models are considered. $L_x$ and $L_z$ are the streamwise and the spanwise domain sizes, respectively. $\delta$ is the channel half width, and $N_x$, $N_y$ and $N_z$ are numbers of grid points in the streamwise, the wall-normal, and the spanwise directions, respectively. $\Delta x^+$ and $\Delta z^+$ are the streamwise and the spanwise grid sizes in wall units, respectively. $\Delta y_{min}^+$ and $\Delta y_{c}^+$ are the wall-normal grid sizes at the wall and the centerline, respectively.}
		\label{tab:LES_channel_cases}
	\end{center}
\end{table} 

ANN-SGS models are trained from the scratch using the fDNS dataset of turbulent channel flow by the same method in Section~\ref{sec:Artificial neural network for subgrid-scale modelling}, and listed in table~\ref{tab:LES_channel_cases}. Each ANN-SGS model is named following the same rules as in Section~\ref{sec:Artificial neural network for subgrid-scale modelling}. The characters C and H at the end of the model names denote channel flow and homogeneous isotropic turbulence, respectively, and represent flow with which each ANN-SGS model is trained. For example, S-180C uses $\overline{S}_{ij}$ as the input, whereas SL-180C uses $\overline{S}_{ij}$ and $L_{ij}$ as inputs, and both models are trained with turbulent channel flow at $Re_{\tau} = 180$. SL-106H180C is trained with both channel flow at $Re_{\tau} = 180$ and homogeneous isotropic turbulence at $Re_{\lambda} = 106$. SL-106H is only trained with homogeneous isotropic turbulence at $Re_{\lambda} = 106$.

In order to investigate the generalisability of the ANN-SGS models, the models are tested for LESs of turbulent channel flow at various conditions listed in table~\ref{tab:LES_channel_cases}. The Reynolds number and the grid size of LES180 case are the same as those of the training data (\emph{i.e.}, $Re_{\tau} = 180$, $N_x\times N_y \times N_z = 48\times48\times48$), and the remaining four cases correspond to untrained Reynolds number and grid resolution conditions.

\begin{figure}
	\centering
	\begin{subfigure}[b]{0.49\textwidth}
		\includegraphics[width=\textwidth]{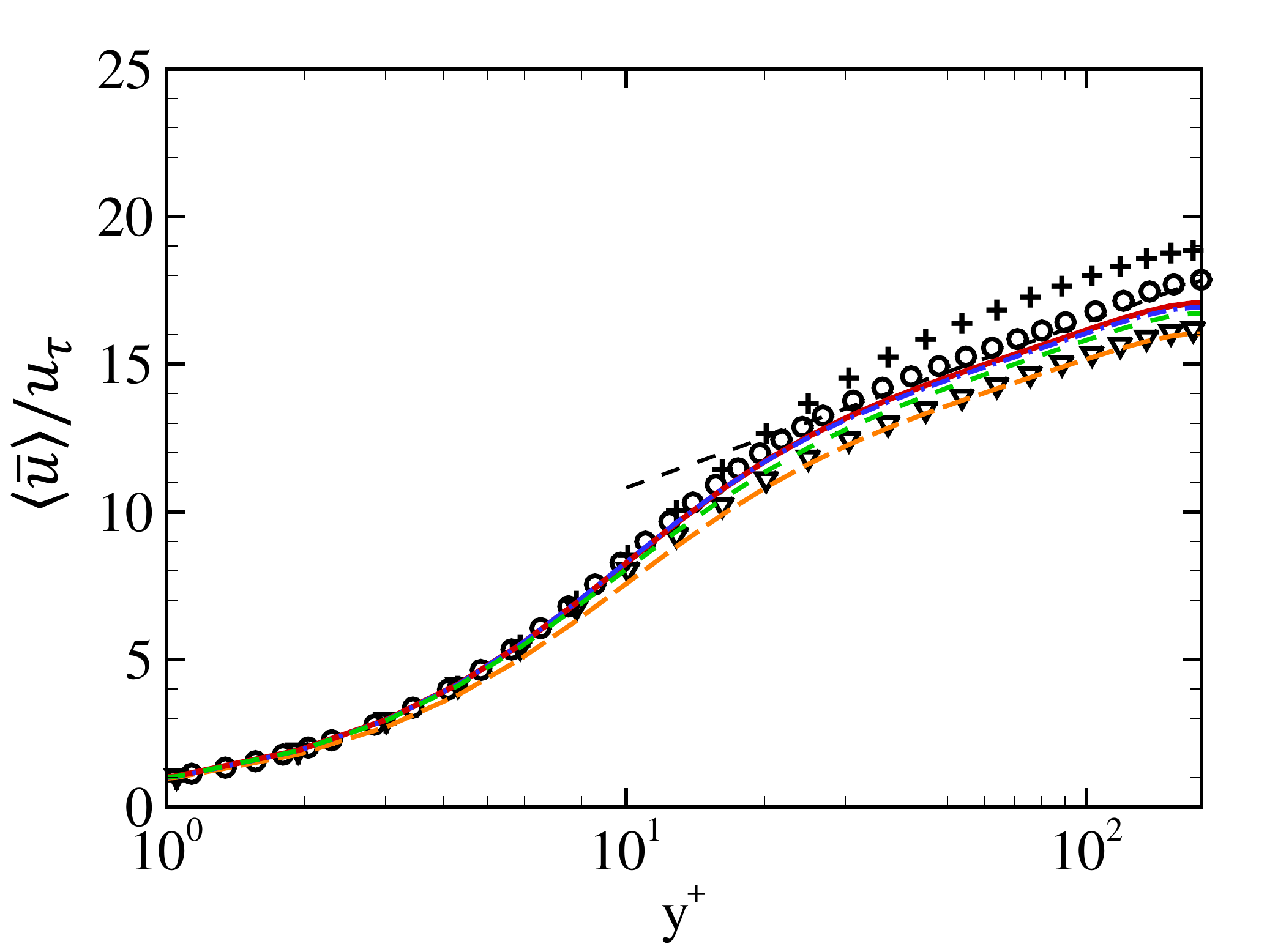}
		\caption{}
		\label{}
	\end{subfigure}
	\begin{subfigure}[b]{0.49\textwidth}
		\includegraphics[width=\textwidth]{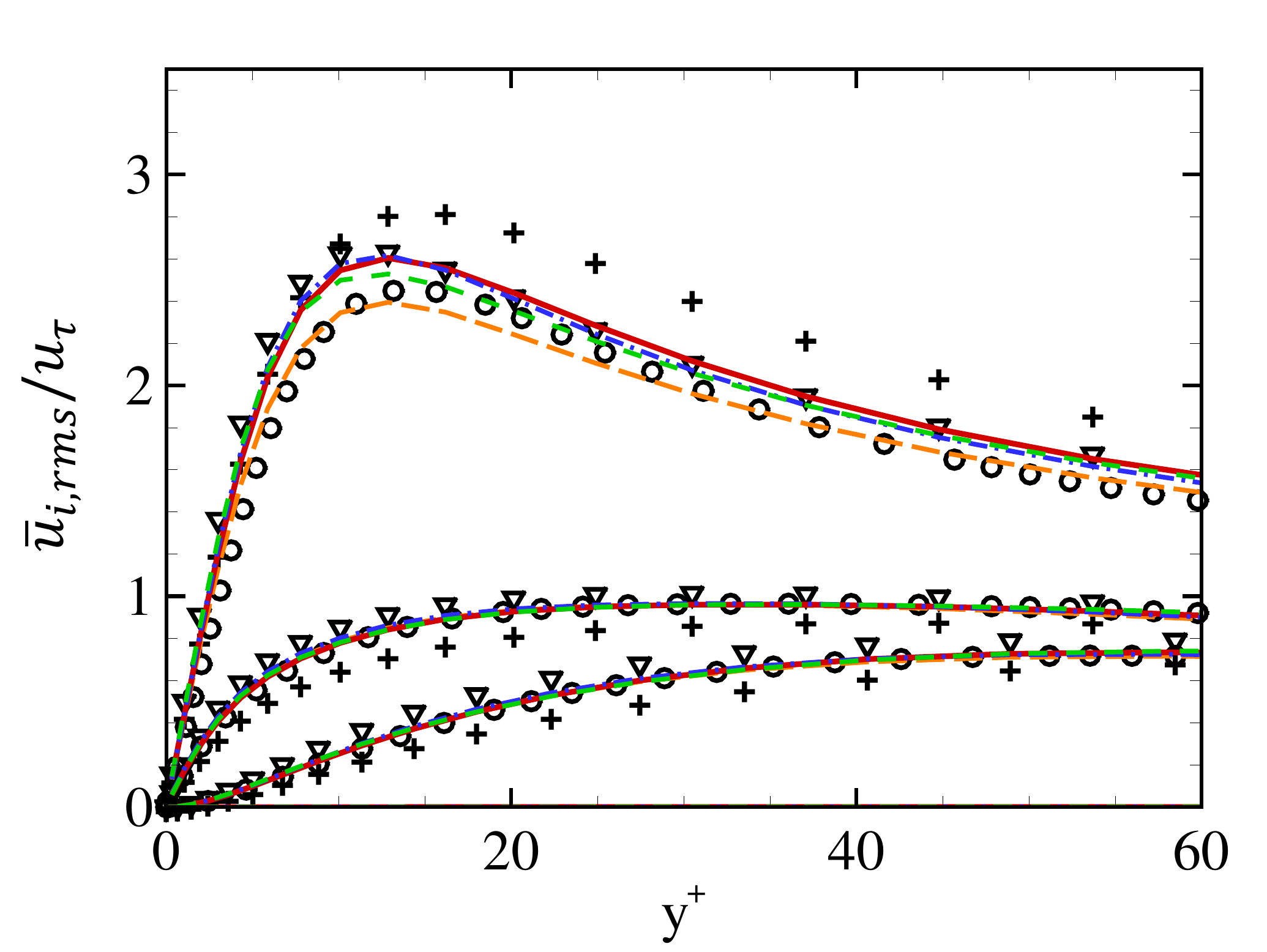}
		\caption{}
		\label{}
	\end{subfigure}
	\begin{subfigure}[b]{0.49\textwidth}
		\includegraphics[width=\textwidth]{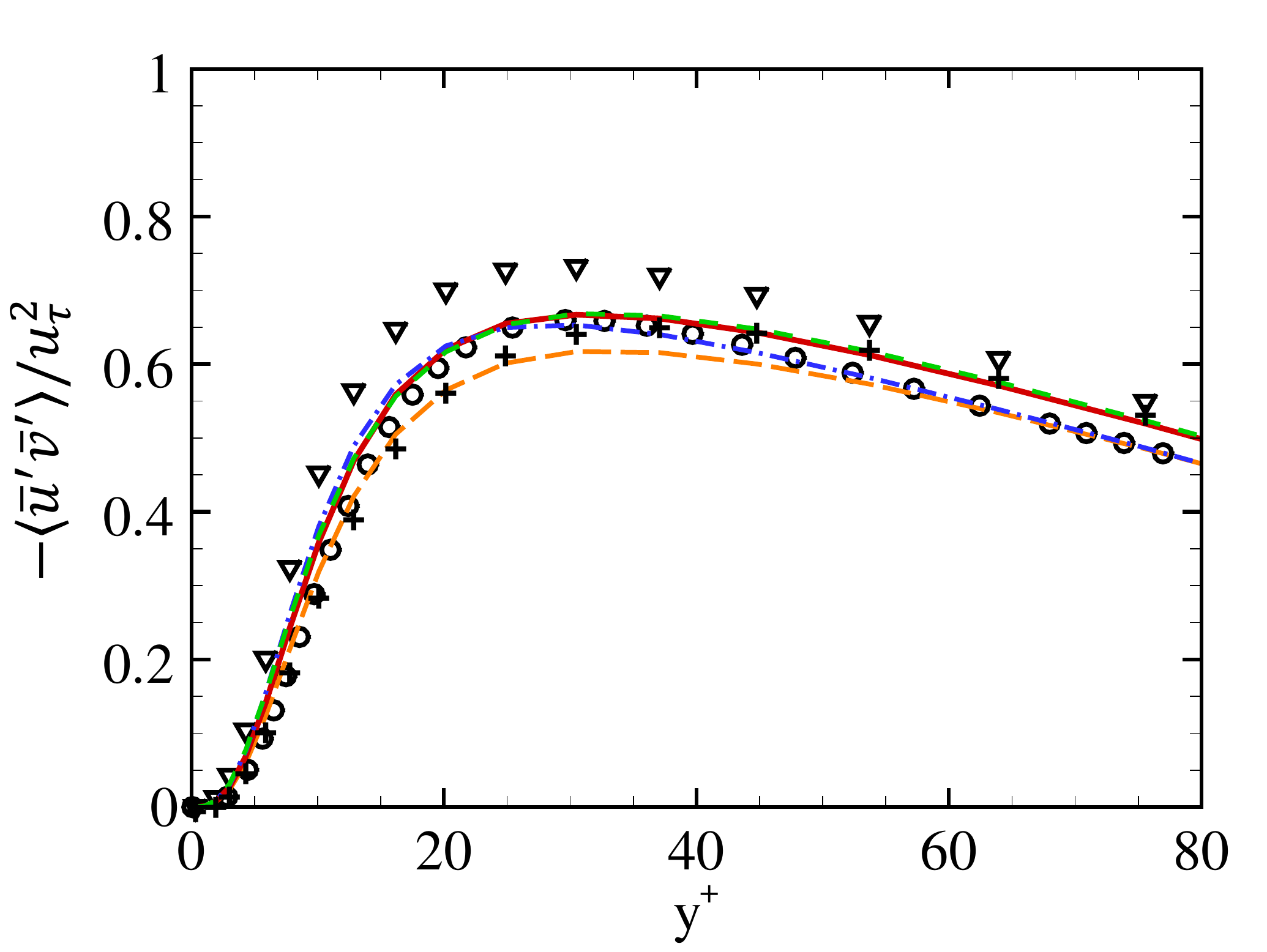}
		\caption{}
		\label{}
	\end{subfigure}
	\caption{Results from fDNS and LESs of turbulent channel flow at $Re_\tau=180$ with grid resolution of $48\times48\times48$ (LES180). (a) The mean streamwise velocity; (b) root-mean-squared (rms) velocity fluctuations; (c) the mean Reynolds shear stress $\left<\mybar{u}'\mybar{v}'\right>$, where $\left< \cdot \right>$ denotes averaging over the $x-z$ plane and time. $\bigcirc$, fDNS; +, DSM; $\nabla$, no-SGS; {\color{red}\sampleline{}}, SL-180C; {\color{green}\sampleline{dash pattern=on 6pt off 2pt}}, S-180C; {\color{blue}\sampleline{dash pattern=on .7em off .2em on .2em off .2em}}, SL-106H180C; 
	{\color{YellowOrange}\sampleline{dash pattern=on 9pt off 2pt}}, SL-106H; {\color{black}\sampleline{dash pattern=on 6pt off 2pt}}, the law of the wall $\left<\mybar{u}\right>/u_{\tau} = 0.41^{-1}\log{y^{+}}+5.2$.}
    \label{fig:channel_LES180}
\end{figure}

\begin{figure}
	\centering
	\begin{subfigure}[b]{0.49\textwidth}
		\includegraphics[width=\textwidth]{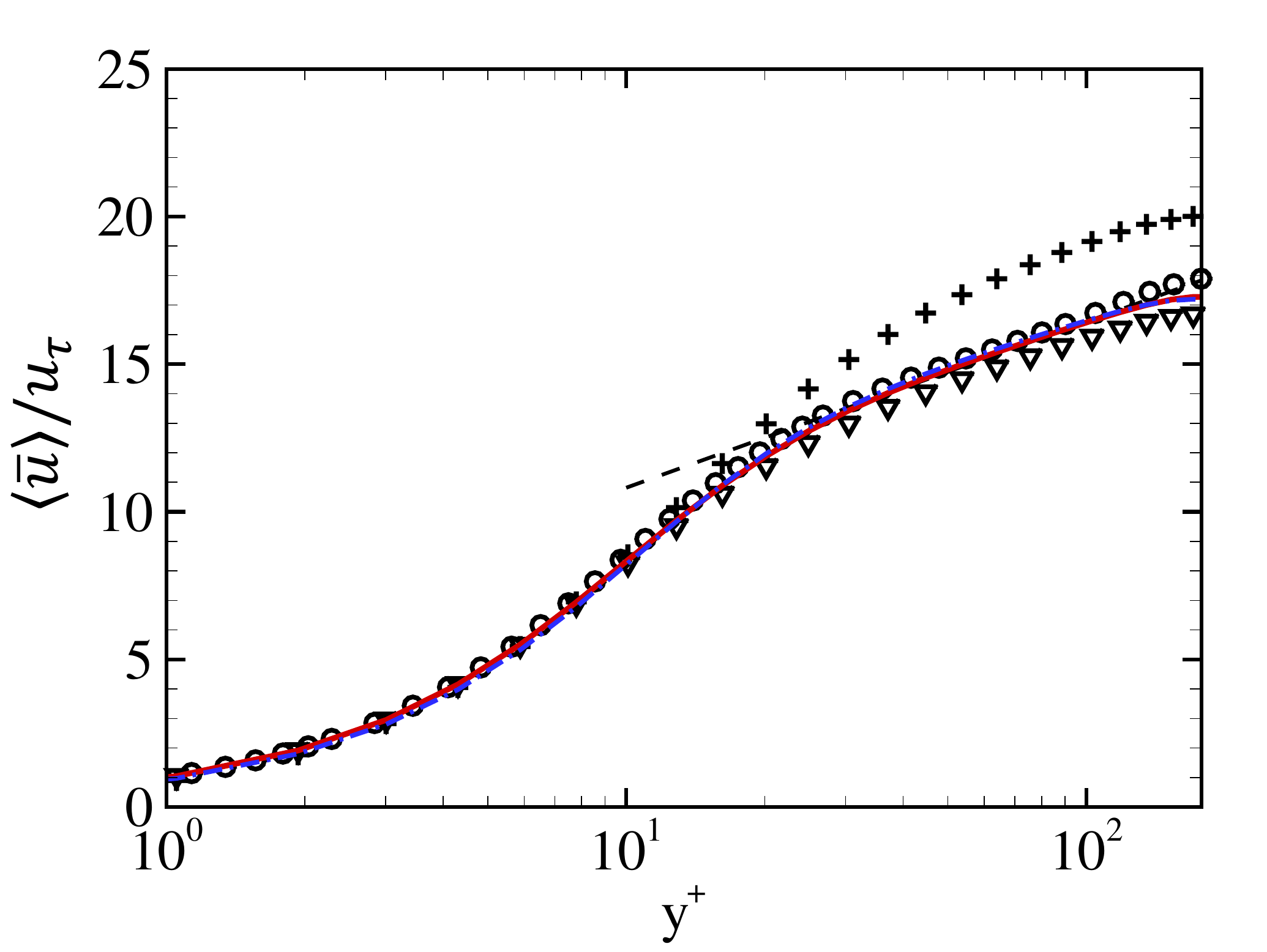}
		\caption{}
		\label{}
	\end{subfigure}
	\begin{subfigure}[b]{0.49\textwidth}
		\includegraphics[width=\textwidth]{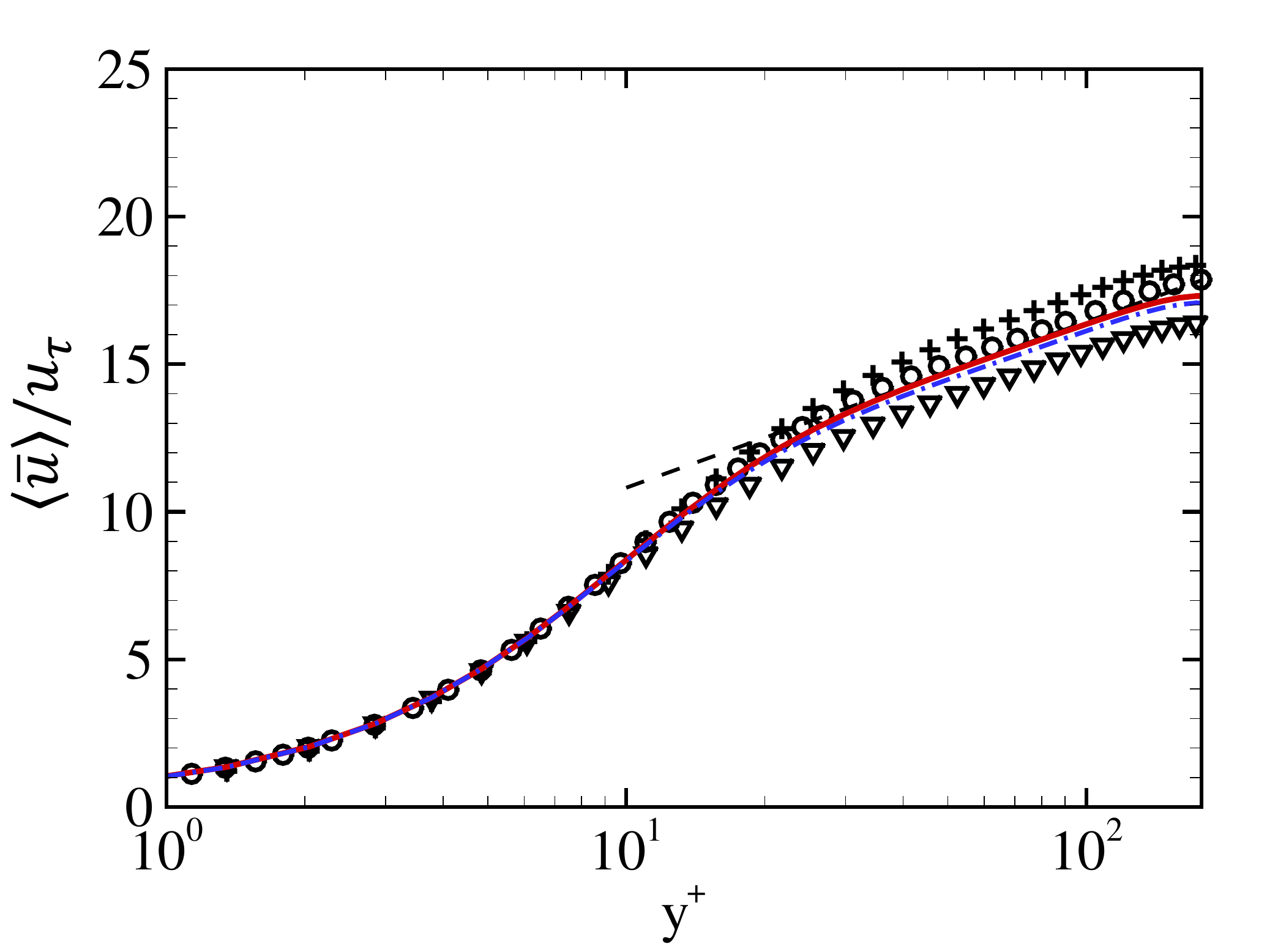}
		\caption{}
		\label{}
	\end{subfigure}
	\begin{subfigure}[b]{0.49\textwidth}
		\includegraphics[width=\textwidth]{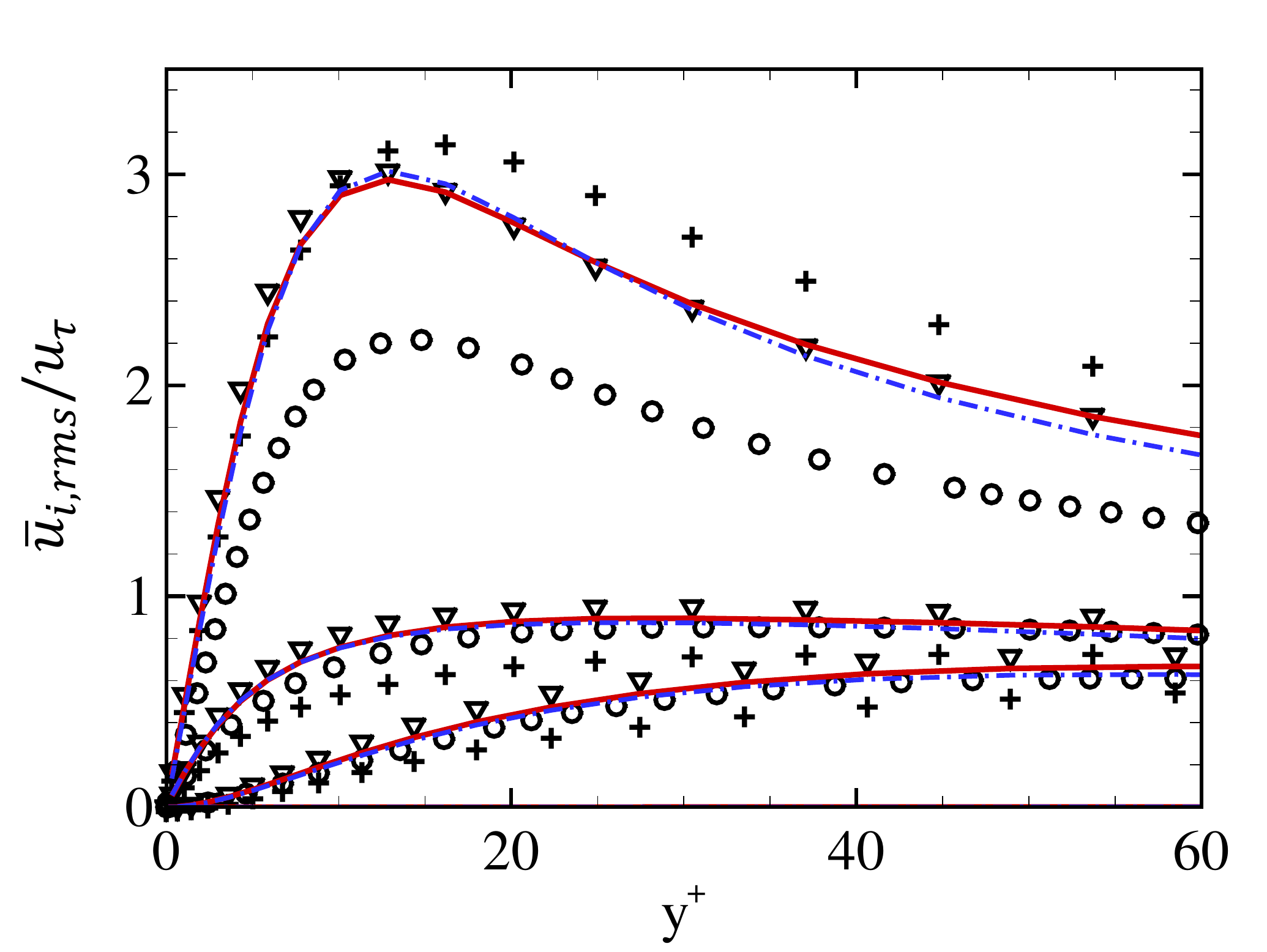}
		\caption{}
		\label{}
	\end{subfigure}
	\begin{subfigure}[b]{0.49\textwidth}
		\includegraphics[width=\textwidth]{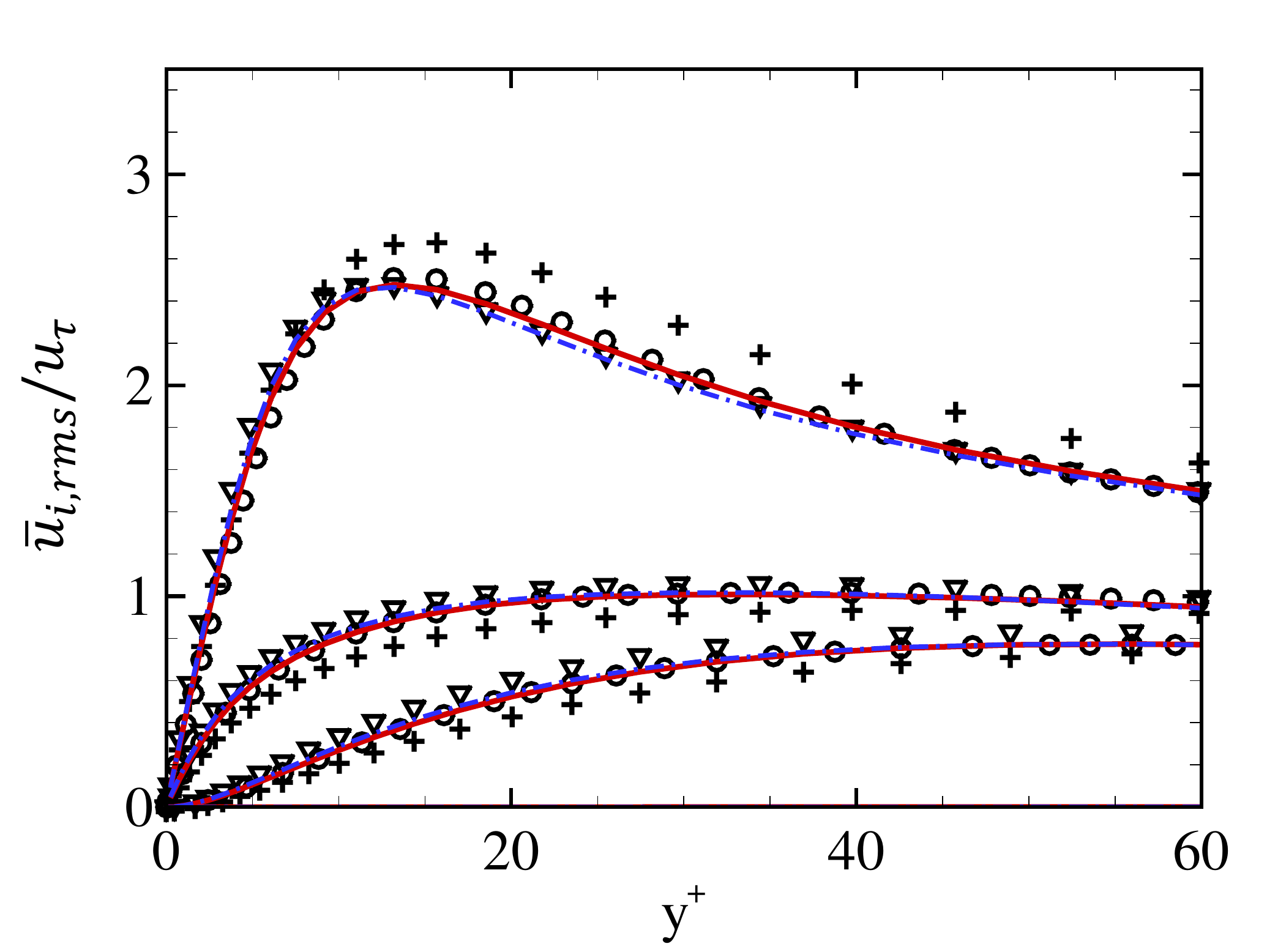}
		\caption{}
		\label{}
	\end{subfigure}
	\begin{subfigure}[b]{0.49\textwidth}
		\includegraphics[width=\textwidth]{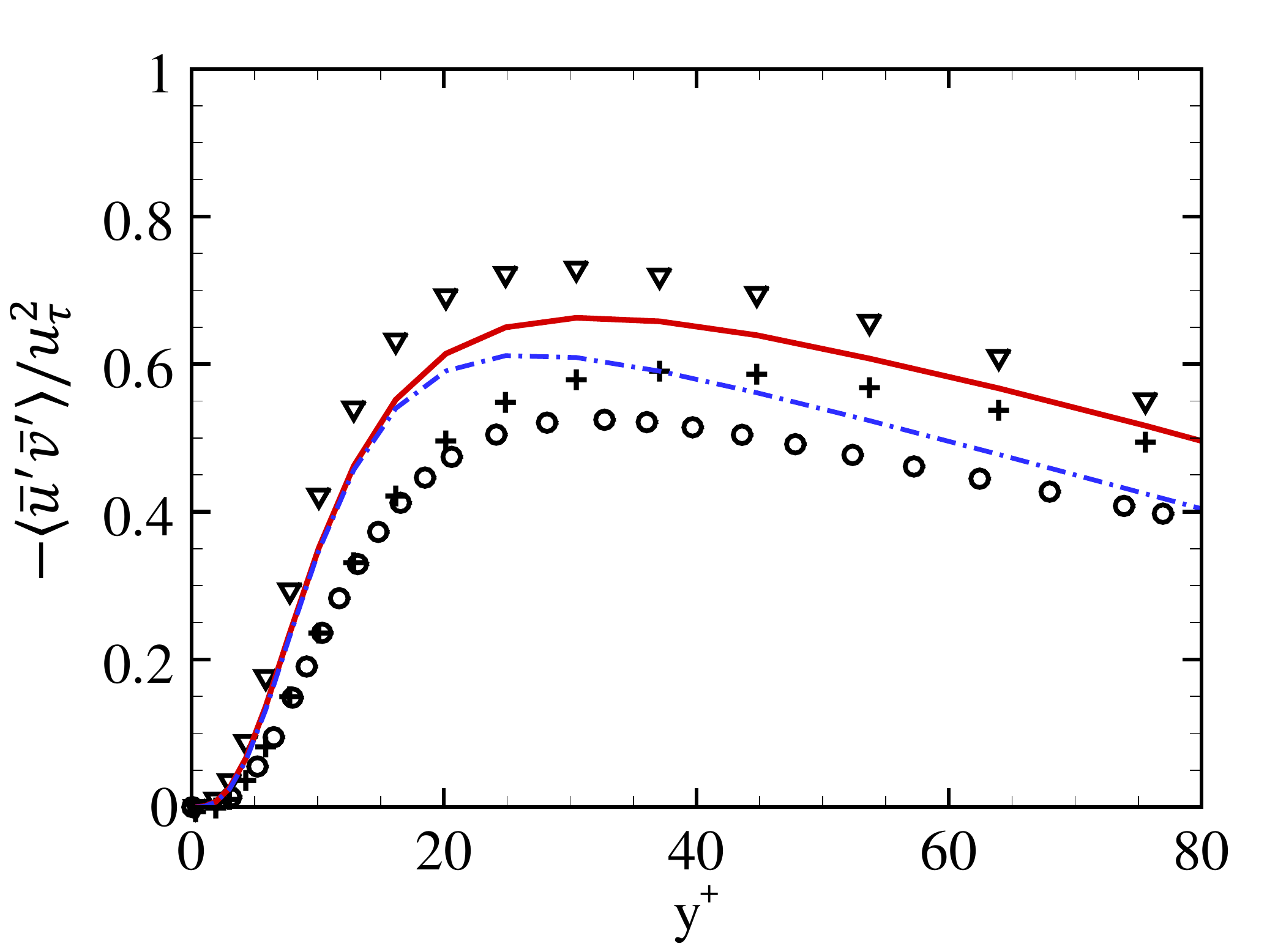}
		\caption{}
		\label{}
	\end{subfigure}
	\begin{subfigure}[b]{0.49\textwidth}
		\includegraphics[width=\textwidth]{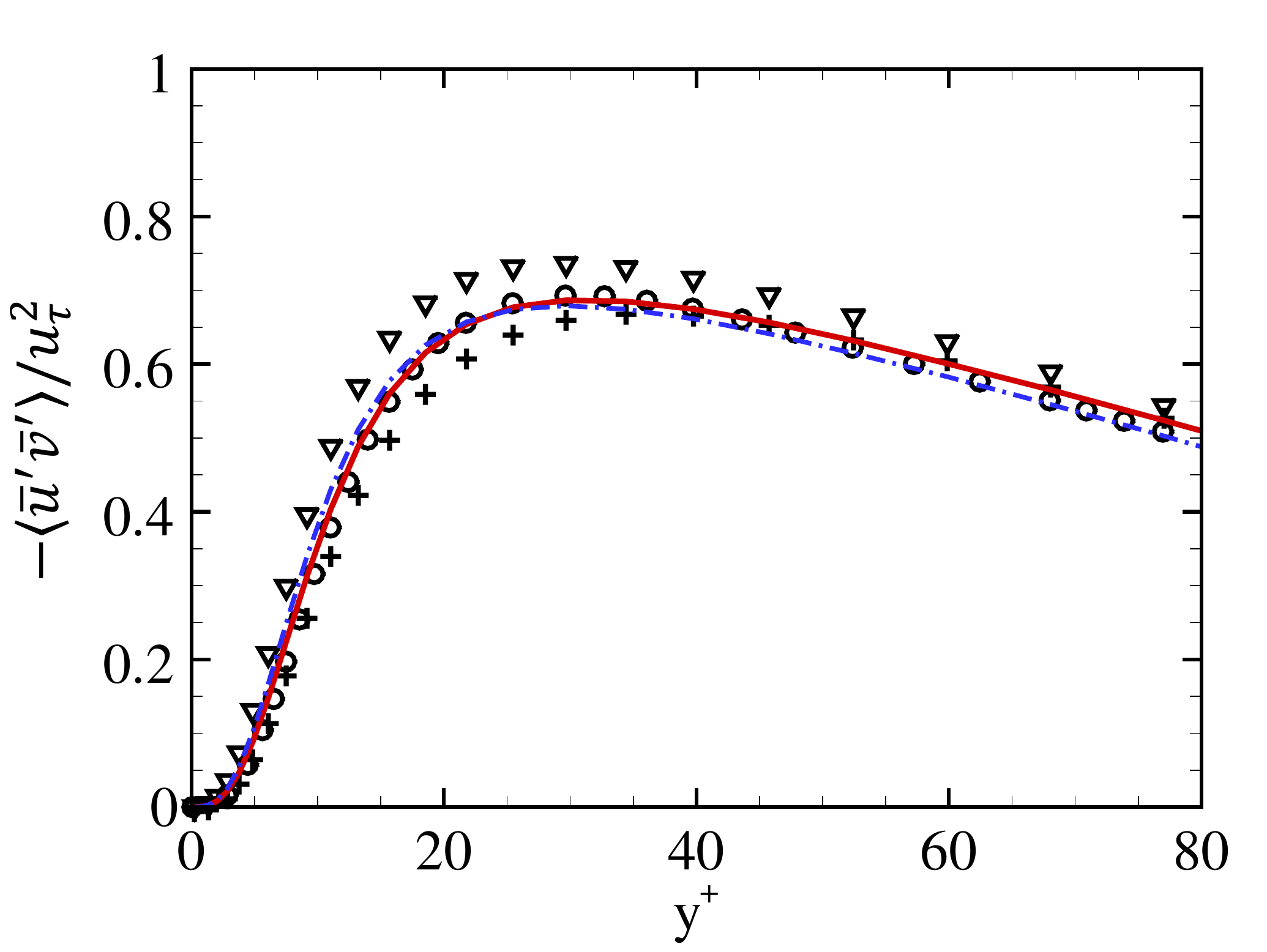}
		\caption{}
		\label{}
	\end{subfigure}
	\caption{Results from fDNS and LESs of turbulent channel flow at $Re_\tau=180$ with grid resolution of $32\times48\times32$ ((a), (c), (e); LES180c) and $64\times64\times64$ ((b), (d), (f); LES180f). (a), (b) The mean streamwise velocity; (c), (d) root-mean-squared (rms) velocity fluctuations; (e), (f) the mean Reynolds shear stress $\left<\mybar{u}'\mybar{v}'\right>$, where $\left< \cdot \right>$ denotes averaging over the $x-z$ plane and time. $\bigcirc$, fDNS; +, DSM; $\nabla$, no-SGS; {\color{red}\sampleline{}}, SL-180C; {\color{blue}\sampleline{dash pattern=on .7em off .2em on .2em off .2em}}, SL-106H180C; {\color{black}\sampleline{dash pattern=on 6pt off 2pt}}, the law of the wall $\left<\mybar{u}\right>/u_{\tau} = 0.41^{-1}\log{y^{+}}+5.2$.}
    \label{fig:channel_LES180cf}
\end{figure}

\begin{figure}
	\centering
	\begin{subfigure}[b]{0.49\textwidth}
		\includegraphics[width=\textwidth]{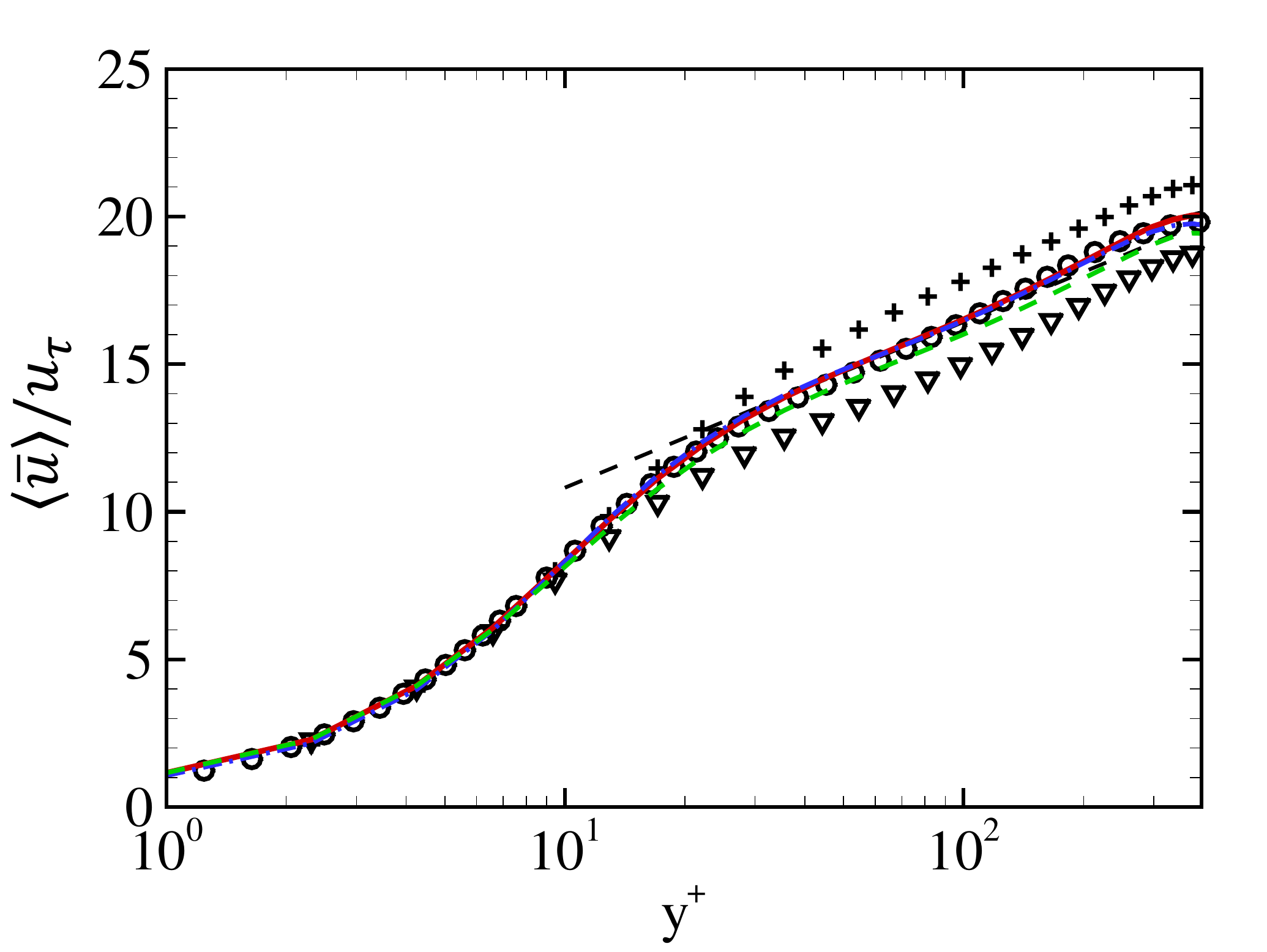}
		\caption{}
		\label{}
	\end{subfigure}
	\begin{subfigure}[b]{0.49\textwidth}
		\includegraphics[width=\textwidth]{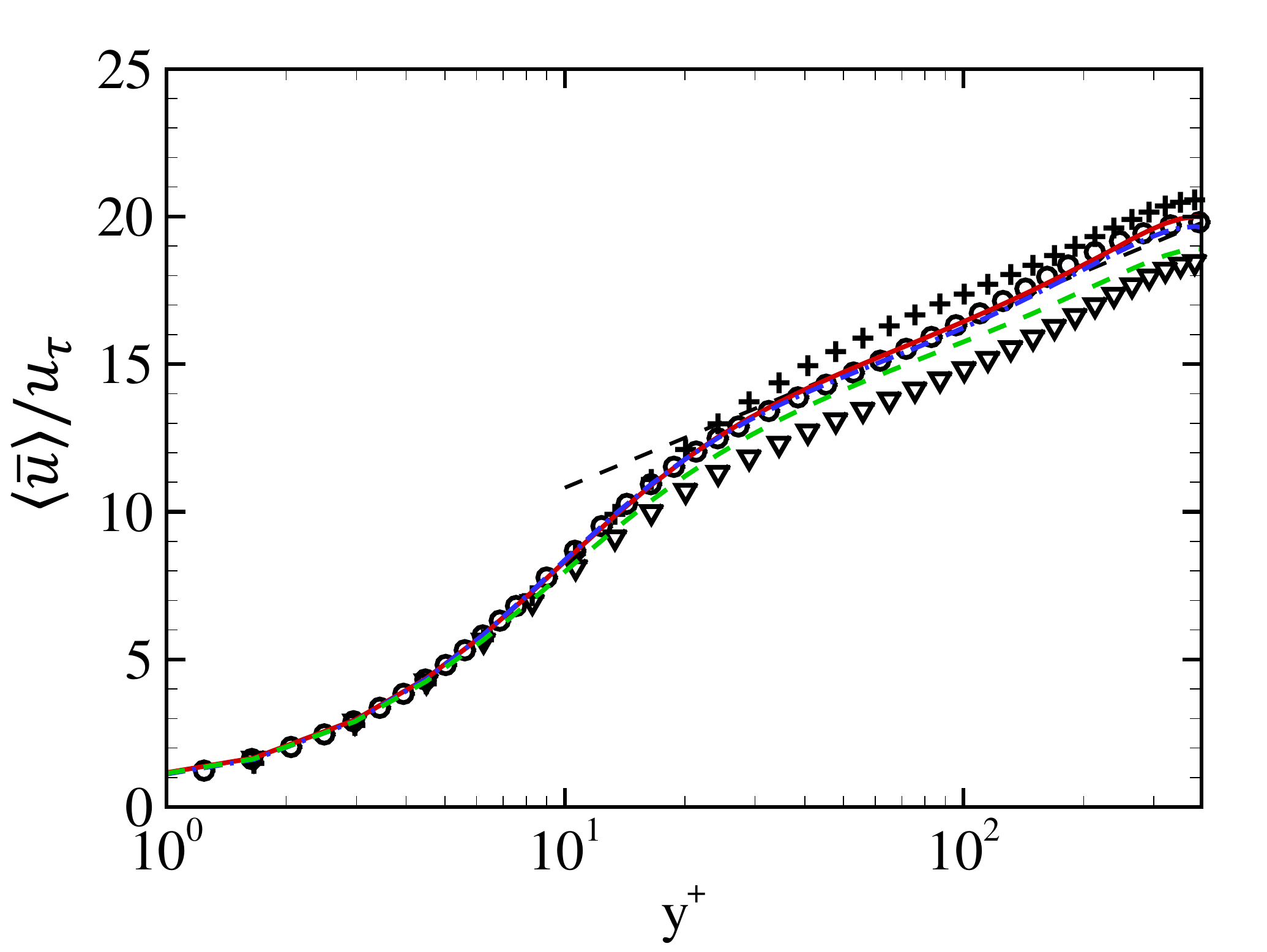}
		\caption{}
		\label{}
	\end{subfigure}
	\begin{subfigure}[b]{0.49\textwidth}
		\includegraphics[width=\textwidth]{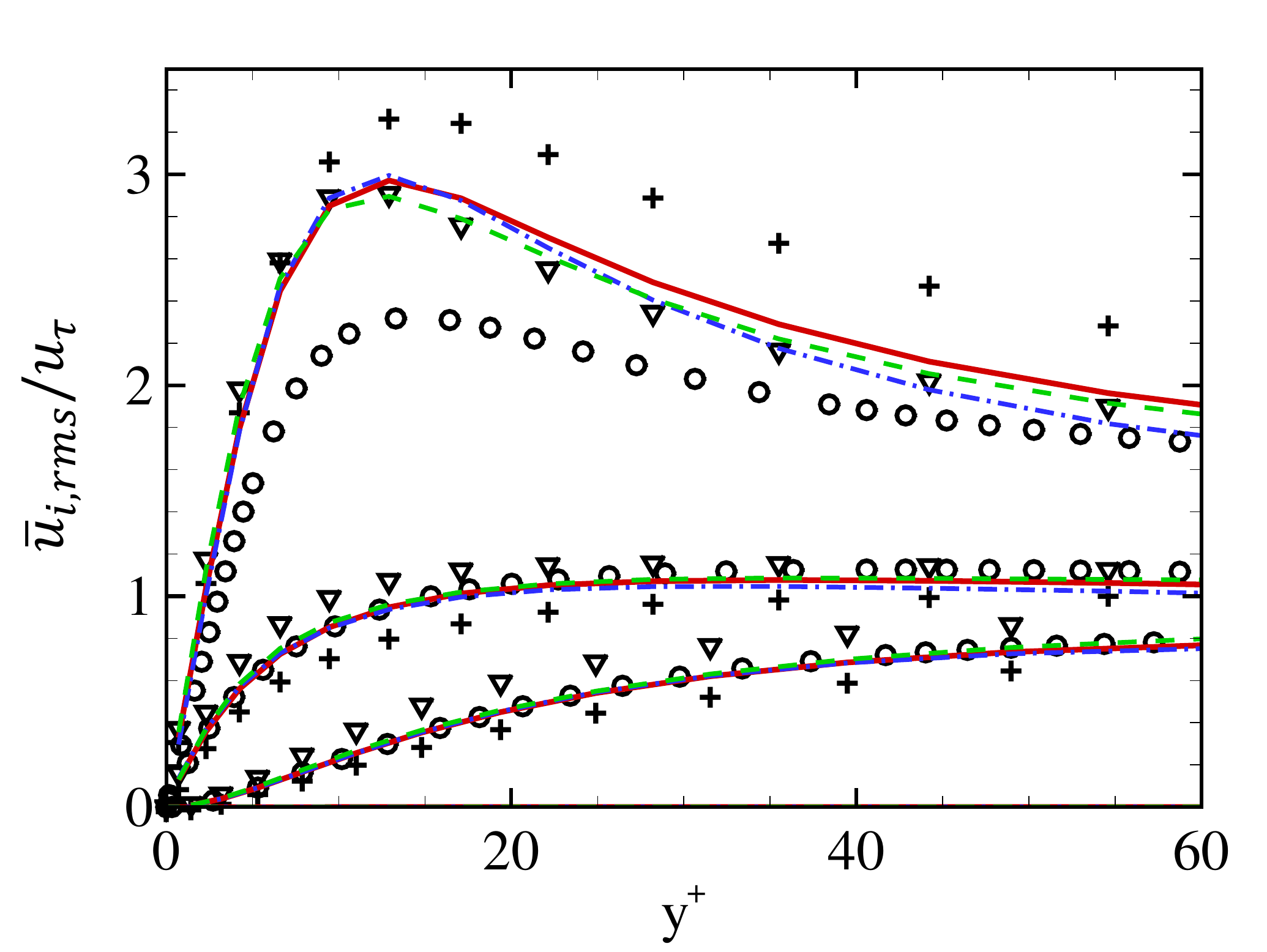}
		\caption{}
		\label{}
	\end{subfigure}
	\begin{subfigure}[b]{0.49\textwidth}
		\includegraphics[width=\textwidth]{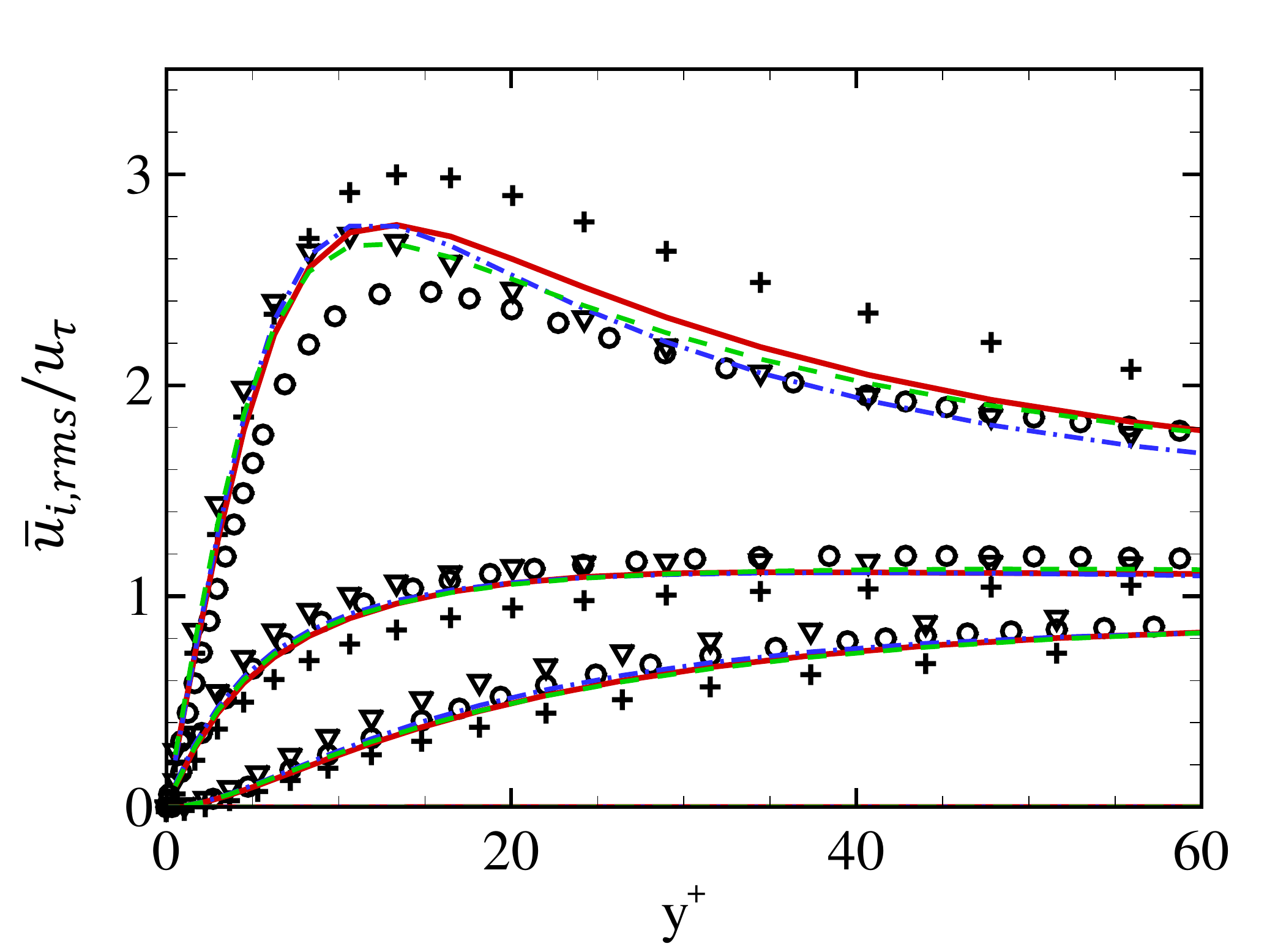}
		\caption{}
		\label{}
	\end{subfigure}
	\begin{subfigure}[b]{0.49\textwidth}
		\includegraphics[width=\textwidth]{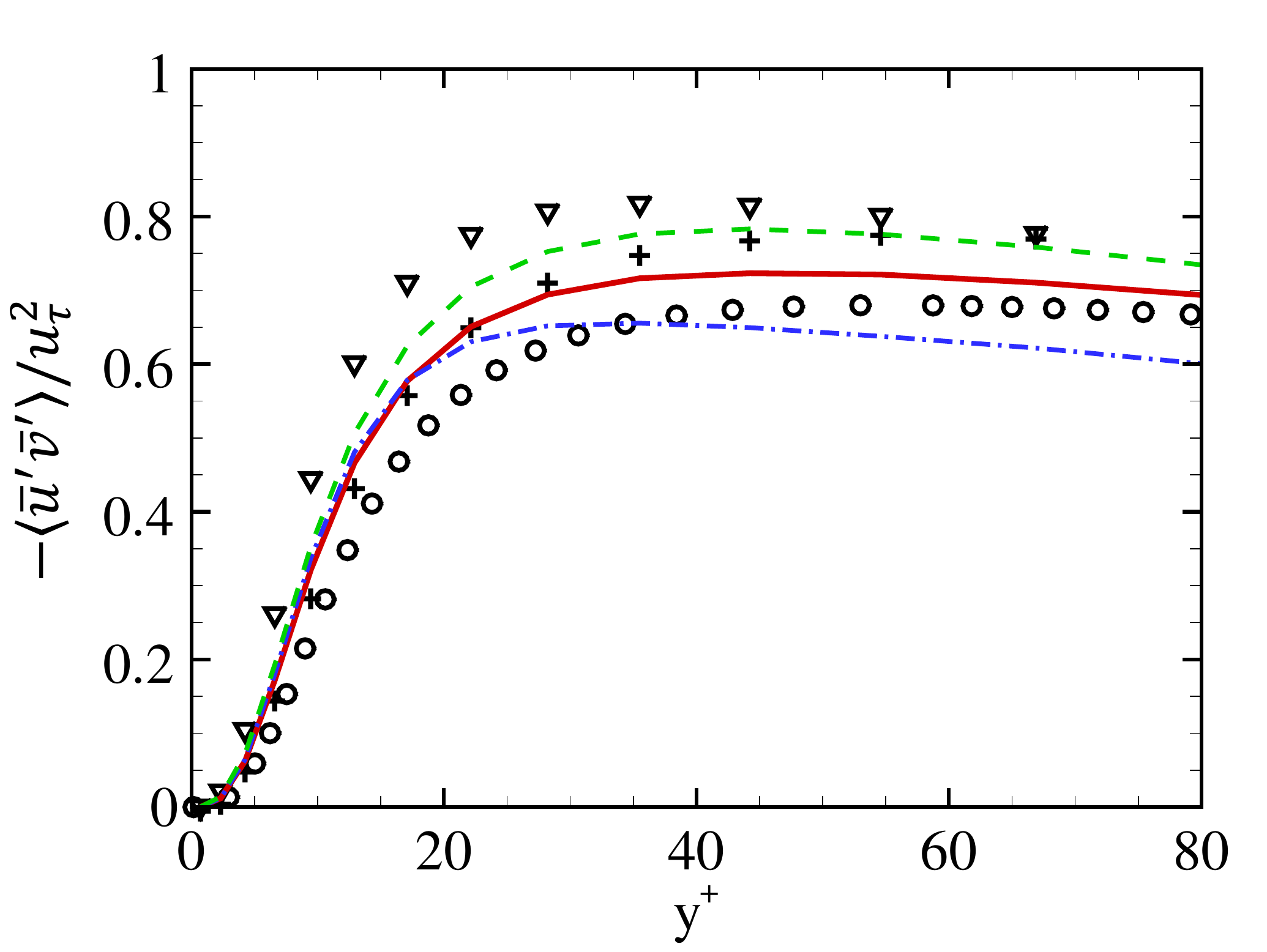}
		\caption{}
		\label{}
	\end{subfigure}
	\begin{subfigure}[b]{0.49\textwidth}
		\includegraphics[width=\textwidth]{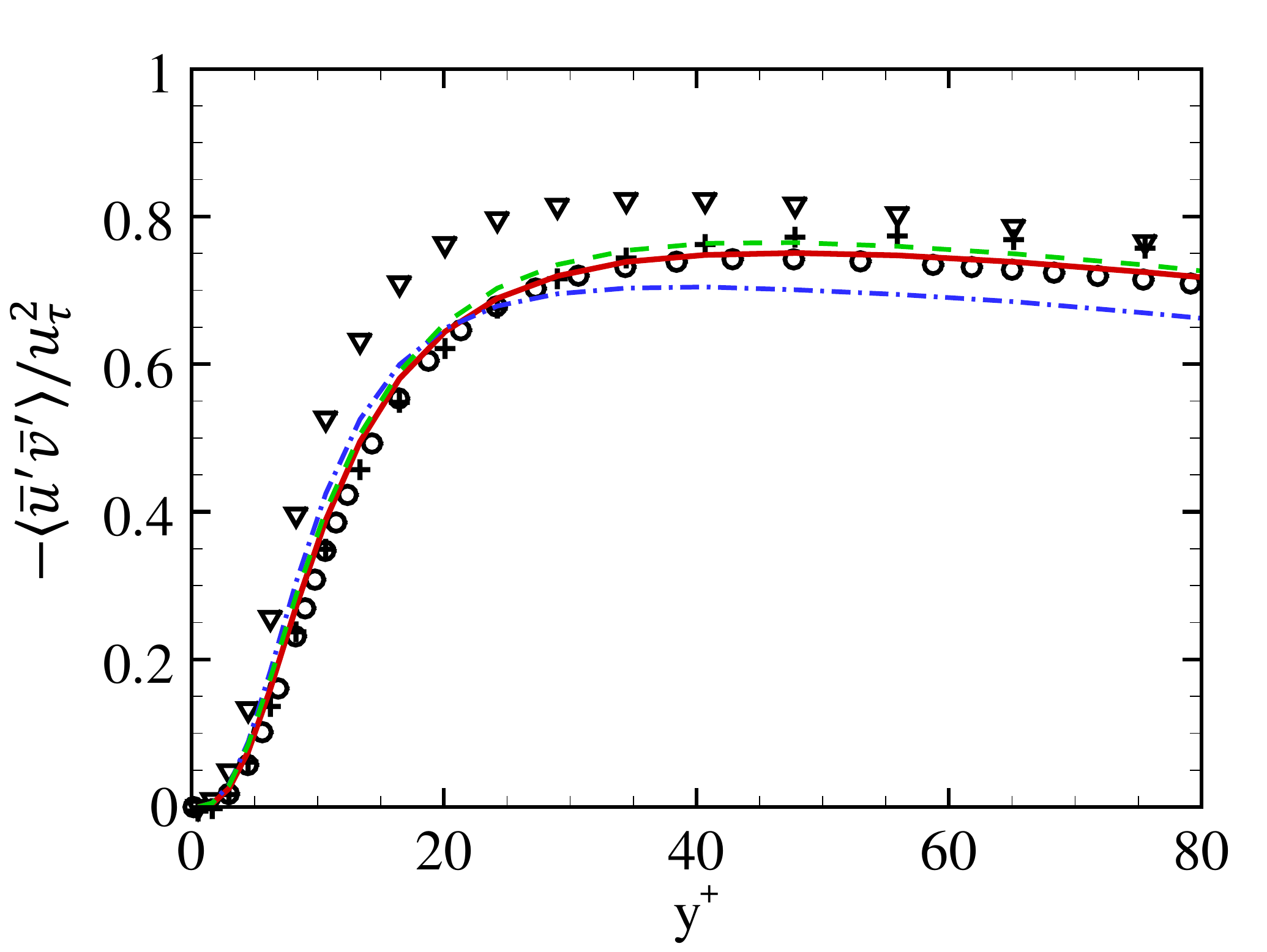}
		\caption{}
		\label{}
	\end{subfigure}
	\caption{Results from fDNS and LESs of turbulent channel flow at $Re_\tau=395$ with grid resolution of $48\times48\times48$ ((a), (c), (e); LES395) and $64\times64\times64$ ((b), (d), (f); LES395f). (a), (b) The mean streamwise velocity; (c), (d) root-mean-squared (rms) velocity fluctuations; (e), (f) the mean Reynolds shear stress $\left<\mybar{u}'\mybar{v}'\right>$, where $\left< \cdot \right>$ denotes averaging over the $x-z$ plane and time. $\bigcirc$, fDNS; +, DSM; $\nabla$, no-SGS; {\color{red}\sampleline{}}, SL-180C; {\color{green}\sampleline{dash pattern=on 6pt off 2pt}}, S-180C; {\color{blue}\sampleline{dash pattern=on .7em off .2em on .2em off .2em}}, SL-106H180C; {\color{black}\sampleline{dash pattern=on 6pt off 2pt}}, the law of the wall $\left<\mybar{u}\right>/u_{\tau} = 0.41^{-1}\log{y^{+}}+5.2$.}
    \label{fig:channel_LES395}
\end{figure}

Figure~\ref{fig:channel_LES180} shows results of the LES180 case. DSM overpredicts the mean streamwise velocity and the streamwise velocity fluctuations $\mybar{u}'_{rms}$, while underpredicts the wall-normal velocity fluctuations $\mybar{v}'_{rms}$ and the spanwise velocity fluctuations $\mybar{w}'_{rms}$. DSM also shows slight underprediction of the mean Reynolds shear stress at $y^+<30$. LES without a SGS model (no-SGS) underpredicts the mean streamwise velocity and overpredicts the Reynolds shear stress. However, SL-180C more accurately predicts the mean velocity, root-mean-squared (rms) velocity fluctuations, and the mean Reynolds shear stress than DSM and no-SGS. S-180C performs similar to SL-180C in terms of the velocity fluctuations and the Reynolds shear stress. However, S-180C shows slight underprediction of the mean streamwise velocity, compared with SL-180C. SL-106H180C shows almost identical performance to SL-180C, while showing slight better prediction of the mean Reynolds shear stress at $y^+>40$. Interestingly, SL-106H is stable and accurately predicts the velocity fluctuations, despite it trained only for homogeneous isotropic turbulence. The Reynolds shear stress of SL-106H shows similar error near $y^+=25$ and smaller error at $y^+>45$, compared to that of DSM. However, SL-106H underpredicts the mean streamwise velocity, which is similar to the result of no-SGS. SL-106H may inaccurately predict the SGS stress at wall-bounded flow since the mean of SGS shear stress (off-diagonal components) vanishes~\citep{kang2003} in homogeneous isotropic turbulence. Therefore, training an ANN-SGS model for turbulent channel flow dataset seems to be essential to accurately predict the mean streamwise velocity.

Figure~\ref{fig:channel_LES180cf} shows results of LES180c and LES180f cases to test the performance of ANN-SGS models on untrained grid resolution. For coarser grid resolution (LES180c), DSM shows larger errors in prediction of the mean velocity and velocity fluctuations. Although DSM provides more accurate prediction of the Reynolds shear stress than SL-180C and SL-106H180C (figure~\ref{fig:channel_LES180cf}(e)), the mean streamwise velocity and the velocity fluctuations predicted by SL-180C and SL-106H180C are significantly better than those of DSM (figures~\ref{fig:channel_LES180cf}(a) and (c)).

For finer grid resolution (LES180f), DSM slightly overpredicts the mean streamwise velocity.  DSM also shows a larger error than SL-180C and SL-106H180C in the velocity fluctuations and the Reynolds shear stress. In contrast, all results of SL-180C and SL-106H180C are in excellent agreement with those of fDNS (figures~\ref{fig:channel_LES180cf}(b), (d) and (f)). SL-180C and SL-106H180C perform better than DSM regardless of grid resolution.

The developed ANN-SGS models are tested for LESs at a higher Reynolds number $Re_{\tau}$ of 395, and results are shown in figure~\ref{fig:channel_LES395}. DSM overpredicts the mean streamwise velocity at LES395 as shown in figure~\ref{fig:channel_LES395}(a). Although DSM shows a smaller error at LES395f than LES395, DSM shows slight overprediction of the mean streamwise velocity (figure~\ref{fig:channel_LES395}(b)). In contrast, S-180C underpredicts the mean streamwise velocity for both LES395 and LES395f cases, and the error becomes larger at finer grid resolution (figures~\ref{fig:channel_LES395}(a) and (b)). On the other hand, the mean velocity profiles of SL-180C and SL-106H180C are more accurate than those of DSM and S-180C. In addition, the mean velocity profiles of SL-180C and SL-106H180C are less sensitive to grid resolution compared to DSM and S-180C. 

\begin{figure}
	\centering
	\begin{subfigure}[b]{0.49\textwidth}
		\includegraphics[width=\textwidth]{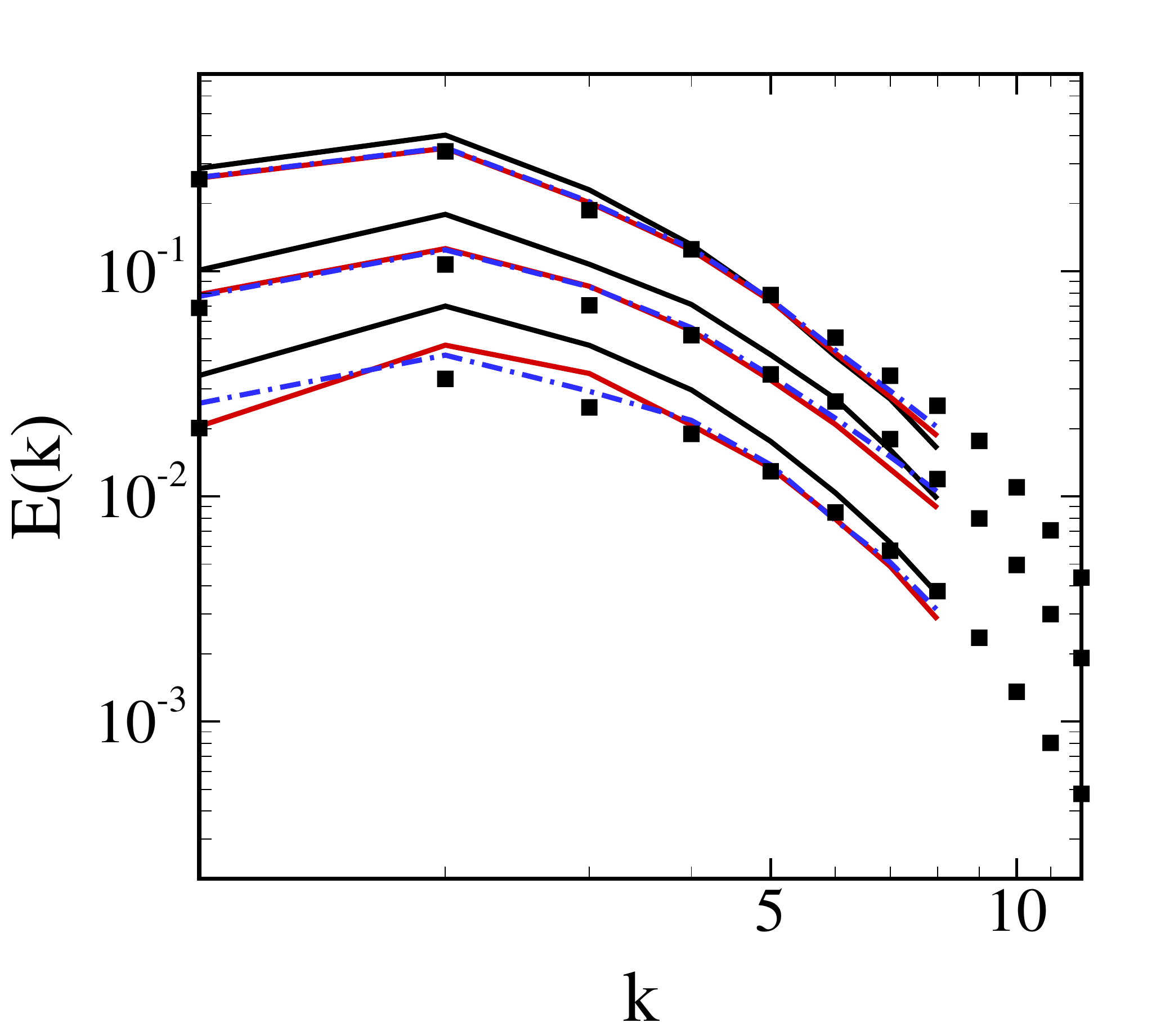}
		\caption{}
		\label{}
	\end{subfigure}
	\begin{subfigure}[b]{0.49\textwidth}
		\includegraphics[width=\textwidth]{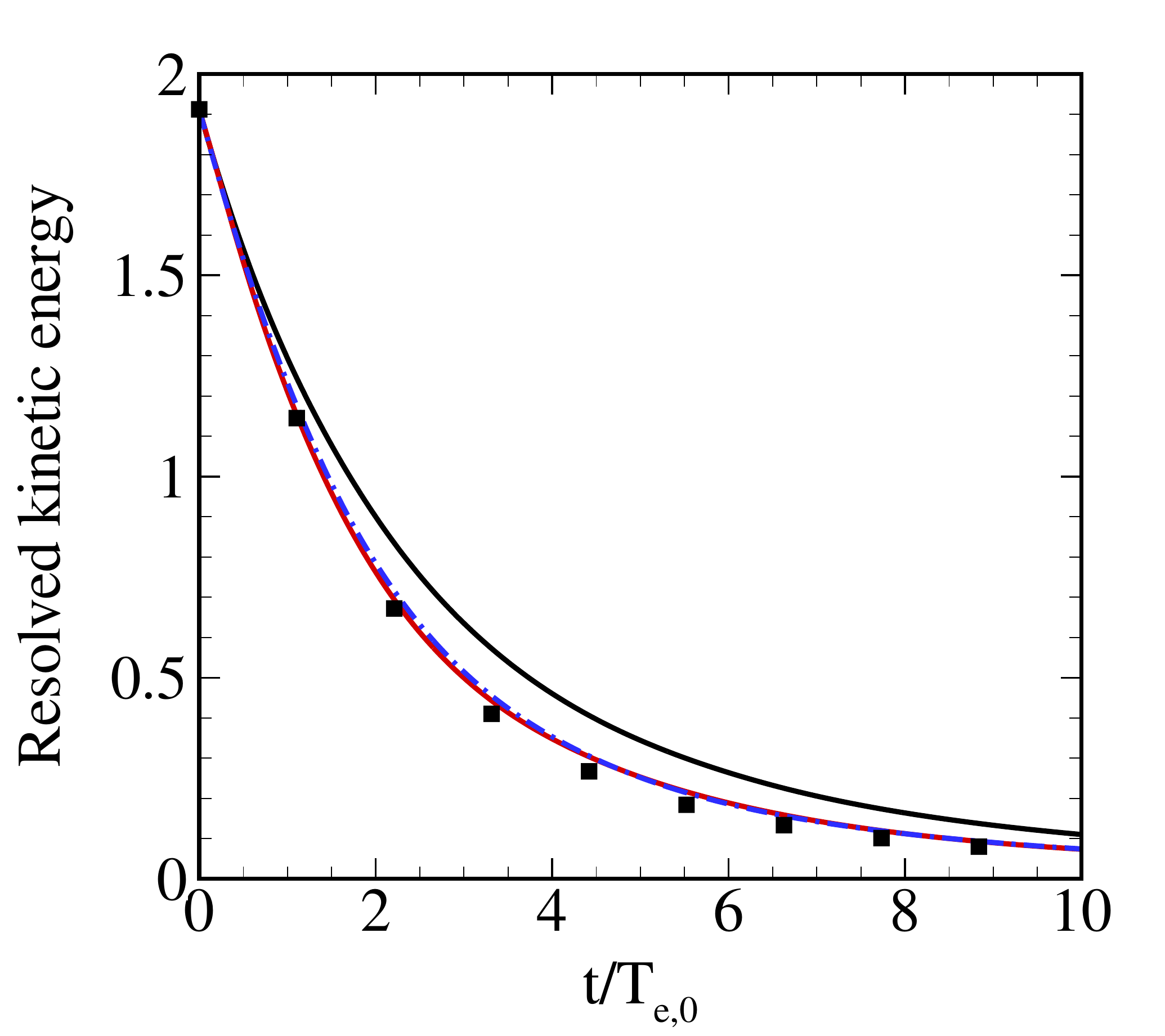}
		\caption{}
		\label{}
	\end{subfigure}
	\caption{Results from fDNS and LESs of decaying isotropic turbulence at the initial Reynolds number $Re_\lambda$ of 106 with grid resolution of $48^3$. (a) Energy spectra at $t/T_{e,0} = 1.1$, $3.3$ and $6.6$; (b) temporal evolution of the resolved kinetic energy. $\blacksquare$, fDNS; {\color{black}\sampleline{}}, DSM; {\color{blue}\sampleline{dash pattern=on .7em off .2em on .2em off .2em}}, SL-106H180C; {\color{red}\sampleline{}}, SL-106H.}
    \label{fig:DHIT106ir_channel}
\end{figure}

In figures~\ref{fig:channel_LES395}(c) and (d), DSM overpredicts $\mybar{u}'_{rms}$ while it underpredicts $\mybar{v}'_{rms}$ and $\mybar{w}'_{rms}$. In contrast, S-180C, SL-180C, and SL-106H180C show more accurate predictions of $\mybar{u}'_{rms}$, $\mybar{v}'_{rms}$ and $\mybar{w}'_{rms}$ than DSM. In figures~\ref{fig:channel_LES395}(e) and (f), DSM and S-180C similarly overpredict the Reynolds shear stress, while the Reynolds shear stress of SL-180C show good agreement with that of fDNS. As SL-180C provides more accurate prediction of the mean streamwise velocity and the Reynolds shear stress than S-180C, it can be concluded that the use of $L_{ij}$ in addition to the $\overline{S}_{ij}$ as an input improves the performance of the ANN-SGS  model for turbulent channel flow, similarly to the results of homogeneous isotropic turbulence. 
In LESs of turbulent channel flow (figures~\ref{fig:channel_LES180}, \ref{fig:channel_LES180cf} and~\ref{fig:channel_LES395}), SL-106H180C shows similar performance to SL-180C, especially in predicting the mean velocity and the velocity fluctuations. 

Additionally, it is tested whether SL-106H180C provides accurate solutions for LES of decaying isotropic turbulence, and the results are shown in figure~\ref{fig:DHIT106ir_channel}. SL-106H180C shows almost identical performance to SL-106H in prediction of the energy spectra and the resolved kinetic energy. Therefore, the present ANN-SGS mixed model which is trained with isotropic turbulence and turbulent channel flow at a certain Reynolds number and grid resolution conditions (SL-106H180C) is found to be capable of accurately and stably predicting both turbulent channel flow and isotropic turbulence flow at untrained Reynolds number and grid resolution conditions.

\section{Conclusions}\label{sec:Conclusion}

In the present study, an ANN-based mixed SGS model applicable to LES of turbulent flow under untrained Reynolds number and grid resolution conditions has been developed. The predictive capabilities of two different ANN-SGS models which use either the resolved strain-rate tensor ($\overline{S}_{ij}$) as the only input (ANN-SGS model) or the combination of the resolved strain-rate tensor and the resolved stress tensor ($\overline{S}_{ij}$ and $L_{ij}$) as the inputs (ANN-SGS mixed model) to predict the six components of the SGS stress tensor have been examined for forced and decaying isotropic turbulence and turbulent flow through a channel.

From \emph{a-priori} tests with forced isotropic turbulence, it was found that the ANN-SGS model with the resolved strain-rate tensor as the only input predicts the SGS stress which is nearly aligned with the given input strain-rate tensor instead of the true SGS stress. In addition, in an \emph{a-posteriori} test for the same flow, the ANN-SGS model showed almost identical performance to that of the algebraic dynamic Smagorinsky model in terms of the energy spectrum and PDFs of the SGS dissipation and the SGS stress. To improve the accuracy of the predicted SGS stress as well as its misalignment  to the true SGS stress, the ANN-SGS mixed model has been devised using the resolved stress tensor as an additional input based on the concept of the algebraic dynamic mixed SGS models.

The ANN-SGS mixed model has been found to predict the SGS stress with higher correlation coefficients to the true SGS stress and to provide more accurate PDFs of the SGS dissipation and SGS stress than the ANN-SGS model. In addition, the ANN-SGS mixed model has been found to predict the SGS stress with closer alignment with the true SGS stress than the ANN-SGS model. At the same time, the present ANN-SGS mixed model has been found to perform better than the algebraic dynamic Smagorinsky model and the ANN-SGS model in an \emph{a-posteriori} test with forced isotropic turbulence in terms of the energy spectrum and the PDF of the SGS dissipation.

The present ANN-SGS mixed model also has additional advantages over the algebraic dynamic SGS models in that the model predicts the SGS stress with less computational cost and does not necessitate \emph{ad-hoc} stabilisation procedures for numerical stability.
Compared with the algebraic dynamic mixed models~\citep{zang1993,vreman1994,anderson1999}, the present ANN-SGS mixed model has been found to  more accurately predict energy spectra and temporal evolution of the resolved kinetic energy of decaying isotropic turbulence without any \emph{ad-hoc} stabilisation procedures. The computational time required for the ANN-SGS mixed model to produce the SGS stress has been found to be notably less than that of the algebraic dynamic mixed models and even that of the dynamic Smagorinsky model.

For generalisation the ANN-SGS models to LES of untrained flows, it has been found that proper selection of normalisation factors such that distributions of the normalised input and output variables remain unchanged as the Reynolds number and grid resolution vary is crucial.   
From \emph{a priori} and \emph{a posteriori} studies with forced and  decaying homogeneous isotropic turbulence, it has been found that the averaged $L_2$ norm of the gradient model term can be the best scaling factor for normalisation of the output SGS stress tensor as $\tau_{ij}^* = \tau_{ij}/\left< \left| \tau^{grad}\right|\right>$ while the magnitudes of the resolved strain-rate tensor and the resolved stress tensor are sufficient for normalisation of the input variables as $\overline{S}_{ij}^* = \overline{S}_{ij} / \left< \left| \overline{S} \right|\right> $ and $\overline{L}_{ij}^* = \overline{L}_{ij} / \left< \left| \overline{L} \right|\right> $ since they maintain the most constant distributions of the input and the output variables for various Reynolds numbers and grid resolution.

The present ANN-SGS mixed model trained even with forced homogeneous isotropic turbulence only has been found to be successfully generalised for LES of untrained transient decaying homogeneous isotropic turbulence. The ANN-SGS mixed model has been found to predict energy spectra and temporal evolution of the resolved kinetic energy more accurately than DSM and the ANN-SGS model. Furthermore, the ANN-SGS mixed model has shown consistently good performance in LESs of decaying homogeneous isotropic turbulence with various untrained initial Reynolds numbers and grid resolution. 

Lastly, it has been investigated whether the developed ANN-SGS mixed model can be applied to wall-bounded turbulent flow. The present ANN-SGS mixed model has been found to be capable of predicting the SGS stress and thereby the mean and fluctuating velocity fields of turbulent channel flow. If the model was trained with turbulent channel flow at a certain Reynolds number and a certain grid resolution, the model accurately predicted turbulent channel flow even at untrained Reynolds number and on untrained grid resolution. The ANN-SGS mixed model trained with forced homogeneous isotopic turbulence only has been also found to be capable of predicting the Reynolds shear stress and velocity fluctuations favorably with marginal under-prediction of the mean velocity profile.
Interestingly, the ANN-SGS mixed model trained with both homogeneous isotropic turbulence and turbulent channel flow has provided consistently accurate and stable solutions for both types of flow. 

\section*{Acknowledgements} 

This work was supported by the National Research Foundation of Korea (NRF) under Grant Number NRF-2021R1A2C2092146 and the Samsung Research Funding Center of Samsung Electronics under Project Number SRFC-TB1703-51.

\section*{Declaration of interests} 

The authors report no conflict of interest.

\appendix

\section{Forcing schemes used at $\Rey_{\lambda}=164$}
\label{app:app1-1}

While \citet{langford1999} reported the use of a negative viscosity forcing for simulating forced homogeneous isotropic turbulence at $\Rey_{\lambda}=164$, both the deterministic forcing and the negative viscosity forcing are used in the present study. The energy spectrum reported in \citet{langford1999} is almost identical to that of the case in which the deterministic forcing is used as shown in figure~\ref{fig:FHIT_ek}(b). Furthermore, the value of $\Rey_{\lambda}$ calculated from the cases with the deterministic forcing and the negative viscosity forcing are found to be $165$ and $123$, respectively, where the former case is close to the reported value of $164$ by \citet{langford1999}.

While the major differences in the energy spectra obtained using different types of forcing schemes are observed to be in the forcing wavenumber range as shown in figure~\ref{fig:FHIT_ek}(b), the characteristics of each forcing scheme can be inferred from (\ref{eq:forcing_vis}) and (\ref{eq:forcing}). The deterministic forcing scheme (\ref{eq:forcing}) acting on each wavenumber shell is weighted by the Fourier coefficient of the velocity at each wavenumber shell. In forced homogeneous isotropic turbulence, higher energy is contained at lower wavenumbers; therefore, such weighting gives higher energy contents at lower wavenumbers in the forcing range which can be seen in the energy spectra of the cases with the deterministic forcing scheme in figures~\ref{fig:FHIT_ek}(b) and (c). The negative viscosity forcing (\ref{eq:forcing_vis}), on the other hand, weights relatively more on higher wavenumber shells in the forcing range due to $|\textbf{k}|^2$, leading to higher energy levels at higher forcing wavenumbers. This is consistent with the energy spectra in figures~\ref{fig:FHIT_ek}(a) and~\ref{fig:FHIT_ek}(b) in the forcing range. This trend was also observed in the energy spectra by \citet{jimenez93} in which a negative viscosity forcing was used.

In order to accurately reproduce the DNS results by \citet{langford1999}, the deterministic forcing scheme is adopted to simulate forced homogeneous isotropic turbulence at $\Rey_{\lambda}=164$.

\section{\label{app:subsec}Energy spectra at $\Rey_{\lambda}=286$}
\label{app:app1-2}
Although the overall energy spectra of the present DNS and those obtained by \citet{chumakov2008} are similar, a slight underprediction by the present DNS is observed. To investigate the reason for the difference, the energy dissipation rate $\varepsilon$ is estimated by 
\begin{equation}
	\varepsilon=\nu^3/\eta^4,
	\label{eq:eta_eqn}
\end{equation}
where $\eta$ is the Kolmogorov length scale. This allows us to estimate the energy dissipation rate for the DNS result of forced homogeneous isotropic turbulence, as $\eta$ is available in both the present study and study by~\citet{chumakov2008}. While it is expected for the well-converged statistics of forced homogeneous isotropic turbulence to have $\varepsilon$ close to the prescribed mean dissipation rate of $0.12$, $\varepsilon$ estimated from $\eta$ of the present DNS result and \citet{chumakov2008} are $0.12$ and $0.16$, respectively. Therefore, the DNS database used in the present study is considered to provide better converged statistics, which explains the difference.

\section{\emph{A posteriori} tests with different inputs}\label{sec:Appendix A}

\begin{figure}
	\centering
	\begin{subfigure}[b]{0.49\textwidth}
		\includegraphics[width=\textwidth]{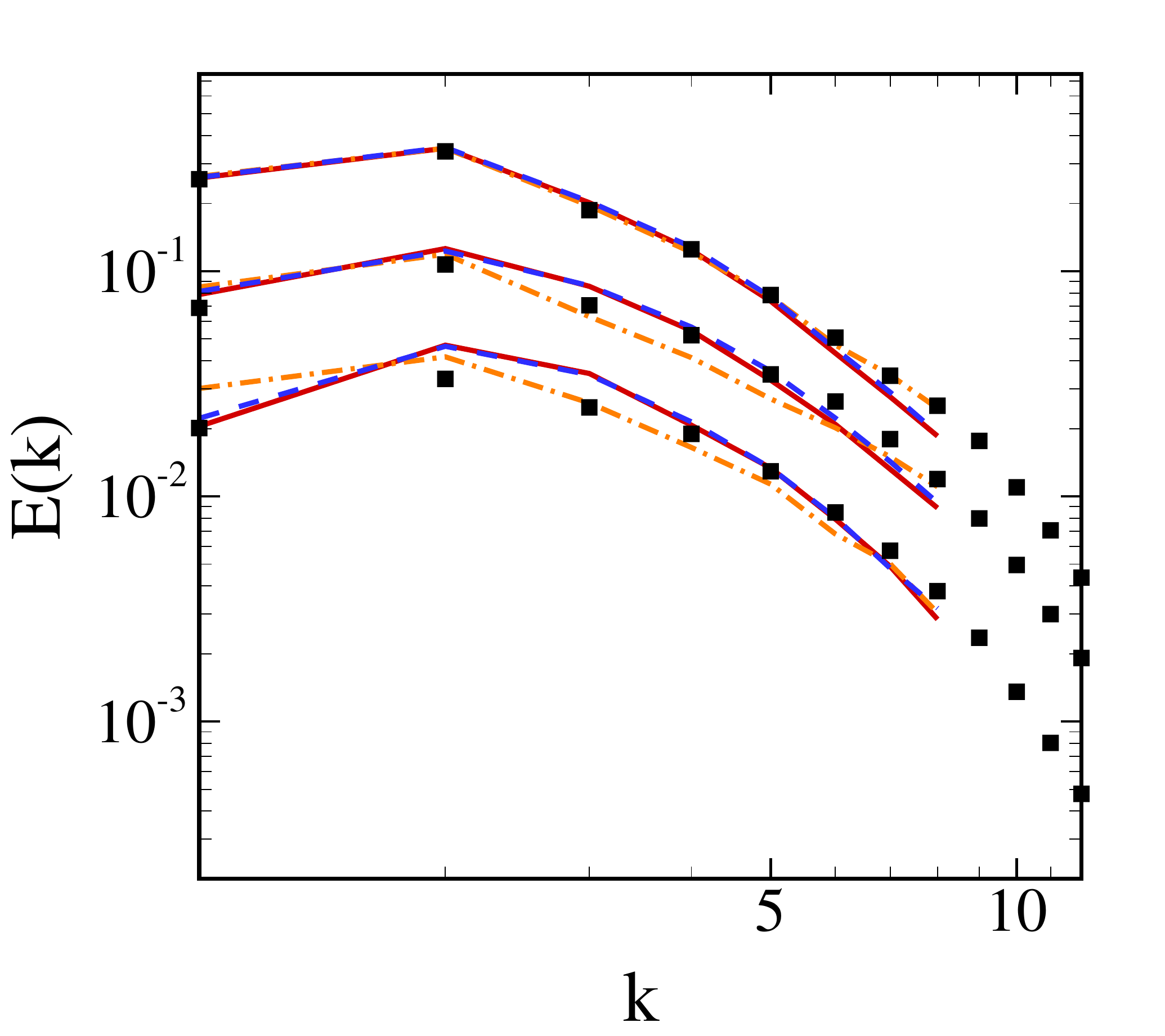}
		\caption{}
		\label{}
	\end{subfigure}
	\begin{subfigure}[b]{0.49\textwidth}
		\includegraphics[width=\textwidth]{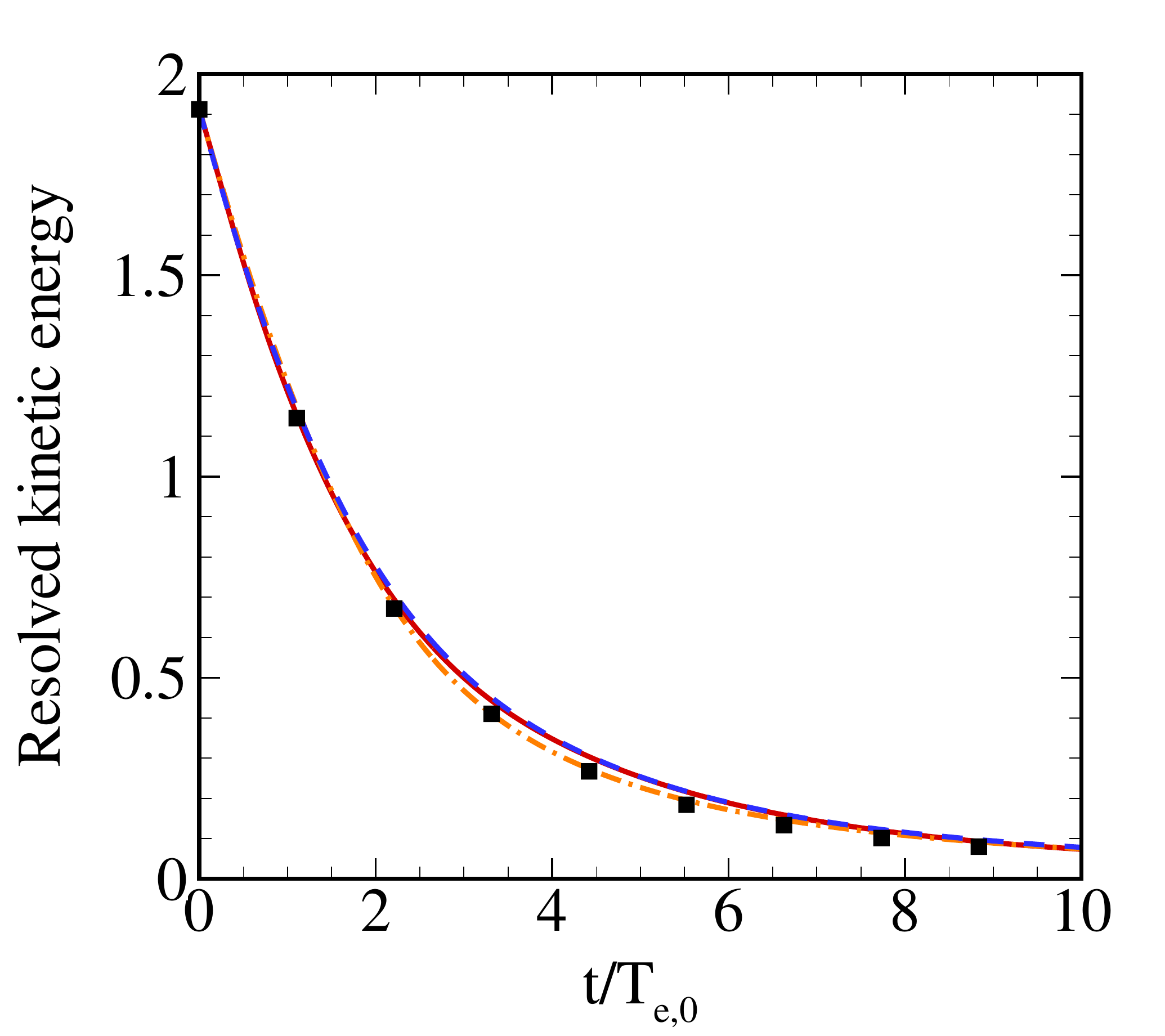}
		\caption{}
		\label{}
	\end{subfigure}
	\caption{Results from fDNS and LESs of decaying isotropic turbulence at the initial Reynolds number $Re_\lambda$ of 106 with grid resolution of $48^3$. (a) Energy spectra at $t/T_{e,0} = 1.1$, $3.3$ and $6.6$; (b) temporal evolution of the resolved kinetic energy. $\blacksquare$, fDNS; {\color{red}\sampleline{}}, ANN-SGS model with $\overline{S}_{ij}$ and $L_{ij}$; {\color{blue}\sampleline{dash pattern=on 6pt off 2pt}}, ANN-SGS model with $\overline{S}_{ij}$ and the modified Leonard term; {\color{YellowOrange}\sampleline{dash pattern=on .7em off .2em on .2em off .2em}}, ANN-SGS model with $\overline{S}_{ij}$ and the gradient model term.}
    \label{fig:appenA_ir}
\end{figure}

\begin{figure}
	\centering
	\begin{subfigure}[b]{0.49\textwidth}
		\includegraphics[width=\textwidth]{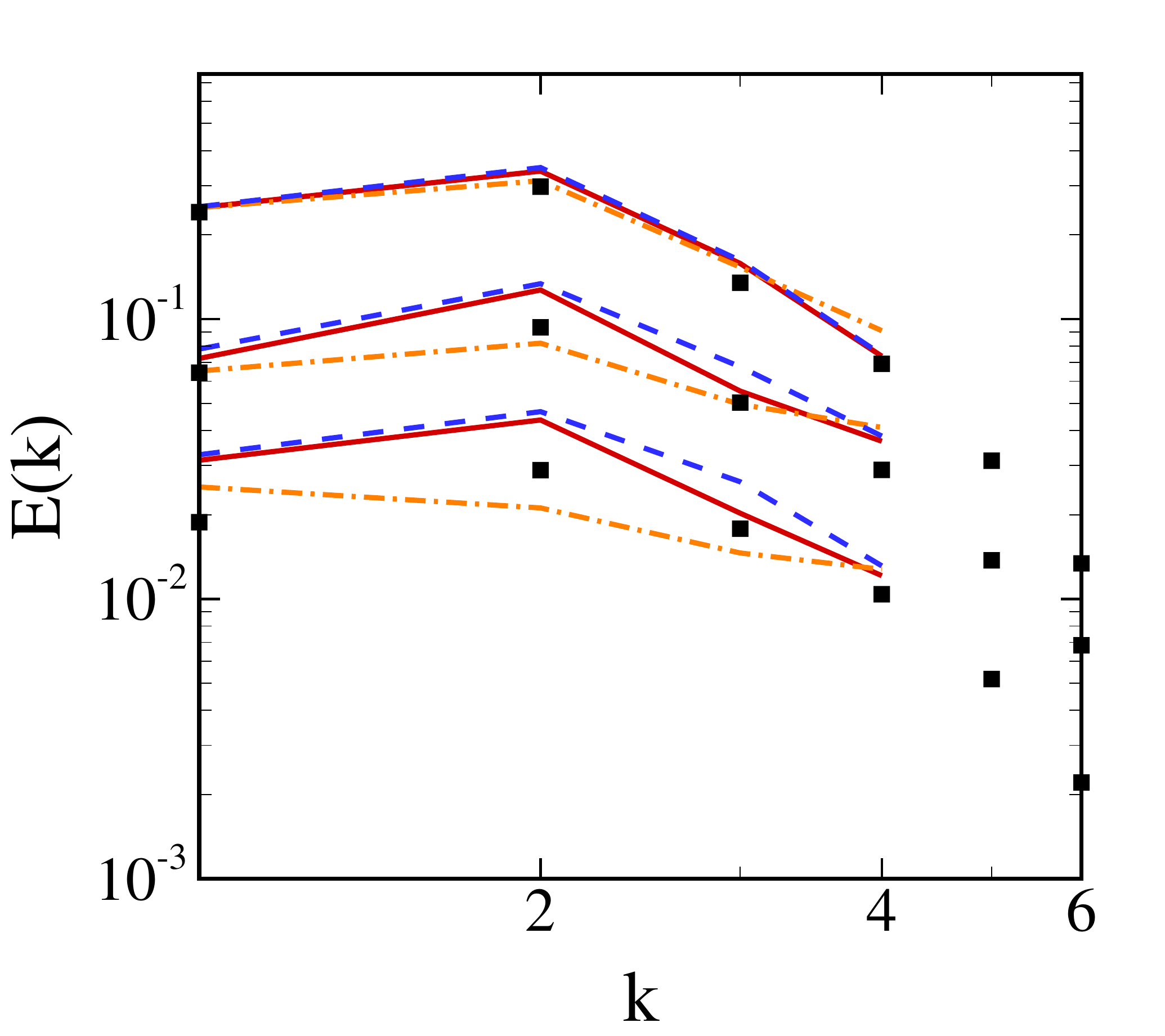}
		\caption{}
		\label{}
	\end{subfigure}
	\begin{subfigure}[b]{0.49\textwidth}
		\includegraphics[width=\textwidth]{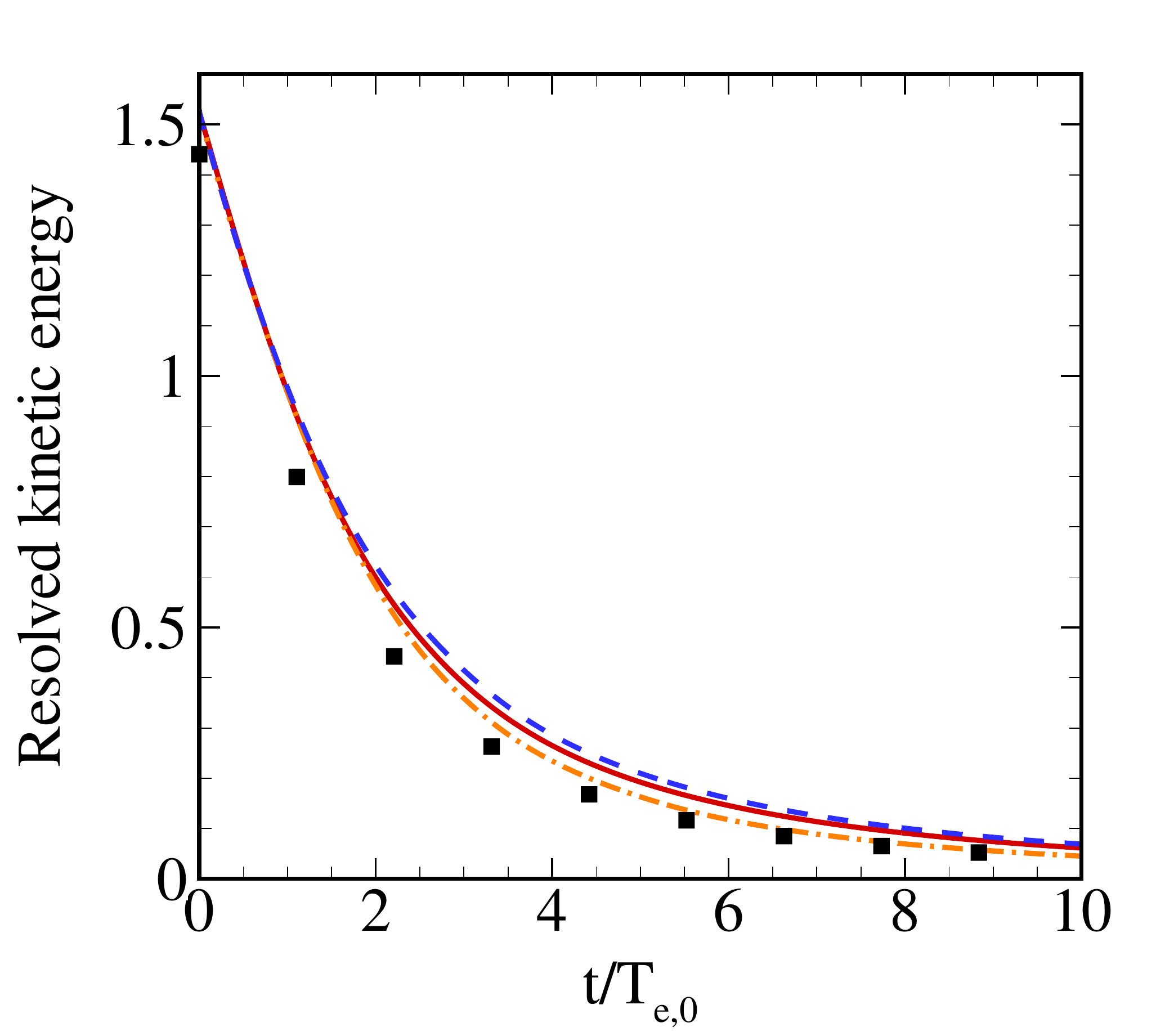}
		\caption{}
		\label{}
	\end{subfigure}
	\caption{Results from fDNS and LESs of decaying isotropic turbulence at the initial Reynolds number $Re_\lambda$ of 106 with grid resolution of $24^3$ (two-times coarser resolution than that of training data). (a) Energy spectra at $t/T_{e,0} = 1.1$, $3.3$ and $6.6$; (b) temporal evolution of the resolved kinetic energy. $\blacksquare$, fDNS; {\color{red}\sampleline{}}, ANN-SGS model with $\overline{S}_{ij}$ and $L_{ij}$; {\color{blue}\sampleline{dash pattern=on 6pt off 2pt}}, ANN-SGS model with $\overline{S}_{ij}$ and the modified Leonard term; {\color{YellowOrange}\sampleline{dash pattern=on .7em off .2em on .2em off .2em}}, ANN-SGS model with $\overline{S}_{ij}$ and the gradient model term.}
    \label{fig:appenA_c}
\end{figure}

In the present study, consideration of the resolved stress $L_{ij}$ as an input in addition to the resolved strain-rate tensor $\overline{S}_{ij}$ is found to significantly improve the performance of the ANN-SGS model. The alternative terms of the resolved stress $L_{ij}$ can be found in literature on various mixed SGS models~\citep{salvetti1995,bardinaphd,liu1994,anderson1999}. Thus, in this section, the modified Leonard term $(\mathcal{L}_{ij}^m = \overline{\overline{u}_i \overline{u}_j} - \overline{\overline{u}}_i \overline{\overline{u}}_j)$~\citep{salvetti1995,bardinaphd,germano86}, the resolved stress $(L_{ij} =\myhat{\mybar{u}_i\mybar{u}_j}-\myhat{\mybar{u}}_i\myhat{\mybar{u}}_j)$~\citep{liu1994}, and the gradient model term $(\frac{1}{12}\mybar{\Delta}^2\frac{\partial \overline{u}_i}{\partial x_k}\frac{\partial \overline{u}_j}{\partial x_k})$ from the Clark model~\citep{anderson1999} are considered as inputs to the ANN, in addition to the resolved strain-rate tensor $\overline{S}_{ij}$. The performance of ANN-SGS models are compared by conducting LESs of decaying isotropic turbulence with grid resolution of $48^3$ and $24^3$. 

Figure~\ref{fig:appenA_ir} shows results from LES of decaying homogeneous isotropic turbulence with grid resolution of $48^3$. The ANN-SGS model with $\overline{S}_{ij}$ and the gradient model term as inputs is found to predict the temporal evolution of the resolved kinetic energy most accurately, but shows the largest error in the prediction of energy spectra at $k = 1$. On the other hand, the performance of the ANN-SGS model with $\overline{S}_{ij}$ and the modified Leonard term as inputs is found to be similar to that of the ANN-SGS model with $\overline{S}_{ij}$ and $L_{ij}$ as inputs in predicting the energy spectra and the decaying kinetic energy. 

Figure~\ref{fig:appenA_c} shows results from LES of decaying homogeneous isotropic turbulence with grid resolution of $24^3$. Temporal evolution of the resolved kinetic energy predicted by the ANN-SGS model with $\overline{S}_{ij}$ and the gradient model term as inputs shows the smallest error (see figure~\ref{fig:appenA_c}(b)). However, this model has significantly large errors in the energy spectra at $k=2$ (see figure~\ref{fig:appenA_c}(a)). Thus, ANN-SGS models are more robust to the change in grid resolution when the resolved stress $L_{ij}$ or the modified Leonard term $\mathcal{L}_{ij}^m$ is provided as an input in addition to $\overline{S}_{ij}$. In the present study, the resolved stress $L_{ij}$ and $\overline{S}_{ij}$ are considered as the best combination of input variables for the ANN-SGS model, as it shows the best performance and ensures robustness.

\section{\emph{A posteriori} test of decaying isotropic turbulence with a random-phase initial condition}\label{sec:Appendix C}

\begin{figure}
	\centering
	\begin{subfigure}[b]{0.49\textwidth}
		\includegraphics[width=\textwidth]{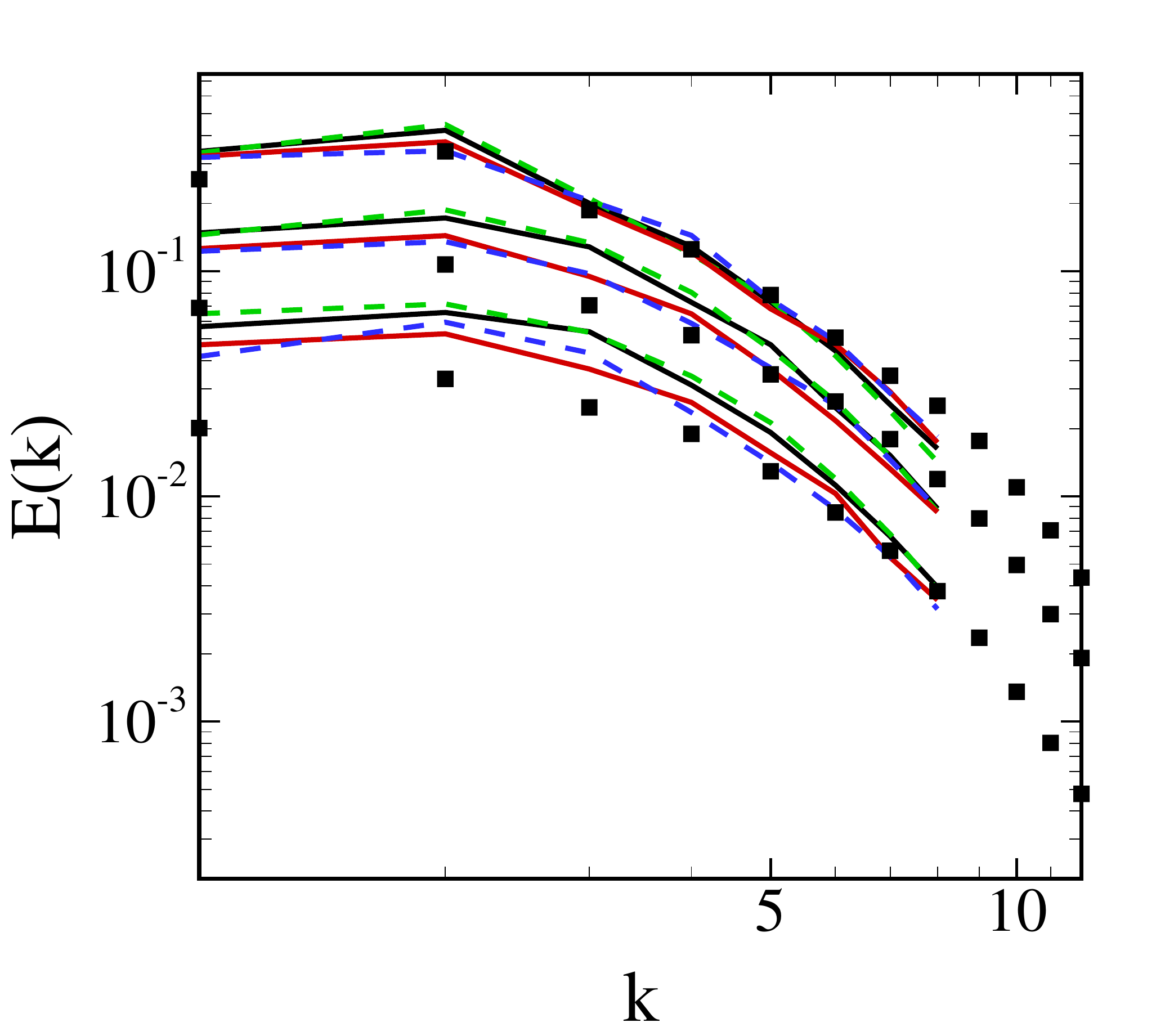}
		\caption{}
		\label{}
	\end{subfigure}
	\begin{subfigure}[b]{0.49\textwidth}
		\includegraphics[width=\textwidth]{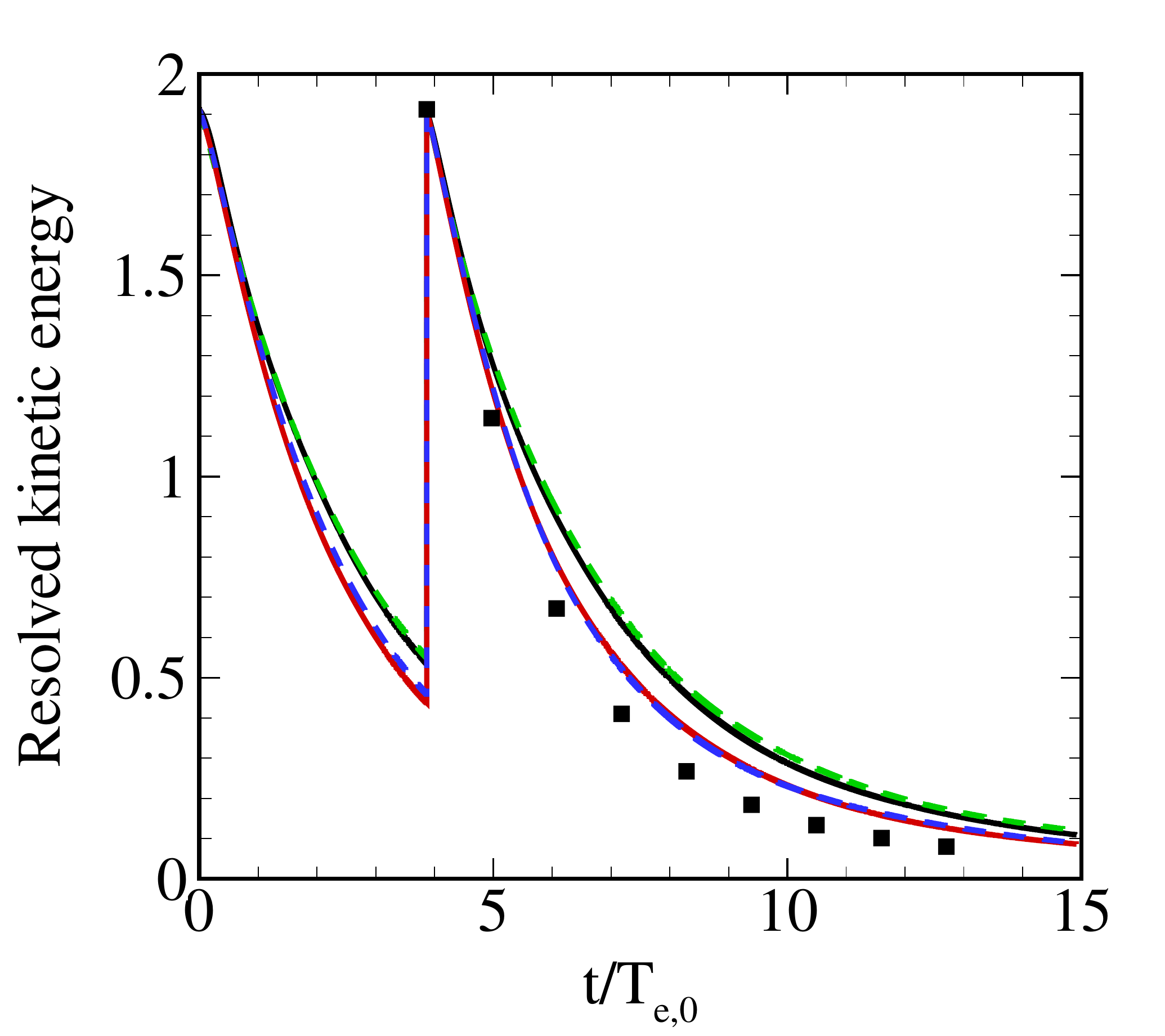}
		\caption{}
		\label{}
	\end{subfigure}
	\caption{Results from fDNS and LESs of decaying isotropic turbulence at the initial Reynolds number $Re_\lambda$ of 106 with a random-phase initial field. (a) Energy spectra at $t/T_{e,0} = 1.1$, $3.3$ and $6.6$; (b) temporal evolution of the resolved kinetic energy. $\blacksquare$, fDNS; {\color{black}\sampleline{}}, DSM; {\color{red}\sampleline{}}, SL-106H; {\color{blue}\sampleline{dash pattern=on 6pt off 2pt}}, SL-286H; {\color{green}\sampleline{dash pattern=on 6pt off 2pt}}, S-106H.
	}
    \label{fig:DHIT106_random_phase}
\end{figure}

To assess the robustness of the present ANN-SGS models to the initial conditions, LES of decaying isotropic turbulence with a random-phase initial condition is performed using the rescaling method of \citet{kang2003}. The three-dimensional energy spectrum from a fully converged flow field of forced isotropic turbulence at $Re_\lambda = 106$ is given as the initial energy distribution. The simulation is started with random-phase Fourier modes until $t/T_{e,0} = 3.87$ when the derivative skewness becomes steady. Thereafter, the velocity field is re-scaled to have the initial energy spectrum of fDNS~\citep{kang2003}. Results with SL-106H, SL-286H, and S-106H are compared with those of DSM and fDNS.

Despite the random-phase initial field as well as the transient development of the flow is not included in the training data, the performance and stability of ANN-SGS models are maintained. SL-106H and SL-286H predict  the energy spectra more accurately, whereas DSM and S-106H show larger errors at $k \leq 5$, as shown in figure~\ref{fig:DHIT106_random_phase}(a). Similarly, SL-106H and SL-286H predict the temporal evolution of the resolved kinetic energy more accurately than DSM and S-106H, as shown in figure~\ref{fig:DHIT106_random_phase}(b).

\bibliographystyle{jfm}
\bibliography{jfm-instructions}

\end{document}